\documentclass[fleqn,usenatbib]{mnras}

\usepackage{threeparttable}
\usepackage{multirow}
\usepackage{hyperref}
\usepackage{lineno}
\usepackage{graphicx}
\usepackage{xcolor}
\usepackage{pdflscape}
\usepackage[utf8]{inputenc} 
\usepackage[T1]{fontenc}    
\usepackage{caption}
\usepackage{subcaption}
\usepackage{url}
\usepackage{amsmath}	
\usepackage{txfonts}
\usepackage{setspace}
\usepackage[normalem]{ulem}

\newcommand\issuenord[2]{ \textbf{[}{\color{orange} ISSUE for #1} \textbf{||} \textbf{#2}\textbf{]}}

\def \lenspop {{\texttt LensPop}}

\def \challengeabbreviation {SL Challenge}
\def \mysep {\sep} 
\def \mysep { -- } 

\def \thetae {\theta_{\rm E}}
\def \zsource     {z_{\rm s}}
\def \zlens       {z_{\rm l}}

\def \sigmav      {\sigma_{\rm v}}
\def \gmag {g}
\def \rmag {r}
\def \imag {i}

\def \lt {<}

\def \bphi{\bar{\phi}}

\def \beq {\begin{equation}}
\def \eeq {\end{equation}}
\def \beqn {\begin{equation*}}
\def \eeqn {\end{equation*}}

\def\aap{Astr. \& Astroph.}
\def\aaps{Astr. \& Astroph., Supp.}

\def\aj{Astron.\ J.}

\def\apj{Astrophys.\ J.}
\def\apjl{Astrophys.\ J.\ Lett.}
\def\apjs{Astrophys.\ J.\ Suppl.}

\def\mnras{Mon.\ Not.\ Roy.\ Astron.\ Soc.}

\def\physrep{Phys.\ Rep.}

\def \figurePredictionsLenspopCorrelationsInternalParameters
{
\begin{figure*}
   \centering
   \includegraphics[width=0.9\textwidth,keepaspectratio]{Figures/data_analysis/deviation_corr.png}
      \caption{
      \label{fig:PredictionsLenspopCorrelationsInternalParameters}
      The fractional deviation from true values in
      source redshift $\zsource$ from as function of Einstein Radius $\thetae$ (left) and  lens redshift $\zlens$ as function of velocity dispersion $\sigmav$ (right). 
      The green shadows are the percentile corresponding to $68\%$, $95\%$, and $99\%$ confidence intervals, and the blue shadows are the $68\%$ confidence on scatter of medians predicted values. 
      The dashed lines represent the $0.0$ (black), $0.1$ (red), $0.2$ (magenta) deviations.
      \issuenord{Jason, Clecio}{I'm trying to keep the discussion of confidence intervals consistent, and putting in percentages in stead of sigma's. Does the language on "scatter" make sense here?}
      }   
   \end{figure*}
}

\def \figurePredictionsLenspopExternalParametersColor
{
\begin{figure*}
   \centering
   \includegraphics[width=0.9\textwidth,keepaspectratio]{Figures/data_analysis/frac_dev_Colors.png}
      \caption{
      \label{fig:PredictionsLenspopExternalParametersColor}
      $\gmag-\rmag$ color (top left) distribution and  fractional deviation from true values as function of $\rmag-\imag$ color for $\theta_E$ (bottom left), $\zlens$ (top right),$\sigmav$ (bottom right). 
      The green shadows are the percentile corresponding to $68\%$, $95\%$, and $99\%$ confidence intervals.  
      The dashed lines represent the $0.0$ (black), $0.1$ (red), $0.2$ (magenta) deviations.
      }   
   \end{figure*}
}

\def \figurePredictionsLenspopExternalParametersSNR
{
\begin{figure*}
   \centering
   \includegraphics[width=0.9\textwidth,keepaspectratio]{Figures/data_analysis/frac_dev_SNR.png}
      \caption{
      \label{fig:PredictionsLenspopExternalParametersSNR}
      Signal to Noise distribution (top left) and  fractional deviation from true values as function of signal to noise for $\theta_E$ (bottom left), $\zlens$ (top right),$\sigmav$ (bottom right). 
      The green shadows are the percentile corresponding to $68\%$, $95\%$, and $99\%$ confidence intervals.  
      The dashed lines represent the $0.0$ (black), $0.1$ (red), $0.2$ (magenta) deviations.
      }   
   \end{figure*}
}

\def \figurePredictionsLenspopComparisonDataSplit
{
\begin{figure*}
\centering
\includegraphics[width=0.9\textwidth]{Figures/data_analysis/train9010.png}
    \caption{
    \label{fig:PredictionsLenspopComparisonDataSplit}
    True vs predicted values of $\thetae$ and the residuals for a $90\%/10\%$ (left) $10\%/90\%$ training/test split.
    The green shadows are the percentile corresponding to $68\%$, $95\%$, and $99\%$ confidence intervals, the blue shadows are the $68\%$-confidence interval scatter of the median predicted values.
    \issuenord{Jason, Clecio}{I'm trying to keep the discussion of confidence intervals consistent, and putting in percentages in stead of sigma's. Does the language on "scatter" make sense here?}
    }
\end{figure*}
}

\def \figurePredictionsLenspopComparisonMultipleVsSingle
{
\begin{figure}
   \centering
   \includegraphics[width=0.5\textwidth,keepaspectratio]{Figures/data_analysis/sum_plot_multivsone.png}
      \caption{
      \label{fig:PredictionsLenspopComparisonMultipleVsSingle}
      The fractional deviation on the full testing sample using models that predicts a single parameter (red circles) and the model that predicts all four parameters at once (Blue star), where the dashed lines represent the $0.0$ (black), $0.1$ (red), $0.2$ (magenta) deviations (top). 
      The percentage of objects with more than $0.15$ fractional deviation (bottom).
      }   
   \end{figure}
}

\def \figurePredictionsSLChallengeCorrelationsInputOutputBaselineAndComparisontoTransferLearning
{
\begin{figure*}
   \centering
   \includegraphics[width=0.9\textwidth,keepaspectratio]{Figures/data_analysis/50vs100_500vs500.png}
      \caption{
      \label{fig:PredictionsSLChallengeCorrelationsInputOutputBaselineTransferLearning}
      Model efficacy for transfer learning experiments on the lensing system parameter, the Einstein radius $\thetae$.
      Comparisons of true vs predicted values of the Einstein radius $\thetae$ for different models over the range of $\thetae$ in the \lenspop\ data set (\S\ref{sec:data}). 
      Lines represent median values, while contours represent 68\% confidence intervals. 
      \textbf{Left:} A model that was trained on \lenspop\ data and retrained with 50 images from the \challengeabbreviation\ data set (red) significantly outperforms a model that was trained from scratch on 100 images from the \challengeabbreviation\ data set (blue), producing tighter constraints on $\thetae$ across the entire range of $\thetae$, but most notably for lower values of $\thetae$. 
      \textbf{Right:} The same models, now trained on 500 images from the \challengeabbreviation\ data set. 
      Both models improve,  but the pre-trained model (red) still significantly outperforms the model that was trained from scratch (blue), producing predictions that are much more accurate, demonstrating the potential of transfer learning in lens analyses.
      }
         
   \end{figure*}
}

\def \figurePredictionsSLChallengeCorrelationsInputOutputBaselineAndComparisontoTransferLearning
{
\begin{figure*}
   \centering
   \includegraphics[width=0.9\textwidth,keepaspectratio]{Figures/data_analysis/50-500-3000_4000vs4000.png}
      \caption{
      \label{fig:PredictionsSLChallengeCorrelationsInputOutputBaselineAndComparisontoTransferLearningB}
      Comparisons of true vs predicted values of the Einstein radius $\thetae$ for different models over the range of $\thetae$ in the \challengeabbreviation\ data set.
      Lines represent median values, while contours represent 68\% confidence intervals. 
      \textbf{Left:} Models that were pre-trained on \lenspop\ data and retrained on 50 (red), 500 (blue) and 3000 (green) images from the \challengeabbreviation\ data set. 
      The model that was retrained on 50 images was unable to accurately predict $\thetae > 3''$ due to a lack of enough training images. 
      Models that were trained on 500 and more images were able to constrain $\thetae > 3''$ much better. 
      \textbf{Right:} Comparing a model that was pre-trained on \lenspop\ data and retrained on 4000 images from the \challengeabbreviation\ data set (red) with a model that was trained from scratch on that same data set (blue). 
      Both models performed similarly for $\thetae > 3''$; however, the model that was pre-trained on \lenspop\ data performed noticeably better at lower values of $\thetae$, producing tighter constraints for $\thetae < 1.8''$.
      }
         
   \end{figure*}
}

\usepackage[colorinlistoftodos,prependcaption,textsize=tiny]{todonotes}

\title[Galaxy morphologies in S-PLUS]{An Extended Catalogue of galaxy morphology  using Deep Learning in Southern Photometric Local Universe Survey Data Release 3} 
\author[C.R. Bom \& A. Cortesi et al.]{
C. R. Bom,$^{1, 2}$\thanks{E-mail: debom@cbpf.br (CRB)}
A. Cortesi$^{3}$,
U. Ribeiro$^{1}$,
L. O. Dias$^{1}$,
K. Kelkar$^{4}$,
A.V. Smith Castelli$^{5,6}$, 
\newauthor
L. Santana-Silva$^{7}$,
V. Silva $^{3}$,
T. S. Gon\c{c}alves$^{3}$,
L. R. Abramo$^{8}$,
E. V. R. Lima$^{9}$,
\newauthor
F. Almeida-Fernandes$^{9}$,
L. Espinosa$^{9}$,
L. Li$^{9}$, 
M. L. Buzzo$^{10}$,
C. Mendes de Oliveira$^{9}$,
\newauthor
L. Sodr\'{e} Jr.$^{9}$, 
A. Alvarez-Candal$^{11}$,
M. Grossi$^{3}$,
E. Telles$^{11}$,
S. Torres-Flores$^{12}$,
S. V. Werner$^{13}$,
\newauthor
A. Kanaan$^{14}$, 
T. Ribeiro$^{15}$,
W. Schoenell$^{16}$\\
$^{1}$ Centro Brasileiro de Pesquisas F\'isicas, Rua Dr. Xavier Sigaud 150, CEP 22290-180, Rio de Janeiro, RJ, Brazil\\
$^{2}$Centro Federal de Educa\c{c}\~ao Tecnol\'ogica Celso Suckow da Fonseca, Rodovia M\'ario Covas, lote J2, quadra J, CEP 23810-000,  Itagua\'i, RJ, Brazil\\
$^{3}$Valongo Observatory, Federal University of Rio de
Janeiro, Ladeira Pedro Antonio 43, Saude Rio de Janeiro,
RJ, 20080-090, Brazil\\
$^{4}$ Instituto de Física Y Astronomía  - U de Valparaíso \\
$^{5}$ Instituto de Astrof\'sica de La Plata, CONICET–UNLP, Paseo del Bosque s/n, B1900FWA, Argentina\\
$^{6}$ Facultad de Ciencias Astr\'onomicas y Geof\'isicas, UNLP, Paseo del Bosque s/n, B1900FWA, Argentina\\
$^{7}$NAT-Universidade Cruzeiro do Sul / Universidade Cidade de São Paulo, Rua Galvão Bueno, 868, 01506-000, São Paulo, SP, Brazil\\
$^{8}$Departamento de Física Matemática, Instituto de Física, Universidade de São Paulo, R. do Matão 1371, 05508-090, São Paulo, SP, Brazil \\
$^{9}$Universidade de S\~ao Paulo, IAG, Rua do Mato 1225, Sao
Paulo, SP, Brazil\\
$^{10}$Centre for Astrophysics and Supercomputing, Swinburne University, John Street, Hawthorn VIC 3122, Australia \\
$^{11}$Observatório Nacional R. Gen. Jos\'e Cristino, 77, 20921-400, Rio de Janeiro, Brazil \\
$^{12}$ Departamento de Astronomia, Universidad de La Serena, Av. Cisternas 1200, La Serena, Chile \\
$^{13}$School of Physics and Astronomy, University of Nottingham, Nottingham, NG7 2RD, UK \\
$^{14}$ Departamento de F\'isica, \\ Universidade Federal de Santa Catarina, Florian\'opolis, SC, 88040-900, Brazil \\
$^{15}$ Departamento de Astronomia, Instituto de F\'isica, Universidade Federal do Rio Grande do Sul (UFRGS), Av. Bento Goncalves 9500.\\
$^{16}$NOAO, P.O. Box 26732, Tucson, AZ 85726
}

\begin{document}
\maketitle




\begin{abstract}

The morphological diversity of galaxies is a relevant probe of galaxy evolution and cosmological structure formation. However, in large sky surveys, even the morphological classification of galaxies into two classes, like late-type (LT) and early-type (ET), still represents a significant challenge. 
In this work we present a Deep Learning (DL) based morphological catalog built from images obtained by the Southern Photometric Local Universe Survey (S-PLUS) Data Release 3 (DR3). 
Our DL method achieves an precision rate of 98.5$\%$ in accurately distinguishing between spiral, as part of the larger category of late type (LT) galaxies, and elliptical, belonging to  early type (ET) galaxies. Additionally, we have implemented a secondary classifier that evaluates the quality of each galaxy stamp, which allows to select only high-quality images when studying properties of galaxies on the basis of their DL morphology. From our LT/ET catalog of galaxies, we recover the expected color--magnitude diagram in which LT galaxies display bluer colors than ET ones. Furthermore, we also investigate the clustering of galaxies based on their morphology, along with their relationship to the surrounding environment. As a result, we deliver a full morphological catalog with $164314$ objects complete up to
$r_{petro}<18$, covering $\sim 1800$ deg$^2$, including a significant area of the Southern hemisphere that was not covered by previous morphology catalogues.

\end{abstract}


\begin{keywords} 
galaxies: fundamental parameters \mysep 
galaxies: structure \mysep 
techniques: image processing \mysep
catalogues 
\end{keywords}


\section{Introduction}
\label{sec:introduction}

Galaxy structure was one of the first properties of galaxies that was ever directly observed and studied. Initially thought to be `nebulae', it soon became evident that these objects showed distinct structural features like spiral arms or a smooth elliptical envelope (\citealp{Zwicky1940,Vaucouleurs1959,Herschel1864,vandenbergh1998}). Decades of studying galaxy shapes and structures thus resulted in several classification schemes, among which the `Hubble tuning fork' system of classifying galaxies based on their observed visual characteristics has been widely used. Collectively known as galaxy `morphologies', galaxies can broadly be divided into two main categories namely early and late type galaxies. Late type galaxies are formed by spiral (S) and irregular/peculiar (Irr) galaxies. The spiral galaxies' branch, bifurcates into barred and un-barred systems. The early type galaxies are composed of elliptical and lenticular galaxies. Elliptical galaxies display an increasing ellipticity, from round  (E0) to flat (E7) systems. Lenticular galaxies lie at the apex of the Hubble tuning fork due to their hybrid strucure, presenting a bulge and a disk, as spiral galaxies, but without spiral arms.

Such morphological diversity often reflects the presence of different and composite stellar populations \citep{Sanchez2007} and  kinematics \citep{Edelen1969,Wang2020}. For example, S galaxies are characterized by the presence of a star-forming disk with blue spiral arms, which indicate rotationally supported stellar kinematics. 
E galaxies have, in general, more smooth featureless morphologies resulting from a lack of star formation. E galaxies present a range of kinematic profiles, being the E0
pressure-supported systems or slow rotators, while intermediate elliptical galaxies (E1/E7) present and increasing contribution of rotation to the total kinematic budjet  \citep{Capellari2011, Bernardi2019}.

Furthermore, galaxy morphologies are found to be tightly correlated to the color bimodality observed in galaxy populations, thereby resulting in the existence of the younger blue star-forming galaxies with late-type (S) morphologies, and the older red passively evolving galaxies with early-type (E/S0) morphologies \citep{baldry2004}. However, 
we are increasingly discovering that several sub-populations of galaxies do not neatly follow this dichotomy, i.e. red spirals and blue ellipticals exist \citep{bamford2009galaxy}, and likely arise from a variety of physical processes, some of which may be environmentally driven \citep[e.g., ][]{vulcani2015}. 

Thus, the evolution of galaxy morphology has always been in tandem with the growth of galaxies' large-scale environment and their masses over cosmic time \citep{desai07,Calvi2012,Crossett2014,Sarkar2020,Wu2020}.
Indeed, using a dichotomous ’bulge/disk’ definition for the Hubble-type morphologies, the redshift range $1 < z < 2$ is found to be abundant with bulge+disk systems \citep[e.g.][]{margalef16},
while massive galaxies in the local universe are majorly bulge-dominated \citep{Buitrago2013}.

Furthermore, higher redshift galaxies predominantly show peculiar/disturbed/irregular morphologies deviant from the classical morphologies observed at the Local Universe \citep[e.g.][]{mortlock13}, suggesting that galaxies have undergone remarkable structural transformation over cosmic time \citep[see also review by ][]{conselice2014}. Undeniably, galaxy morphology is a crucial evolutionary key in tracing and understanding galaxy evolution throughout cosmic times (e.g. \citealt{Shao2015}).

Ample opportunities are now being presented to investigate galaxy morphologies through multi-band sky surveys, giving us hundreds of thousands of galaxies while exploring large volumes of the sky at the same time \citep[e.g., SDSS; ][]{York2000}. The diverse methods employed by such sky surveys vary from human classification of specialists \citep{Nair2010,Ann2015}, to citizen science \citep{lintott2008,lintott2010,Willett2013,Simmons2017}, or from numerically estimating morphology from  galaxy properties \citep{Spiekermann1992,Storrie-Lombardi1992,Walmsley2020} to novel techniques like Principal Component Analysis \citep[PCA; ][]{Kelly2004,Wjeisinghe2010}, most of which heavily rely on image quality either due to resolution and/or sensitivity of the observations \citep[e.g., ][]{Povic2015}. However, migrating to automated methods of classifying galaxies is now necessary to deal with the huge data volumes resulting from such current and upcoming surveys e.g., the Legacy Survey of Space and Time \citep[LSST; ][]{Tyson2002,Axelrod2006} by the Vera C. Rubin Observatory \& sky surveys with the Nancy Grace Roman Space Telescope \citep[][]{WFIRST}.

Machine Learning (ML) is a powerful automated tool for extracting useful information from complex and varied imaging data
sets, and assist in decision-making processes such as classification trees. The use of ML thus is limited not only for galaxy morphologies \citep[][]{tohill2023} but also to detect gravitational lenses, interacting galaxies, to classify quasars \citep[][]{Freeman2013,Shamir2013,Holincheck2016,Bom2017,Ostrovski2017,Ma2019,Knabel2020,zaborowski2022identification}, and more recently to detect outliers in astronomical images \citep[][]{margalef2020}. These applications highlight the wide-ranging capabilities of ML in astrophysical research, enabling researchers to explore and understand diverse phenomena in the cosmos. In the last decade, a sub-field of ML known as Deep Learning (DL)
has emerged as the main technique for computer vision applications \citep[][]{lu2017simultaneous,abdel2014convolutional,vecchiotti2018convolutional}, music classification \citep[][]{choi2017convolutional}, and medical prognostics \& diagnostics \citep[][]{li2018remaining,hannun2019cardiologist}. 

DL is applied model development for processing complex, minimally reduced (or even raw) data from different sources, and extract relevant features that can then be effectively linked to other properties of interest. In particular, Deep Neural Networks (DNNs) are high-performance data-driven models that are capable of exceeding humans in classification tasks \citep[][]{challenge01}. In astronomy, several recent works have exploited this to show
that DNNs can indeed be successfully used to identify not only the morphological features in raw
images with minimal human intervention \citep[][]{glazebrook2017,lanusse2018cmu,jacobs2019,madireddy2019,cheng2019,petrillo2017,petrillo2019a,petrillo2019b,Farias2020,hausen2020morpheus, bom2022developing},but also outliers in astronomical images \citep[][]{margalef2020}.

In this paper, we present the morphological classification of galaxies into LT and ET, using the new Southern Photometric Local Universe Survey DR3 (S-PLUS; \citealt{mendes_de_oliveira2019}). 
As a follow-up to \cite{Bom2021} hereafter BOM21, our main aim is to apply a high-performance DL algorithm to the imaging data, to obtain a novel and reliable morphological catalogue in the Southern Hemisphere, with a complementary coverage to other morphological catalogues. Furthermore, we also develop the first Deep Network to evaluate the quality of the stamps and clean spurious detections. Finally, we take advantage of the high precision photometric redshifts derived using  the 12 bands in S-PLUS to explore the dependence of morphology on the environment and color, used as a proxy for the galaxy stellar population properties. We compare the classification presented in this work with \cite{Vega-Ferrero2021}, and we discuss the implications arising by studying differently classified objects on the current understanding of galaxy morphological categories.   

This paper is organized as follows, in section \ref{sec:data} we describe the data from iDR3 used in this work,  the
sample selection, and auxiliary data used, such as the photometric
redshift. In section 3, we present the Deep Learning method
used for galaxy morphology classification, and the novel-
ties in its implementation since S-PLUS DR1 morphology paper \cite[BOM21,][]{Bom2021}. In section 4, we present the  results of the model, including deep
learning performance. We also show the relation
between environmental density and morphology, and we analyse the distribution of the different morphological classes in a (g-r) colour versus $M_{r}$ absolute magnitude diagram. In section 5, we present our summary
and discuss the results.

\section{Data}
\label{sec:data}
\subsection{Southern Photometric Local Universe Survey}

\begin{figure*}
    \centering
    
    \includegraphics[width=0.98\linewidth]{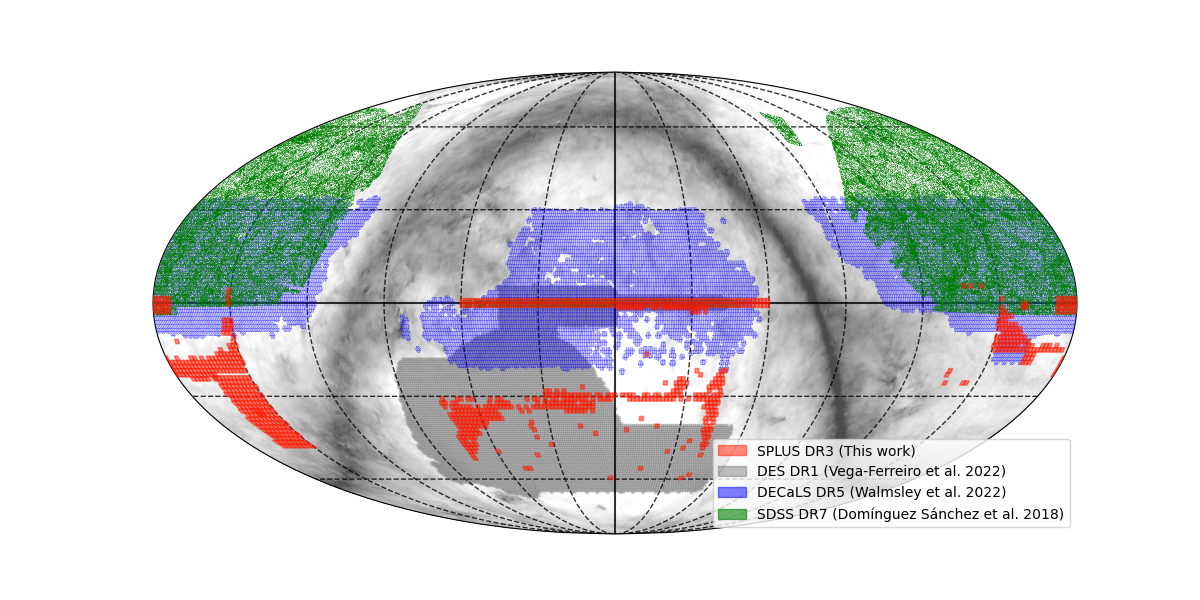}
    
    \caption{S-PLUS DR3 footprint (used in this work) and the footprint of some recent galaxies morphology catalogues available in the literature.}
    
    \label{fig:footprint}
    
\end{figure*}

The Southern Photometric Local Universe Survey (S-PLUS) is performed with a robotic 86-cm telescope located at the Cerro Tololo Interamerican Observatory to cover $\sim 9300{\rm \ deg}^2$ of the sky in 12 optical bands.  S-PLUS uses a wide field optical camera with a field-of-view of 2 deg$^2$ and a plate scale of 0.55\arcsec\ pixel$^{-1}$. The optical filters \citep[the so-called Javalambre filter system, with 5 SDSS-like bands and 7 narrow bands][]{Cenarro2019} are quite unique for the southern hemisphere and are optimal for source classification, given its better definition of the spectral energy distribution of the observed objects, than the usual 4 or 5-band systems. The narrow bands are designed to be centered on important stellar features, for instance, the OII line, Ca H+K, H$\delta$ and H$\alpha$. The survey reaches a typical limiting magnitude of r$<$21 AB mag for the broad bands and r$<$20 AB mag for the narrow bands \citep{mendes_de_oliveira2019}. 

The third public data release of S-PLUS (DR3) covers $\sim 2000{\rm \ deg}^2$ over the Southern Sky. It includes the areas covered in the previous Data Releases such as the Stripe 82. However, the images were reprocessed, with a new reduction and calibration of the data being done from DR2 to DR3, as described in \citet{Almeida-Fernandes2022}. In figure \ref{fig:footprint} we present the area covered by DR3 in comparison with other surveys with available morphological catalogues. The area of the Stripe 82 (at the equator) has overlaps with a number of surveys, in optical and other wavelengths, and it has been used as a benchmark for checking the data reduction and calibration procedures. Other important area covered by the DR3 is the Hydra supercluster (the long vertical red rectangle at the far left of Figure \ref{fig:footprint}).  

\subsubsection{Sample selection}
\label{sec:sample selection} 
We use the full DR3 catalogue containing $\sim$ $50$ millions of sources. During the DR1 morphological classification, we selected the objects only by Petrosian magnitude in $r$ band (${\rm r}_{\rm petro}) < 17 ~{\rm AB~mag}$ and probability of being a galaxy ${\rm prob}_{\rm gal} \geq 0.6$~\citep[for further information see][]{splus_star_gal}. However, we had a visual inspection phase to remove undesired spurious detection (see BOM21). The current catalogue covers an area of $1800$ deg$^2$, which makes the visual inspection unfeasible in a reasonable time scale with limited human resources. Therefore, we define more stringent cuts and include four extra constraints compared to BOM21. Additionally,  we added an automated selection phase by Neural Network that is detailed in Section \ref{sec:Deep Learning Classification}. Thus, 
we apply the following selection criteria to define our galaxy sample from the full catalogue of the S-PLUS DR3: 

\begin{align}
    {\rm r}_{\rm petro} < 18 ~{\rm AB~mag}\\
    {\rm prob}_{\rm gal} \geq 0.7\\
    0 <= {\rm photoflag}_{\rm r}  <= 3 \\
    {\rm BrightStarFlag} = 0 \\
    {\rm R_{Kron}} >= 3 \\
    {\rm FWHM_n} >= 1.5\\
    \nonumber
\end{align}
\noindent where ${\rm photoflag}_{\rm r}r$ is a photometry quality flag from \textit{SExtractor} \citep{Bertin1996},  ${\rm R_{Kron}}$ is the Kron radius, i.e. the first moment of the surface brightness light profile,   ${\rm FWHM_n}$ is the Full Width at half maximum of the object divided by the median FWHM of all bright non-saturated stellar objects of the field. All those features are available and described in the SPLUS catalogue. The probability of being a galaxy, ${\rm prob}_{\rm gal}$ \citep{Nakazono2021}, and the flag indicating a presence of a bright star nearby, ${\rm BrightStarFlag}$,  are listed in the 'star-galaxy-quasar' and 'masks' Value Added Catalogues (VAC; see SPLUS.cloud for further details ~\footnote{ \url{https://splus.cloud/catalogtools}}). Specifically, $0 <= {\rm photoflag}_{\rm r} r <= 3$ ensure the goodness of \textit{Sextractor} fit in most of the cases of interest. The $BrightStarFlag$ parameter  is very effective in removing bright stars, and allow to clean the few stars which are erroneously assigned a  probability higher than $0.7$ of being galaxies by the star-galaxy-quasar classification \citep{Nakazono2021}. 
The conditions  ${\rm Kron_{Radius}} >= 3$ and  ${\rm FWHM_n} >= 1.5$ are included to select resolved objects (the average $FWHM_{seeing} \simeq 1.2"$).

Following these selection criteria, we obtained a final catalogue of 164314  
objects, for which we created image stamps in the 12 bands, with a size of 200$\times$200 pixels$^2$\footnote{The image cutout tasks can be found in this GitHub repository: \href{https://github.com/lucatelli/splus-tools}{https://github.com/lucatelli/splus-tools}.}.
The final catalogue is mostly composed of reliable stamps, i.e. stamps centered on a galaxy, complete up to $r_{petro} < 18$. Further improvement on the sample selection are described in Section \ref{sec:Non reliable Stamps}.

\subsubsection{Samples definition}
\label{sec:Samples definition}

The supervised Deep Learning (DL) assessment requires to be trained on a sample of objects with known classification, i.e. a \textit{labeled set} ({\it Training/Validation and Test Set - I}), sharing as much as possible, the same properties of the sample where the algorithm will be applied in a second moment ({\it Blind Set - II}).  
In this section we describe the characteristics of the two samples, but we refer to Section \ref{sec:Deep Learning Classification} for more details on the DL algorithm and its  performance.

\begin{figure}
    \centering
    
    \includegraphics[width=0.93\linewidth]{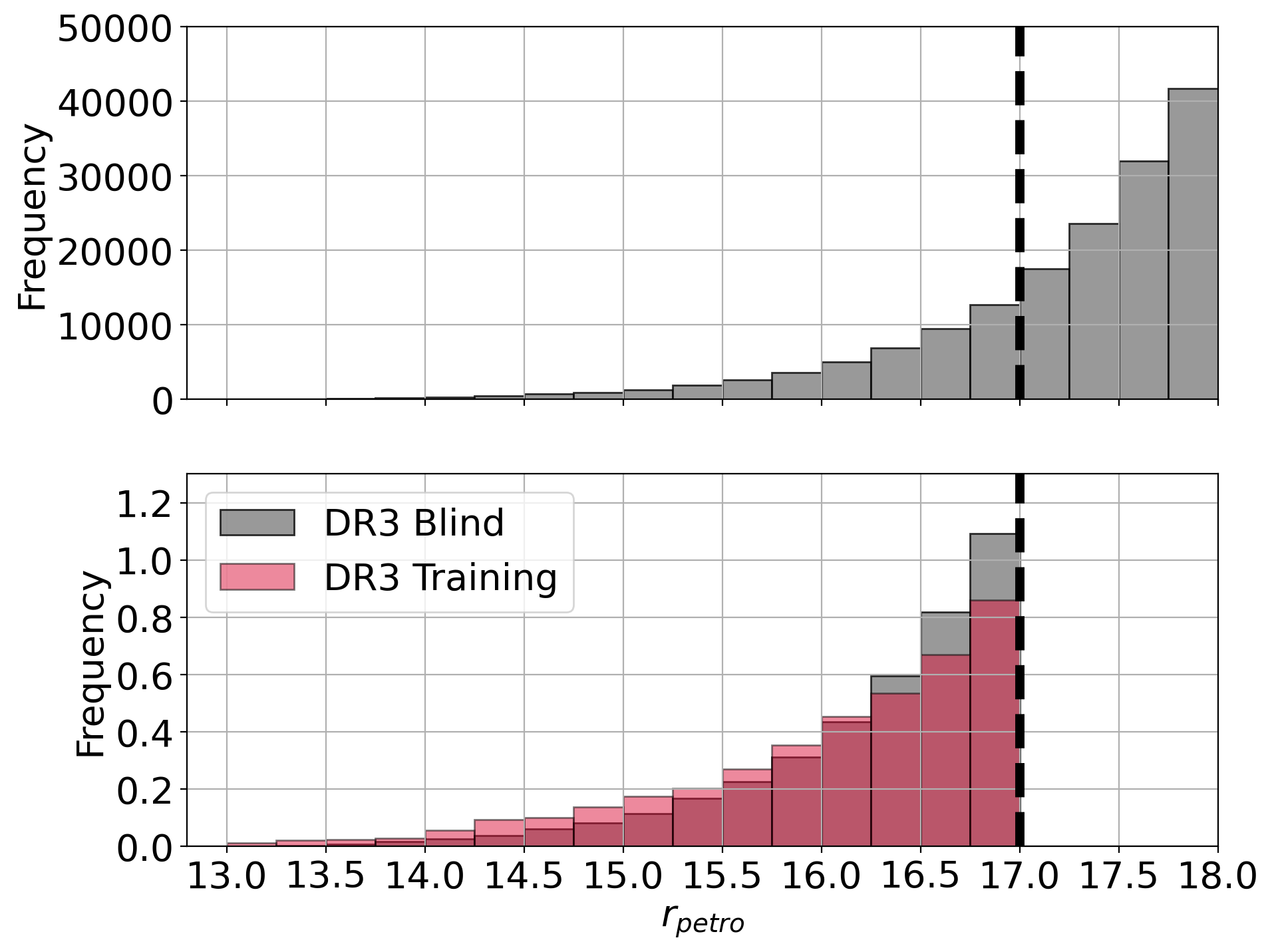}
    
    \caption{Distribution of the r-petrosian magnitudes (r$_{\rm petro}$) for the whole sample, i.e. up to magnitude r$_{\rm petro} \le 18$ ({\it top}) and a normalised histogram  of the magnitude distribution of  {\it I} (Training/Validation set) and {\it II} (Blind set) up to r$_{\rm petro} \le 17$ {\it bottom}, i.e. the limiting magnitude of {\it I} - the training sample.}
    
    \label{fig:histsamplesrpetro}
    
\end{figure}

We used the same objects presented in training and validation and the test scheme used in BOM21, which used Galaxy Zoo 1 unbiased morphological classification into elliptical and spiral galaxies \citep{lintott2008,bamford2009galaxy, lintott2010} as true label. Such choice was possible since S-PLUS DR1 is included in S-PLUS DR3. It is important to note, though, that since the reduction pipeline has been improved between the two data releases, new stamps were created using the novel images, to ensure the homogeneity of the two data sets (I and II). Another relevant difference between the data from DR1 and DR3 is the new photometric calibration applied for the S-PLUS DR3. This calibration consists of fitting synthetic stellar templates to well-known data from other surveys, deriving precise zero-points and magnitudes that were tested on 170 STRIPE82 fields \citep[see][for a detailed description of the method]{Almeida-Fernandes2022}. We obtained the stamps for each object in the 12 bands from the DR3 data access, for both samples.

In total, there are $4232$ objects in  training sample {\it I}, while set {\it II} is composed of $164314$ objects. As presented in the top panel of Figure\,\ref{fig:histsamplesrpetro}, the training sample, i.e. sample {\it I}, is approximately complete only up to $r_{petro} < 17$\footnote{This magnitude limit is required in Galaxy Zoo 1 in order to perform the debiasing process, which requires spectroscopic redshifts, see \citet{bamford2009galaxy} for more details}.
As described in Section \ref{sec:sample selection} in this work we select objects up to $r_{petro} <18$. The implications of this choice in the DL performance are discussed in Section \ref{sec:results}. Both the samples I \& II show similar distribution of $R_{petro}$ for magnitudes $<17$ (bottom panel of Figure \ref{fig:histsamplesrpetro})

\subsubsection{Photometric redshifts}
\label{sec:Photoz}

The S-PLUS DR3 photometric redshifts catalogue uses a DL model based on a Bayesian Mixture Density Network architecture. This specific configuration allows single-point estimates while also providing probability distribution functions (PDFs) for each galaxy. This network is trained on 12-band photometry from S-PLUS, cross-matched with the unWISE \cite[Wide-field Infrared Survey Explorer,][]{unwise}, GALEX \cite[Galaxy Evolution Explorer,][]{GALEX}, and 2MASS \cite[The Two Micron All Sky Survey,][]{2mass}  catalogs (W1/W2, NUV/FUV, and J/H/K magnitudes, respectively). Spectroscopic redshift targets are compiled from various surveys, including SDSS DR16, 2dFRGS, 2dFLenS, 6dFGS, and others. A total of 262,521 objects are used for training/validation, and an independent test set.

\begin{figure}
    \centering
    
    \includegraphics[width=0.93\linewidth]{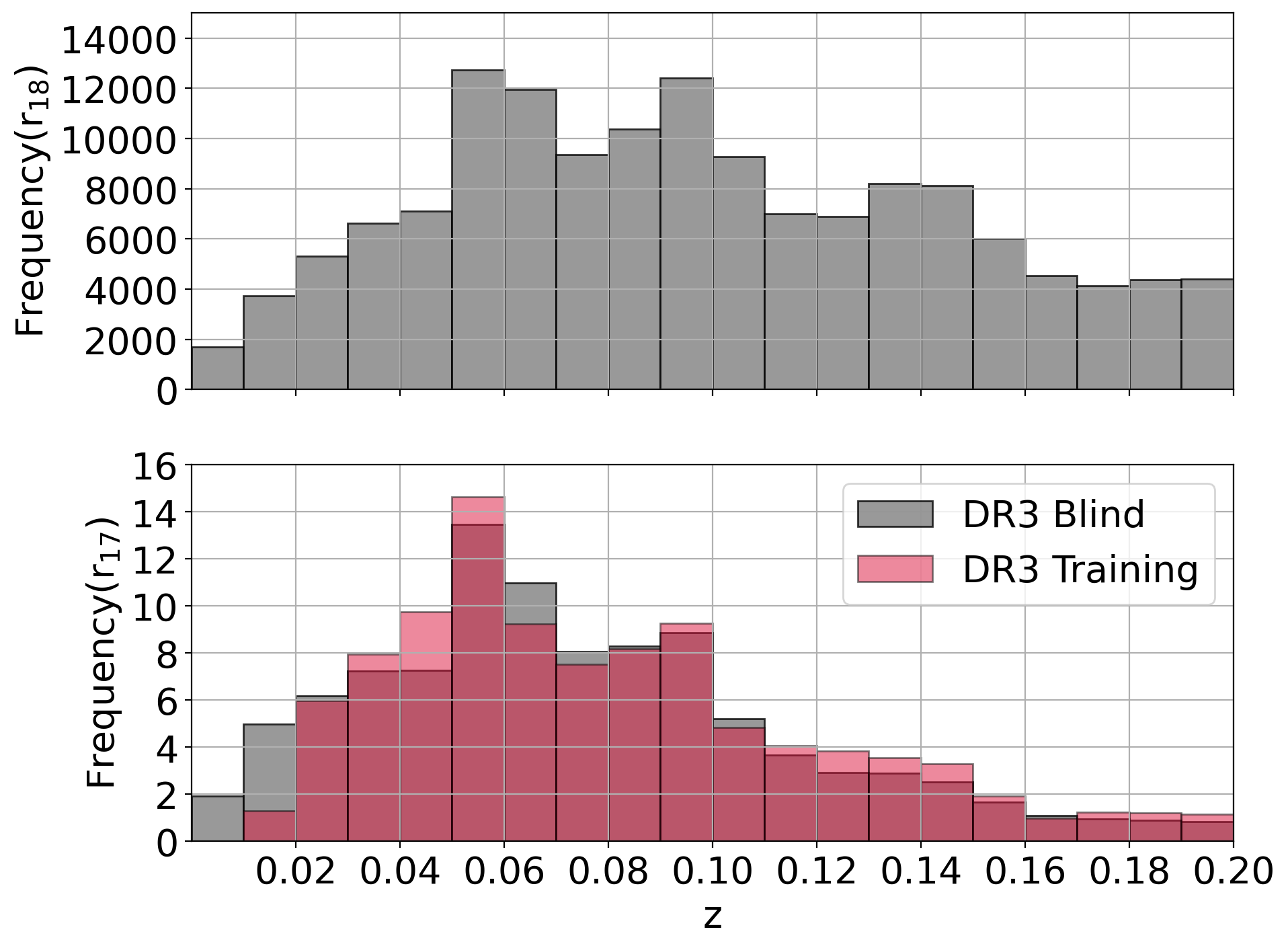}
    
    \caption{Distribution of the Photometric Redshift for the training and blind sample. On top of the distribution only for the blind sample up to r$_{\rm petro} < 18$. On the bottom is the normalized distribution for both blind and training samples up to r$_{\rm petro} < 17$ }.
    
    \label{fig:histoz}
    
\end{figure}

Due to its unique filter system with a set of broad- and narrow-band photometry, the current model is capable of providing accurate photometric redshifts, while also maintaining low bias and negligible outlier fraction. In fact, within the magnitude range of interest of the present work, $r_{petro} \in [14, 18]$, the median normalized bias stands $\sim -0.0015$, the scatter is $\sim 0.015$, and the outlier fraction is below $1\%$. The catalog not only includes single point estimates but also well-calibrated probability distribution functions, enabling users to evaluate the uncertainties associated with each estimate. Further information regarding the methodology and resulting findings can be found in \citet{Lima2022}\footnote{The S-PLUS public data, including the photometric redshifts are also available in \url{splus.cloud}}. Figure \ref{fig:histoz} shows the distribution of the photometric redshift for samples {\it I} and {\it II}. Specifically,  {\it II} is divided into the whole sample, with $r_{petro} < 18$ and a sub-sample   with $r_{petro} < 17$, sharing the same magnitude limit as the training sample.

\section{Deep Learning classification }
\label{sec:Deep Learning Classification}

\subsection{Training, Validation and Test sample}
\label{sec:Training and Validation sample}

 We split the cross-match data between S-PLUS DR3 and Galaxy Zoo I STRIPE82, i.e., Dataset {\it I}, into Training-Validation-Test sets. Dataset {\it I} contains unbiased classification only \citep{lintott2008,bamford2009galaxy,lintott2010}. The data presents an 80 percent threshold on the probability of being a galaxy as true labels distributed in 29 percent of early-type galaxies (ETG) and 71 percent of late-type galaxies (LTG). This distribution reflects the proportion between the two classes,  in the local Universe ($0 < z < 0.2$) as reported by \citep{lintott2010}.

\par
\begin{figure*}
    
    \centering
    \includegraphics[width=0.7\linewidth]{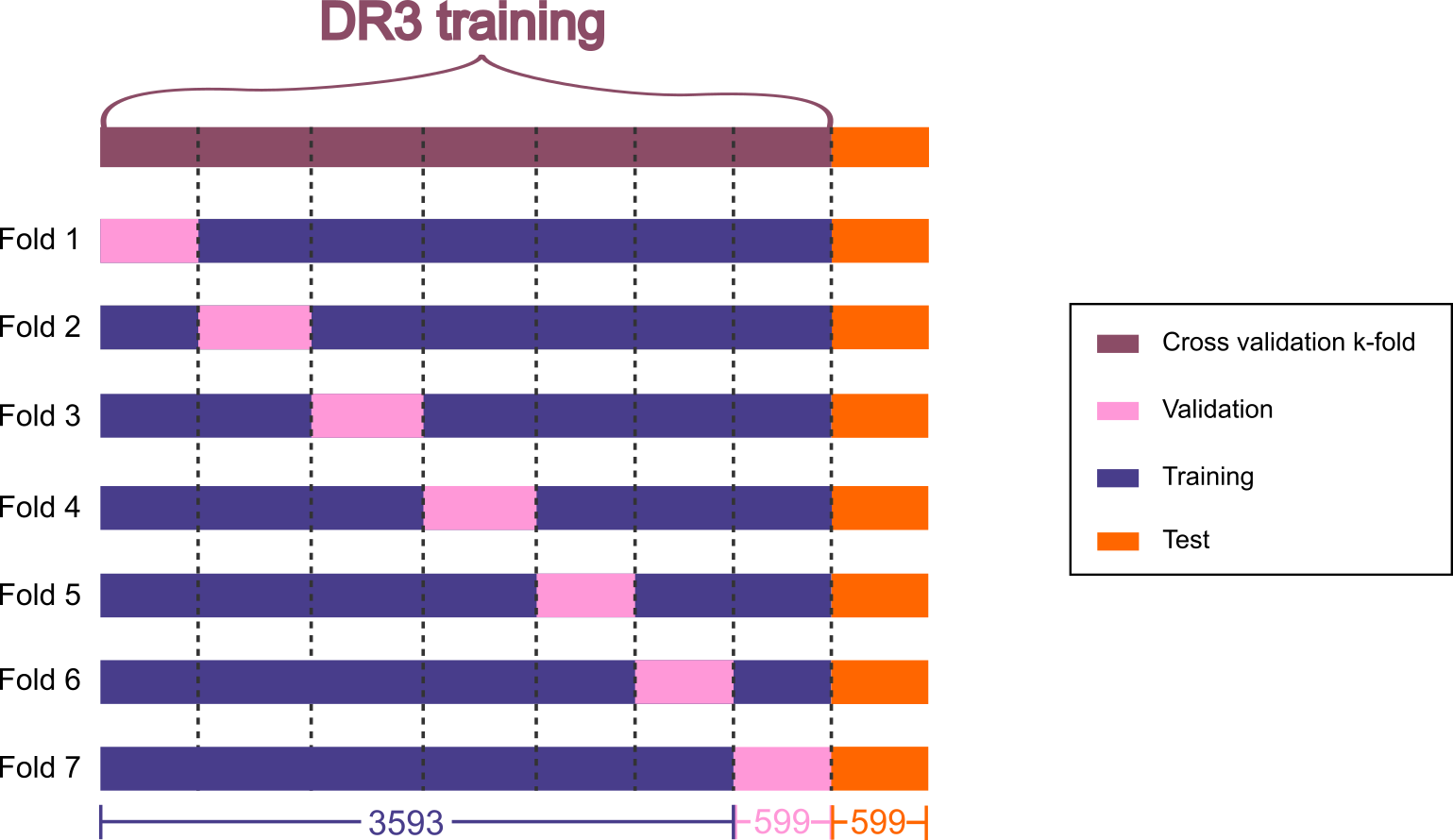}
    \caption{In this figure we show the cross validation k-fold method applied to the training of the LTG / ETG model. Each fold is separated into training, validation and test. That process is made in a way that there is no match between each validation. Additionally, the training for each fold is slightly different, which reduces a possible bias concerning the selection of the objects that composes it. Considering that the technique will define an architecture with certain weights for each fold, the metrics in each training stage can be used to evaluate which fold has the best training set configuration. The numbers at the bottom indicate the size of the training, validation, and test sets in each fold.}
    \label{fig:cross validation}
    
\end{figure*}

 We split the DR3-Training dataset in $7$ folds. These folds are subsamples of the training set used to perform a cross-validation procedure \citep{moreno2012study}. We have evaluated other choices for a number of folds. However, with more folds the validation set is smaller, the validation loss starts to be more unstable, and we found a good trade-off with $7$ folds. Thus, as shown in figure \ref{fig:cross validation}, we define $7$ different training and validations sets, 
 each containing $\sim 85\%$ and $ \sim 15\%$
 of the data, respectively. This separation is made so there is no match between the validation sample for every fold. Additionally, this method guarantees that each object will be used at least once in the test set. 
 We use $599$ objects as a test set for performance evaluation, these are not used for training. As in BOM21, the training set based on debiased GZ1 contains $71\%$ of LTG
and $29\%$ ETGs, and thus is an imbalanced dataset. Therefore, in order to train the Neural Network to prevent our model of being biased towards the most abundant class, we adopt the same data treatment scheme presented in BOM21, applying weights to each class.  For a set of $N$ objects in the training set and if the number of objects in the class $\alpha$ is $N_{\alpha}$, we define the weights as :
\begin{equation}
   w_{\alpha}= \frac{N}{m N_{\alpha}},
   \label{weight_eq}
\end{equation}
where $m$ is the total number of classes.
This is a standard procedure in ML field \footnote{see, e.g.,  \url{https://www.tensorflow.org/tutorials/structured_data/imbalanced_data}}. The weights defined in Equation \ref{weight_eq} are then applied in the objective or loss function minimized during the training phase. This procedure enables each of the classes to have the same impact on the loss function. 

\subsection{Non reliable Stamps}
\label{sec:Non reliable Stamps}

In the DR1 catalog from BOM21, all the stamps were visually inspected prior to the analysis with the DL algorithm aiming to prevent biases in the classification process caused by spurious objects. This became no longer feasible due to the size of the current and future S-PLUS data releases. 
Therefore, in this study, in order to generate a more robust catalog, we implemented a new DL model to separate {\it reliable} from {\it non-reliable} stamps. To this end, we use stamps that were excluded as {\it non-reliable} from the DR1 classification as a training sample for this new DL model.  This approach is advantageous to avoid spurious classifications such as faint galaxies in the same field of nearby saturated stars whose spikes can affect the accuracy of the magnitude estimation of that galaxy. Given the considerable extent of the dataset used in this study, other undesirable objects might include artifacts and problematic stamps in general. Therefore, it is essential to count with a robust  method of distinguishing good images from low-quality images to use as an input for the main ETG/LTG DL model.

\begin{table*}
\small
\caption{\label{tab:sample_description} Sample Description of the samples used in this work. }
\begin{tabular}{llcl}
\hline\hline
Sample & Subsample &Number of objects & Description  \\

\hline 

I & $DR3$-Training & 4192 & ETG and LTG galaxies splitted between training and validation. \\
I &$DR3$-Test & 599  &  ETG and LTG galaxies for performance test.   \\
II & $DR3$-Blind & 46763 & galaxies for blind classification with $r_{petro}\leq 17$.\\
II & $DR3$-Extended & 161635 & galaxies with  $r_{petro} \leq 18$ for blind classification. \\
\hline \hline
\end{tabular}
\\
\end{table*}

\subsection{Deep Learning Model}
\label{sec:DL}
 Following a strategy similar to that in BOM21, in this work we also made use of EfficientNet algorithms \citep{efficientnet}, which are part of the Convolutional Neural Networks (CNN) family-models well-known for having high performance on visual pattern recognition problems in standard image datasets such as ImageNet \citep{deng2009imagenet}. 
This kind of Network is based on an initial model similar to a MobileNet \citep[MnasNet;][]{tan2019mnasnet} and can be also scalable by parametrizing the number of layers if needed to gain in performance by making a more complex network while constraining the number of FLoating-point Operations Per Second (FLOPS). Therefore, each parameter choice defining a model and thus defining a family of models. Additionally, this kind of model can also be easily adapted to classify datasets with different resolutions \citep{bom2022developing}. In this contribution we made use of similar model based on EfficientNet B2 firstly described in \citep{efficientnet}, with the minor adaptations detailed in BOM21. For a diagram presenting all the layers in this model please refer to figure 5 (b) and (c) in \citep{bom2022developing} paper.  

\begin{figure*}
    \centering
    
    \includegraphics[width=0.95\linewidth]{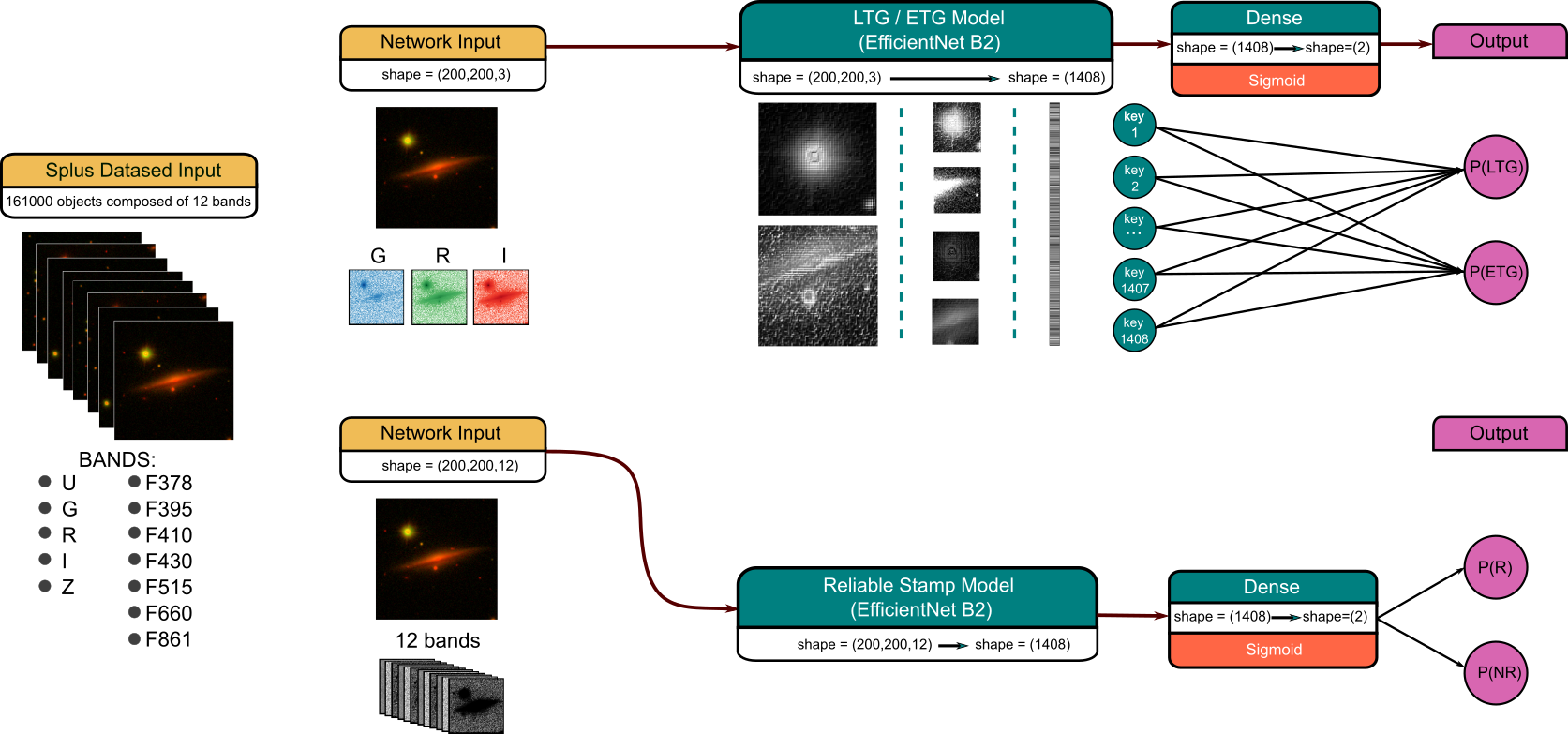}
    
    \caption{Workflow from the stamps taken from S-PLUS data while they passes through the model. Both architectures works in the same way, with the difference that the first one uses only the G, R and I bands available in S-PLUS as the network input. The LTG/ETG Model as well as the Reliable Stamp Model consists of some convolutional layers in the beginning responsible to compact and recognize patterns in the stamp, then, in the end, all that information passes through a dense layer that compacts it into a list containing 1408 keys represented by the bar code in the figure. Both models works with binary classification, then one more dense layer is needed to calculate the probability of each classification given by a sigmoid activation function. }
    
    \label{fig:deep_model}
    
\end{figure*}

Nonetheless, we implemented several innovations compared to the workflow described in BOM21. Firstly, we added a second EfficientNet B2 model to evaluate whether a stamp is reliable for morphological classification. The main goal of this NN is to identify spurious detections, such as crowded stamps where the central galaxy in the stamp is visually indistinguishable, stamps satured by close bright stars, and galaxies that are not completely contained in the stamps. We explore how this non reliable stamp model would be best defined in terms of inputs. After initial tests following Bom21 approach, we used all $12$ bands as inputs in contrast to the ETG/LTG model that shown to be best defined in terms of performance and stability of results by using $g,r, i$ bands only. Although this choice is based on empirical results by adopting the same metrics presented in Bom21, the main difference here is likely due to the nature of patterns we are trying to characterize in the Reliable Stamp model. By visually inspecting the stamps, we find that some of the spurious detections presented large variability of shapes in different bands compared to reliable stamps, and thus are likely to be easily distinguishable by using more bands. For a full discussion of the band choice for finding ETG/LTG please refer to BOM21. A relevant difference in respect to the main ETG/LTG model developed for S-PLUS DR1 is that the probability assigned to a galaxy of being spiral or elliptical is no longer complementary, meaning that the sum of such probabilities is not equal to one, opening a space for a lot of interesting findings like the ones discussed in Section \ref{sec:ES class} and the possibility of pointing objects that do not fill in any category. This was implemented by changing the neural network activation function in the last layer from a {\it softmax} to a {\it sigmoid}. In figure \ref{fig:deep_model} we present a scheme of both DL models, detailing the input bands and also presenting an example of a given stamp flowing towards some of the network convolutional filters. 

\begin{figure*}
    \centering
    
    \includegraphics[width=0.5\textwidth]{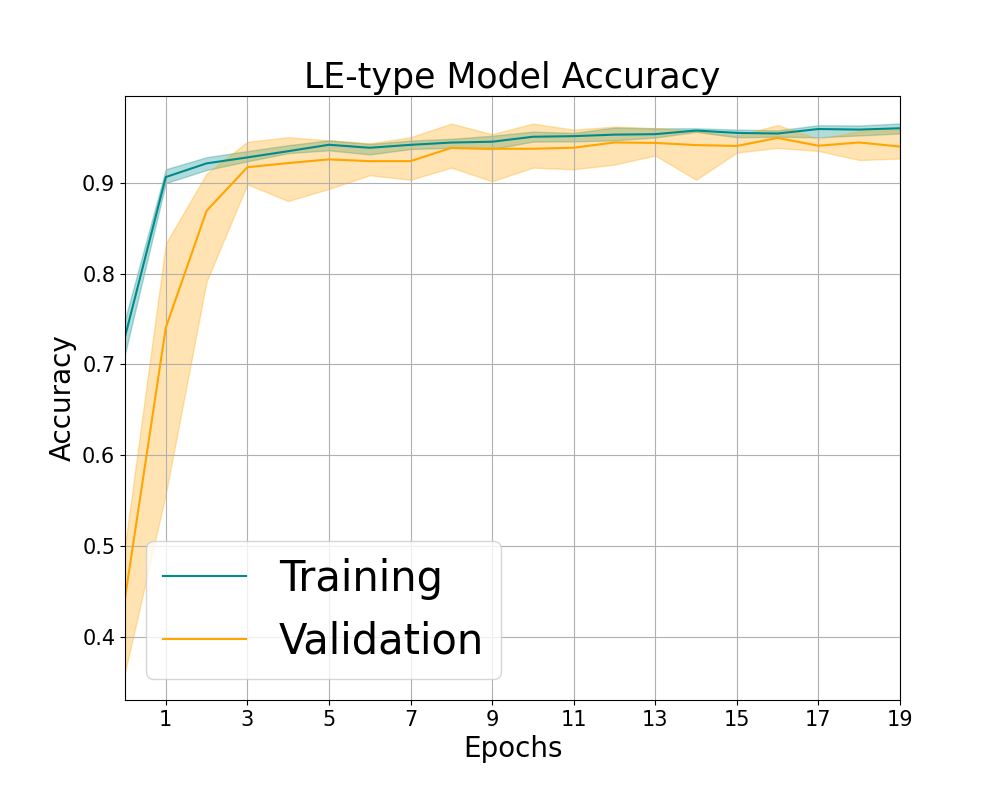}\includegraphics[width=0.5\textwidth]{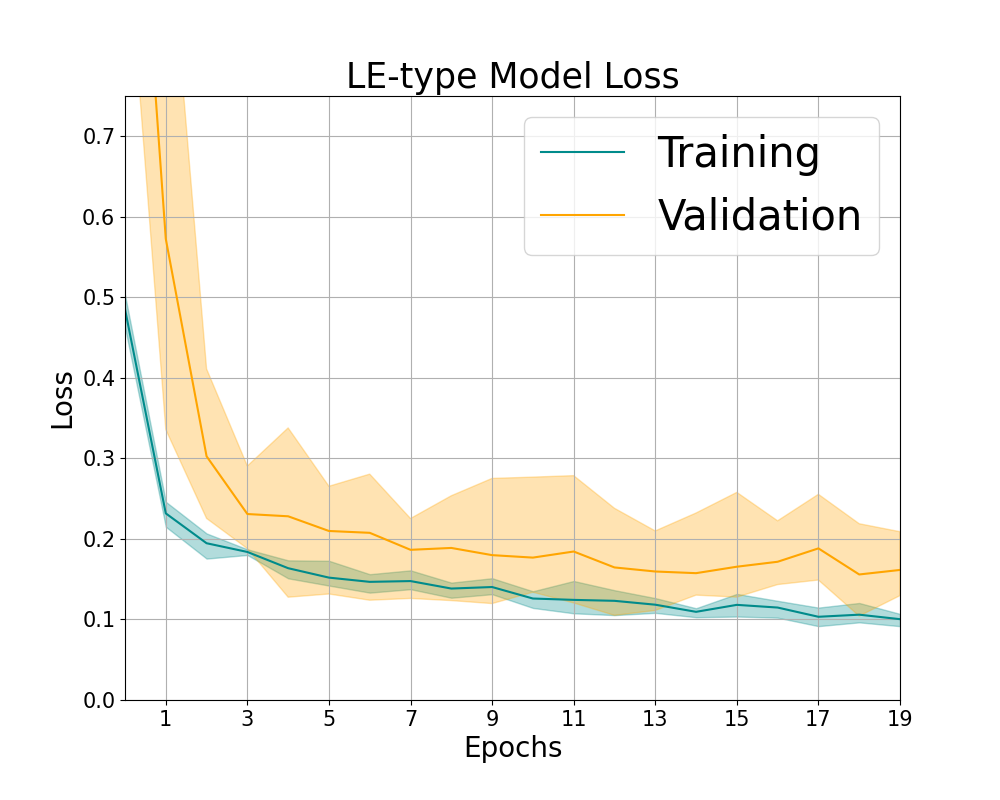}\\
    \includegraphics[width=0.5\textwidth]{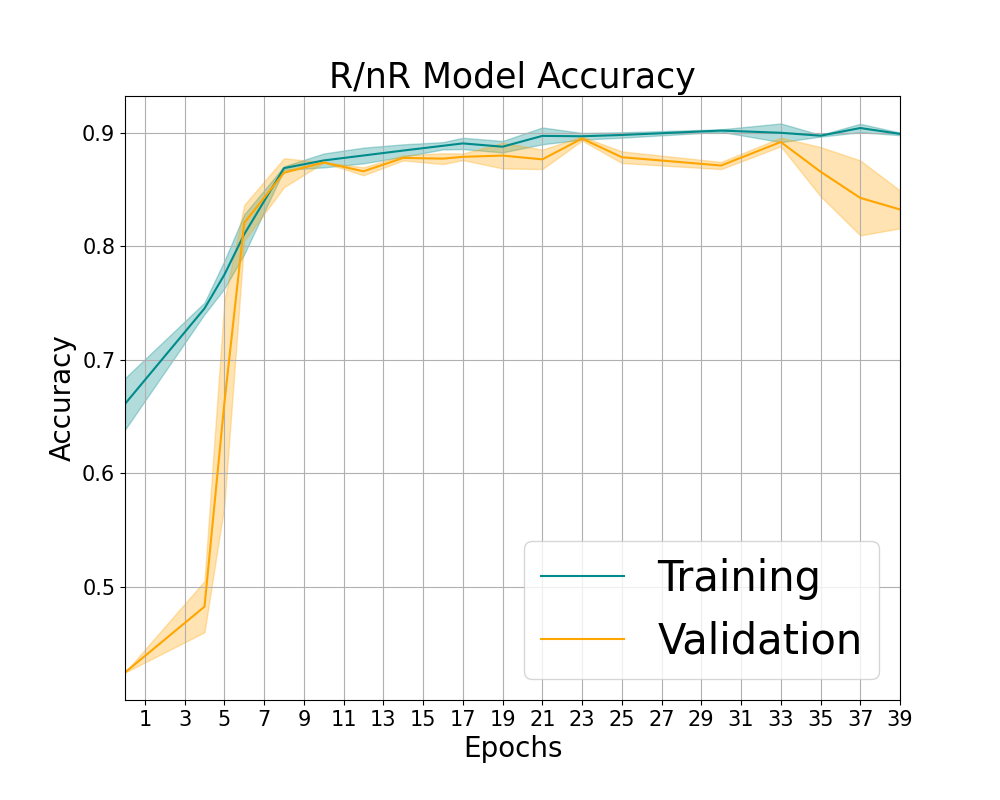}\includegraphics[width=0.5\textwidth]{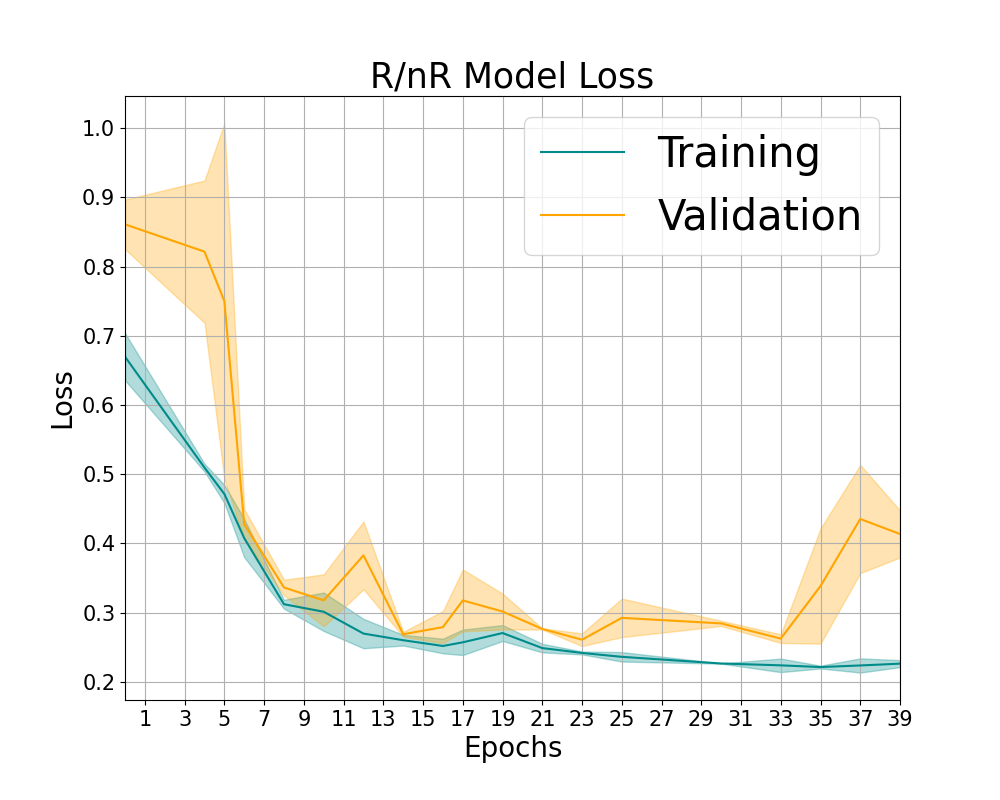}
    
    \caption{Accuracy and Loss in the training of the Late/Early-type (\textit{Top}) model and Reliable Stamp model (\textit{bottom}) as a function of epoch considering all folds. In blue we present these metrics for the training set and in orange the metrics for the validation set. The line in the middle represents the mean value between all 7 folds used in the cross validation k-fold method.} 
    
    \label{fig:train/val}
    
\end{figure*}

\section{Results}
\label{sec:results}


\subsection{Training}

The training process was performed with a Rectified Adam \citep[RADAM,][]{liu2019variance} optimizer and the loss function is a traditional cross-entropy \citep{Goodfellow-et-al-2016}.
In Figure \ref{fig:train/val} we show the Loss and Accuracy achieved in the training procedure considering all 7 folds. The darkest line in the center corresponds to the mean value of those quantities for each epoch and the shaded area corresponds to the standard deviation between folds. In the top of  Figure \ref{fig:train/val} we present the results for ETG/LTG model using $3$ broad-bands as input, similar to BOM21. The training converges fast, around the third-fifth epoch with high accuracy $\sim 0.9$. The additional degree of freedom added compared to BOM21, i.e., the probability of being LTG or ETG set to be independent, does not seem to affect the performance significantly. Considering the errorbars, we did not find significant overfitting over the entire range. However, towards the end of the training, around the $15$th epoch the figure suggests the beginning of slight overfitting. Furthermore, by evaluating the loss function, the reliable/non reliable model, that uses the $12$ band set as the input, presents a more unstable behavior: the convergence is slower, around the epoch $15$. The validation presents some spikes that might be related to a regularization method present in the network. We also notice a tendency of overfitting from epoch $\sim 19$ onwards. The validation accuracy does not reach $0.9$ consistently. However, it is worth noticing that, differently from the $12$ band model ETG/LTG presented in BOM21, the $12$ band model for reliable/non reliable stamps has significant smaller errorbars suggesting that the model is robust, although the overall performance compared to ETG/LTG model as a classifier is expected to be lower.

\subsection{Performance}

As outlined in the previous section, the cross-validation approach establishes a unique network configuration for each fold. Therefore, we may assess our model's performance on every individual fold. For both ETG/LTG or Reliable Stamp Model classification, we applied these individual folds to the test subsample. 

\subsubsection{ETG/LTG Model}
\label{sec:ETG/LTG Model}

\begin{figure}
    \centering

    \includegraphics[width=1\linewidth]{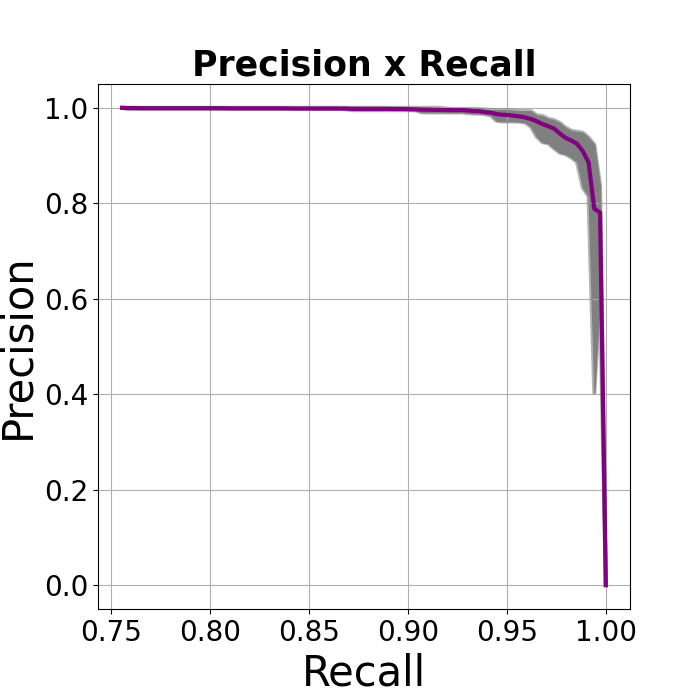}

    \includegraphics[width=1\linewidth]{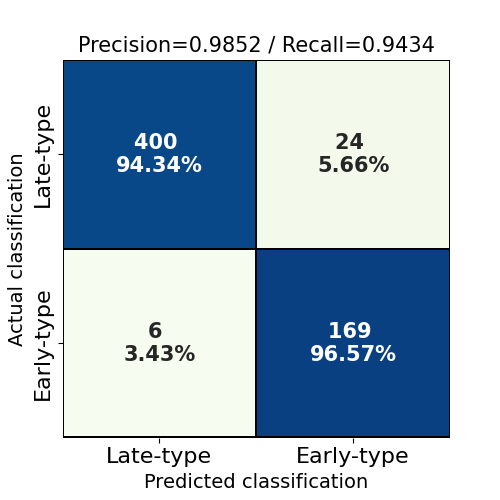}\\

    \caption{Performance in the $DR3$-Test sample for the ETG/LTG model. (Top) The Precision x Recall considering all folds. The purple line was made with the median value for every fold. (Bottom) The confusion matrix for the best fold.}

    \label{fig:Metrics LE}

\end{figure}

\begin{figure}
 
    \includegraphics[width=1\linewidth]{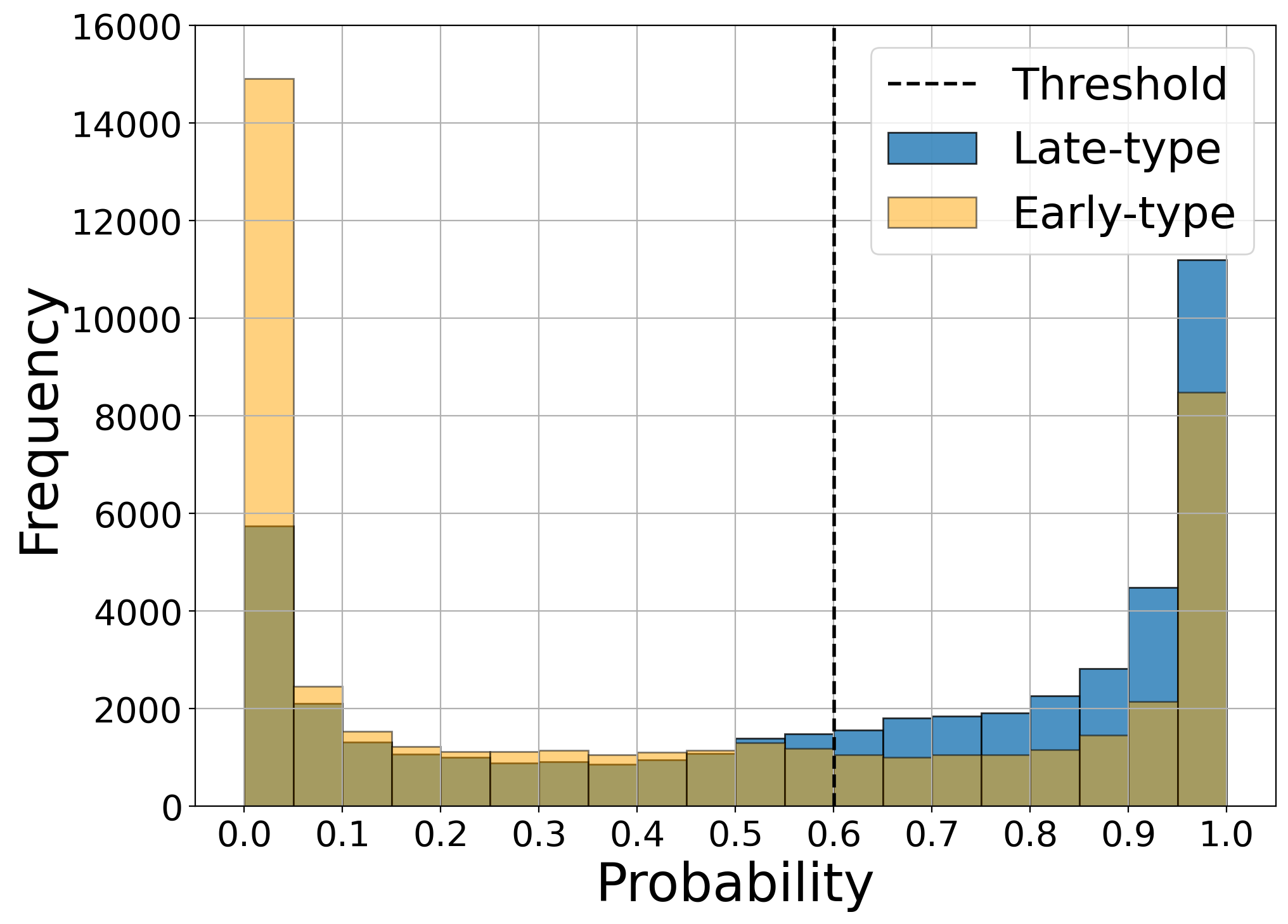} 
    \includegraphics[width=1\linewidth]{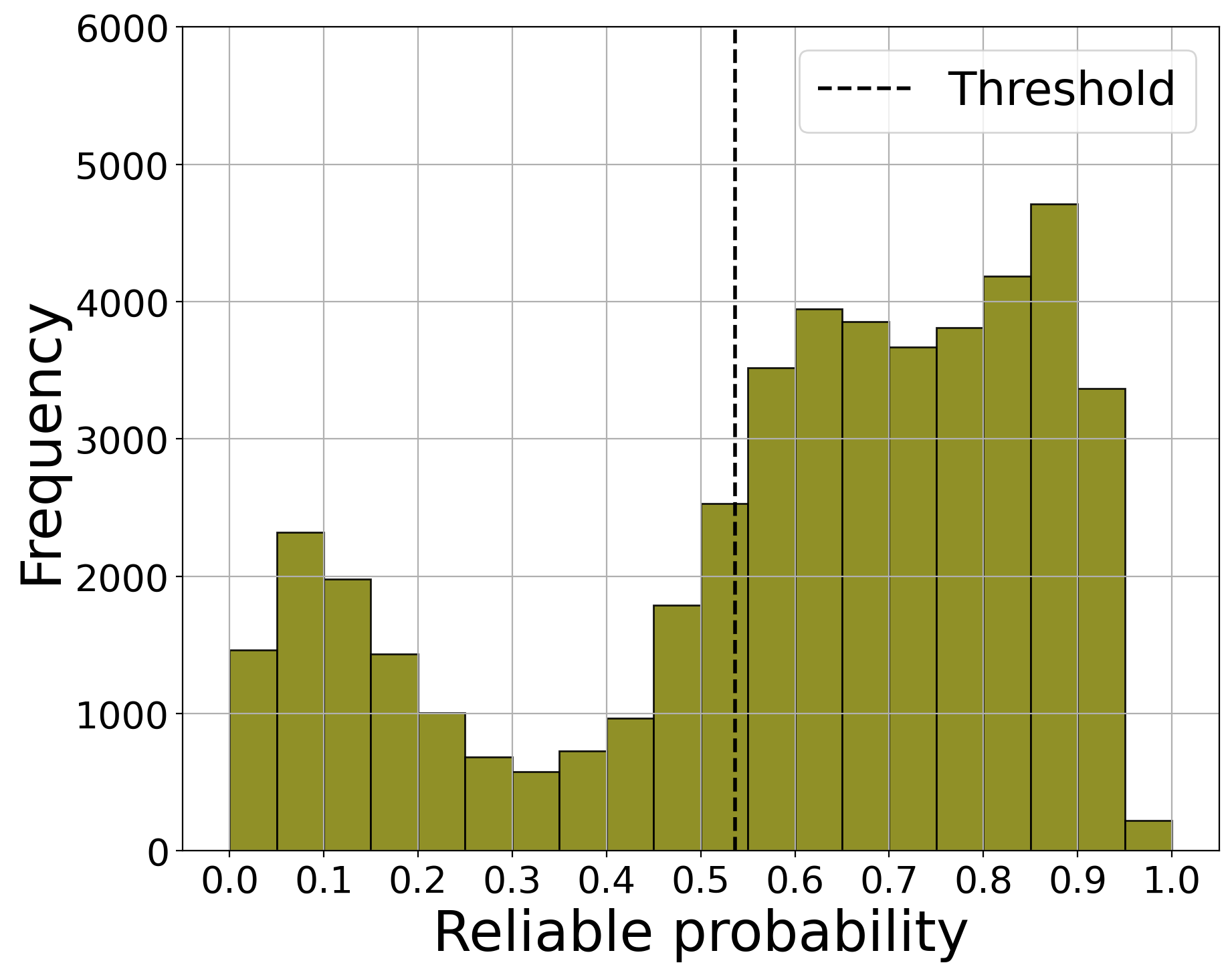}	
    
    \caption{Probability distributions for the classification of the blind set. On top of the distribution for the Late-type and Early-type classification. On the bottom the distribution of being a reliable stamp. In both cases, the dashed line represents the threshold used for the classification itself.}
 
    \label{fig:probs}
  
\end{figure}

We evaluate the performance of our model by evaluating the trade-off of a precision x recall. For a given threshold $t$ that defines which is our ETG if the predicted probability is higher than the $t$, precision or purity measures how many correct predictions were made out of all positive predictions, and recall or completeness presents how many true positives were found among all the actual positives. 
In the bottom of Figure \ref{fig:Metrics LE} we present the median precision-recall for $t$ in the range $[0,1[$  for all folds and its respective standard deviation. Later, we define the best theshold $t_B$ as the $t$ in the precision-recall curve closest to the the point (1,1) which would represent a perfect classifiers, i.e. with both purity and completeness equal to 1. This threshold $t_B$ is set to $\sim 0.60$. To understand the performance outcome with this choice we made use of
a confusion matrix at the bottom of Figure \ref{fig:Metrics LE}. This performance assessment shows the number of correct and incorrect predictions, grouped by each class and therefore presents model performance in a classification task by revealing where it gets confused and makes mistakes. The model demonstrates correct classifications with over $\sim 94\%$ of both ETG and LTG classifications. It is worth mentioning that for this specific performance assessment, we had to assign each galaxy to one category unambiguously. Hence, for this specific analysis we did not take advantage of the fact that the model assign independent probabilities of ETG/LTG. In Figure \ref{fig:probs} top we present the probability distribution of the DR3-Blind set. We notice that the distribution for ETG and LTG classification are well separable with a strong peak around $\sim 0$ and $\sim 1.0$ as one should expect to a two-class classification.

\begin{figure*}
    \centering
    
    \includegraphics[width=1\textwidth]{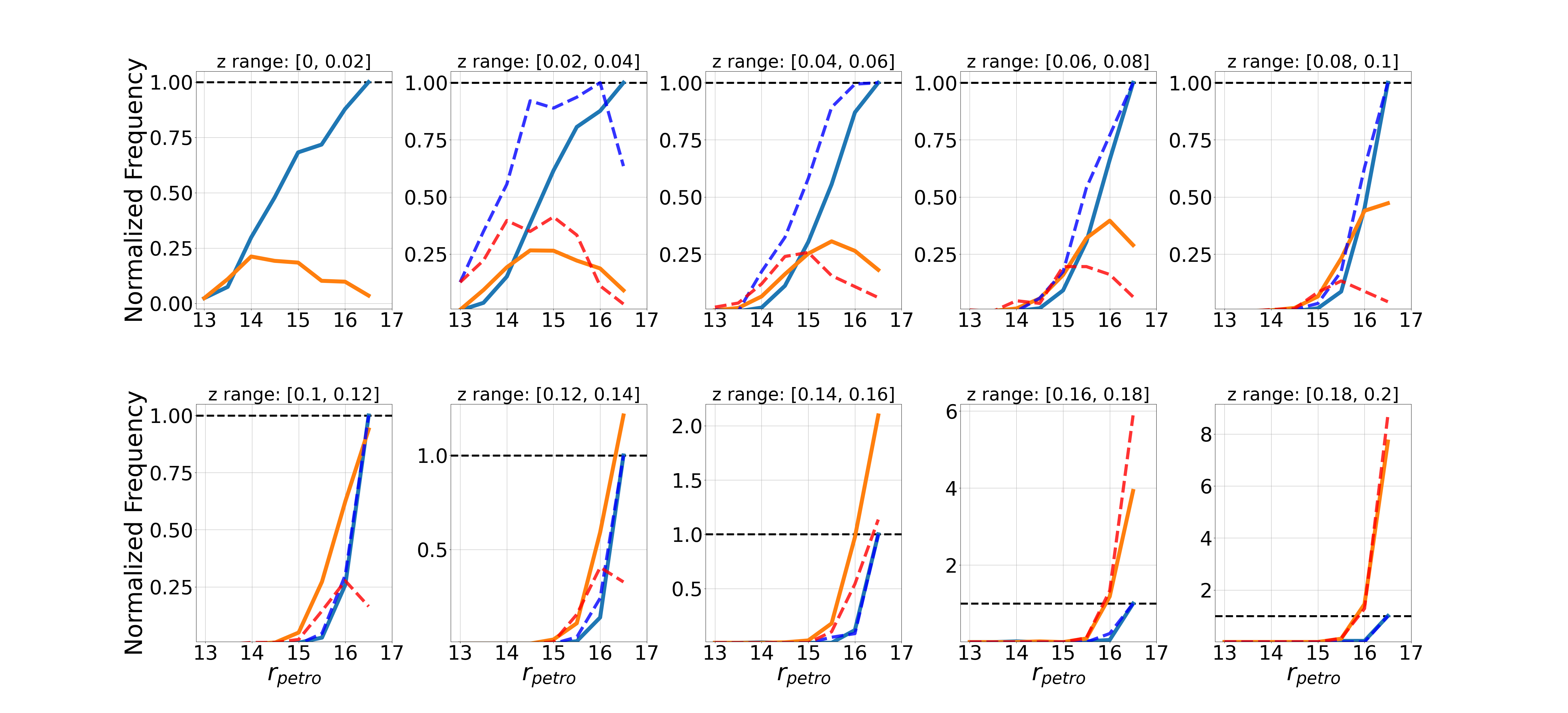}
    
    \caption{
     r-band apparent magnitude distribution for different Photometric Redshift bind distribution f  blind sample (filled line), and  for the training sample (dashed line), for Elliptical (yellow/red) and Spiral (blue/cyan) galaxies. Note that in  the first magnitude bin the training sample is not present ($z< \le 0.02$), since we used Galaxy Zoo data for the training, which are missing in this low mag bin \citep{lintott2008, lintott2010, bamford2009galaxy} 
    }
    
    \label{fig:fracz}
    
\end{figure*}
We present in Figure \ref{fig:fracz} a comparison of the distribution of the photometric redshifts (see Section \ref{sec:Photoz}) of early (orange/red) and late (cyan/blue)  type galaxies for the DR3-Blind/DR3-Training sample, respectively.
It is noticeable that, against  expectations, the number of early type galaxies seems to be larger at higher redshifts than the number of late type galaxies, both for the training and blind data-sets. In fact,  \citet{Buitrago2013} does not finds any strong evolution between the fraction or density of 
spheroid and disk galaxies for $M_{*} > 11$  $M_\odot$ between $0 < z < 0.2$.
We visually inspected galaxies classified as early type  at $z>0.15$ to verify whether the classification is affected by the lack of resolution of the spiral arms. We conclude that the classification is overall correct (see Sections  \ref{sec:comp} where we compare with the morphological classification performed by \citet{Cheng2020} and \citet{Vega-Ferrero2021}) and that the lack of spiral galaxies at high redshift is related to the pre-selection of the stamps since high z spirals tend to have low surface brightness. The training sample used in this work is taken from The Galaxy Zoo project \citep{lintott2008,bamford2009galaxy,lintott2010}, which provides  a debiased morphological classification  \citet{bamford2009galaxy} for galaxies in a redshift range between $z \ge 0.03$  and $z \lt 0.88$, where the lower limit is dictated by the incompleteness at low redshift, while the higher redshift is caused by the loss of objects fainter than $M_r < -20.25$.

\subsubsection{Reliable Stamp Classification}

We used the same analysis scheme to analyze the Reliable Stamp model. The bottom part of Figure \ref{fig:probs} presents the Reliable stamp probability distribution. By comparing the probability distribution of both models we noticed that the reliable model presents wider peaks, which suggests the distribution is not as well separated as in the ETG/LTG model. This conclusion is also indicated by the loss optimization as discussed in Section \ref{sec:DL}. 

\begin{figure}
    
    \centering

    \includegraphics[width=1\linewidth]{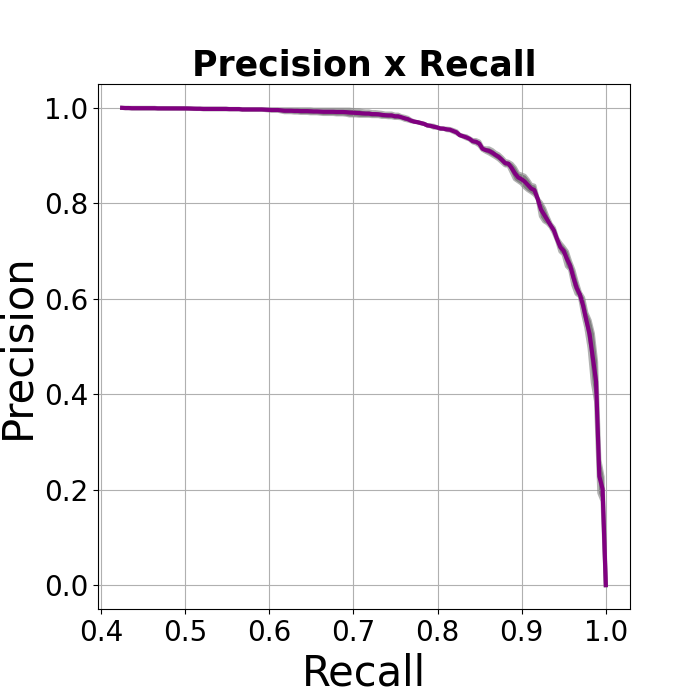}
    \includegraphics[width=1\linewidth]{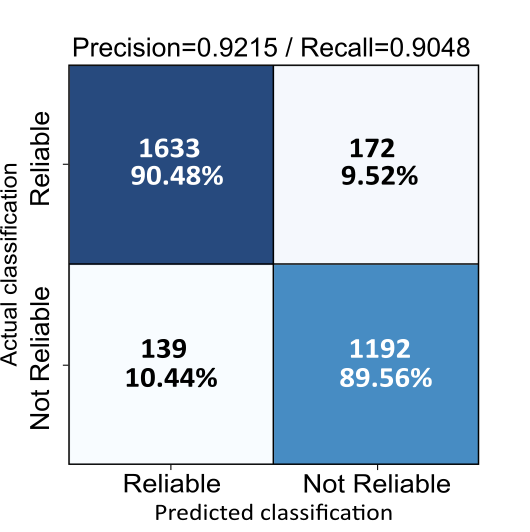}\\

    \caption{Performance in the $DR3$-R-Test concerning the Reliable/Non Reliable model. (Top) the Precision x Recall plot considering all folds. The purple line was made with the mean value for every fold. (Bottom) The confusion matrix for the best fold.}
    \label{fig:Metrics RnR}
    
\end{figure}

In figure \ref{fig:Metrics RnR} we show  the confusion matrix for the best fold and the precision vs recall plot considering all folds. The overall shape in the precision x recall curve is similar to the ETG/LTG model, however, the total area under the curve of the Reliable stamp model is smaller compared to the ETG/LTG model. The confusion matrix presents  $\sim 90\%$ true positives, which is also interesting since there is a vast variability of what is a non reliable stamp. Additionally, by making a visual assessment over the objects classified as non reliable we can find some interesting objects that we believe are worth investigating. We discuss this with more detail in section \ref{sec:extraordinay trash}.

\subsection{Early-Type and Late-Type Galaxies}
\label{sec:ES class}

Galaxies present a wide range of morphologies \citep[e.g.,][]{Buta2011,vandenbergh1998}, from almost spherical ellipticals to grand design spiral galaxies \citep{Grosbol2012}, with the increasing importance of the disk component along the Hubble sequence. At the vertex of the Hubble tuning fork, lie the lenticular galaxies, which present  bulge and disk components as spiral galaxies, but lack spiral arms and relevant star-forming regions. Moreover, the gallery of galaxy types also encompasses irregular galaxies. 
Elliptical and lenticular galaxies are classified as 'early-type', while spirals and irregulars are called 'late-type' galaxies (here ETG and LTG, respectively). 

In a binary classification (early or late-type galaxies), though, we are forcing the galaxies into one of two classes,  while the  classification could be more gradual, reflecting the complexity of galaxy shapes, such as when using the Numerical Hubble types. To account for this, the network architecture in this work was slightly changed when compared to the one used in BOM21, in order to make the probabilities of ETG or LTG not complementary, i.e. not necessarily summing to one.
In fact, these probabilities are generated independently in a way that a galaxy can have a high probability (higher than the DL threshold, see Figure \ref{fig:probs}) of being both ETG and LTG. Galaxies that have a high probability of being both ETG and LTG  are designated here as {\it Amb1}. 
This brings an interesting {\it ambiguity} to the model that can be explored to make the classification more gradual: 
a galaxy now can be classified as neither ETG nor LTG, and will be ascribed to class {\it Amb0}.   

\begin{figure}
    \centering
    \includegraphics[width=0.5\textwidth]{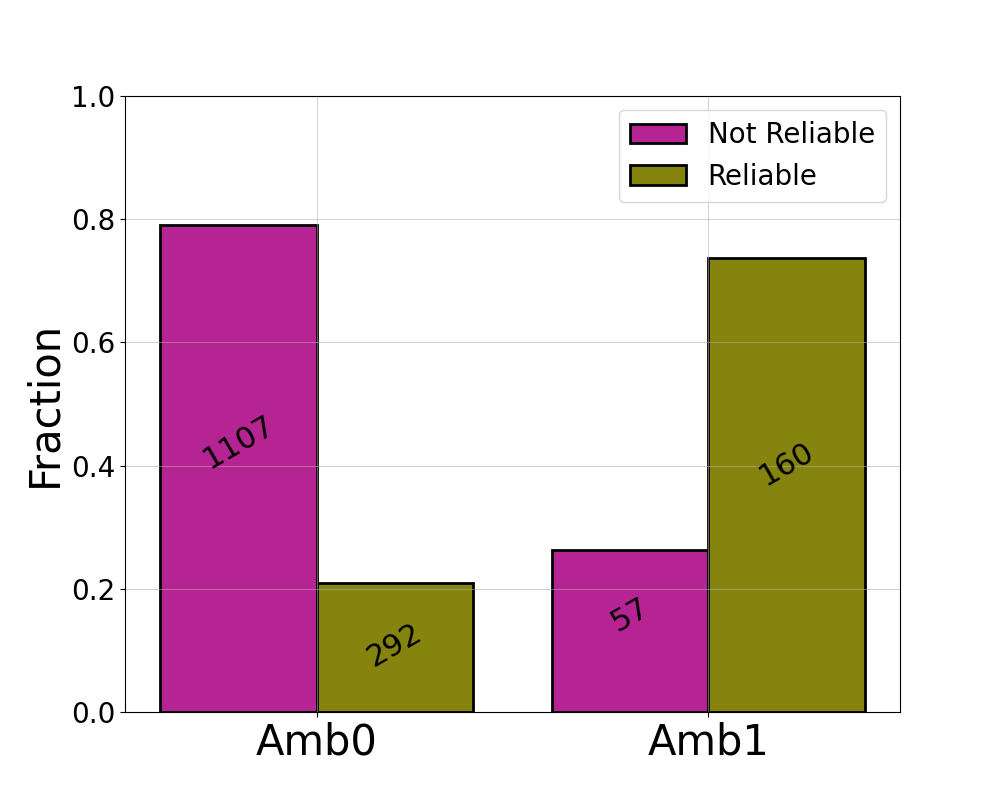} 
    \caption{
     Normalized fraction of galaxies that belong to class Amb0 (left), i.e. galaxies that have a low probability of being ETG or LTG, and of galaxies belonging to the class Amb1 (right), i.e. galaxies with a high probability of being ETG and LTG, classified as no-reliable stamps (blue) and reliable stamps (orange).}
    \label{fig:ambiguity_reliability}
\end{figure}

In our results, as shown in Figure \ref{fig:ambiguity_reliability}, we can see that most of the galaxies that had a low probability of being a ETG or LTG ({\it Amb0}) were also classified as non reliable stamps, while those with higher probability of being either ETG or LTG ({\it Amb1}) were also classified as reliable.
We note here that the galaxies with high probability of being non reliable stamps and with a low probability of being ETG or LTG, are the highest in number (1107), while the majority of galaxies that have high probability of being ETG or LTG (160) are classified as reliable stamps.

\begin{figure*}
\centering
\includegraphics[width=1\linewidth]{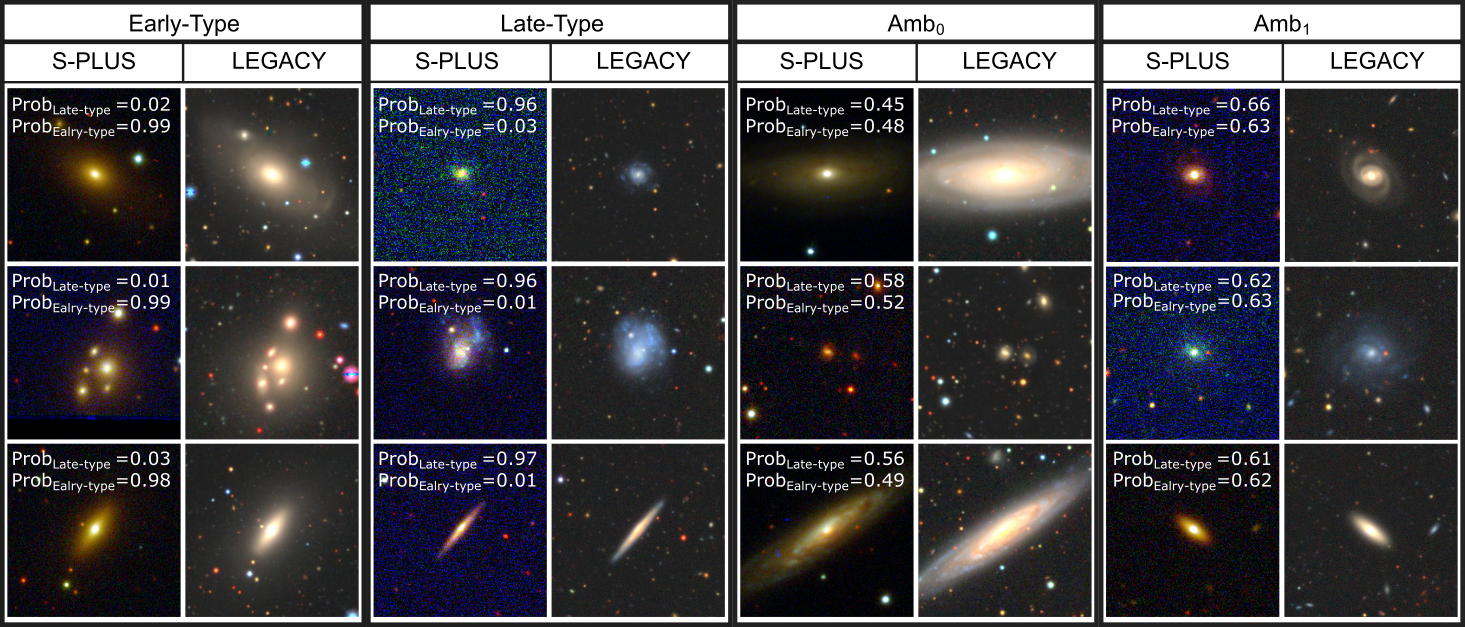}
\caption{In the first two panels we show some examples of stamps that were classified as Early-Type (first panel) or Late-Type (second panel). In the last two panels we have examples of stamps that would fall in the Ambiguous classification. $\textrm{Amb}_\textrm{0}$ are those stamps that had a low probability of being Early-Type and also Late-Type galaxies according to the defined threshold ($\simeq 0.6$). In the other hand, we have $\textrm{Amb}_\textrm{1}$ which are those objects that the model gave a high probability of belonging to both classes. Each panel is made with the same objects taken respectively from S-PLUS and LEGACY survey.}
\label{fig:example_classes}
\end{figure*}

Figure \ref{fig:example_classes} shows examples of reliable stamps, as defined using the 12 S-PLUS images, see Section \ref{sec:Non reliable Stamps}, of galaxies belonging to the four different classes (ETG, LTG, $Amb_0$, $Amb_1$),  from S-PLUS and Legacy surveys. The Legacy data are typically four magnitudes deeper than S-PLUS images and reveal faint outer features, so they can be used to understand the effects of depth and resolution in the ability of the DL method classify objects. In general, galaxies falling in the ETG class are ellipticals (left column, top and middle rows) or lenticulars (left column, bottom row). The LTG objects are either spiral or irregular galaxies (second column, first and middle rows), while the third row shows a disk-dominated lenticular galaxy. In Section \ref{sec:comp} we compare the classification presented in this work with other works. 

Galaxies are classified as {\it Amb0} or {\it Amb1} as the  result of a combination of factors: 

\begin{enumerate}

\item faint/high redshift spiral galaxies can be misclassified as early-type galaxies, due to the  pixel resolution and survey depth, which reflects in the difficulty of identifying the presence of spiral arms.  In turn, they might  present green dots of star formation, rendering them neither ETG nor LTG (see third column, middle panel of Figure \ref{fig:example_classes}); 
 
\item clumpy star-forming galaxies could also be assigned to neither class, due to their un-smooth appearance and the absence of  clear spiral patterns, see third column top and bottom panel of Figure \ref{fig:example_classes};

\item bulge-dominated spirals (see last column, top and middle images), due to the low surface brightness of their spiral arms, clearly visible in the Legacy data, but close to the image noise in S-PLUS data, may have a high probability of being both ETG and LTG galaxies. 

\item lenticular galaxies can also be found in the $Amb_{1}$ class, in particular lenticular galaxies with $B/T \simeq 0.5$ are associated to both classes, due to their hybrid nature. These results will be further discussed in Section \ref{sec:S0formation}.  
\end{enumerate}
In the next section we show that some of the non-reliable stamps (NRS) are actually extraordinary objects.
\subsection{Extraordinary Non Reliable Stamps (NRS)}
\label{sec:extraordinay trash}

\begin{figure}

\centering
\includegraphics[width=1\linewidth]{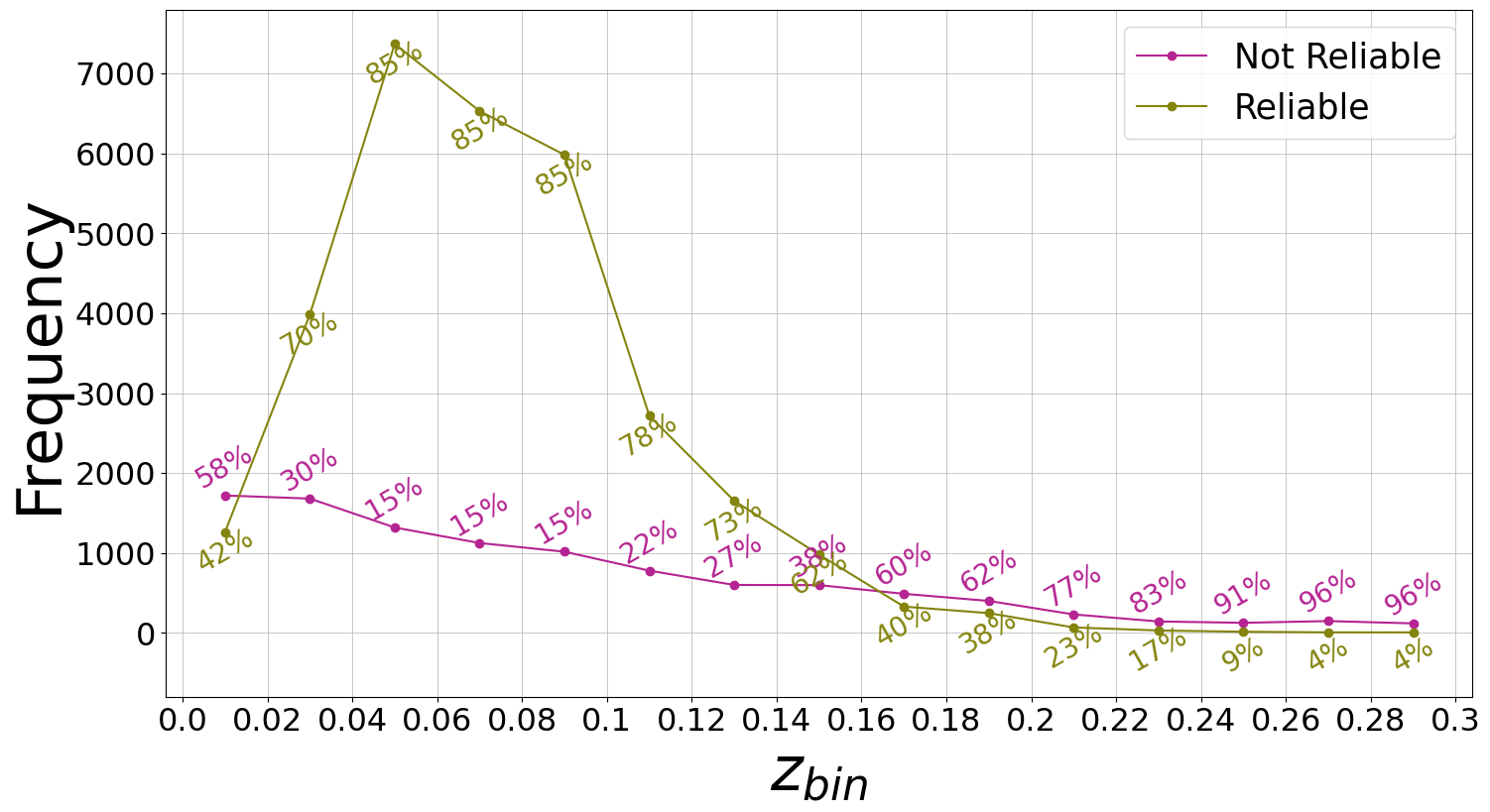}
\caption{Number of NotReliable and Reliable stamps in increasing redshift bins.}
\label{fig:noRelRel_FIG}
\end{figure}

\begin{figure*}
\centering
\includegraphics[width=0.90\textwidth]{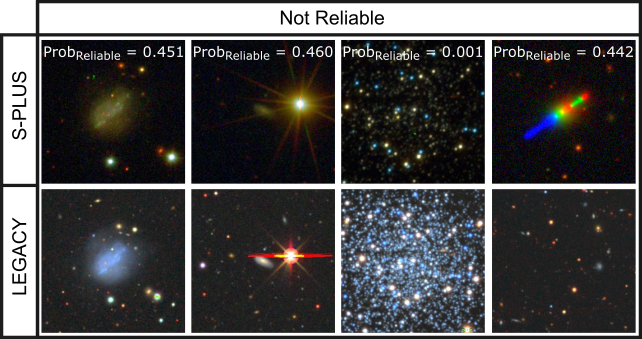}
    \caption{
    \label{fig:trash}
     {Example of not reliable stamps, from S-PLUS data (top) and LEGACY data (bottom). In the last column, it is visible an artifact, the third column present a crowded field, in the second column we find a saturated star compromising the galaxy image, and, finally, in the first column and irregular galaxy.}
    }
\end{figure*}

Figure \ref{fig:trash} shows some example of objects  identified as NRS. Generally, they are objects nearby saturated stars or crowded fields. In fact, even if we select the sample of objects to be analyzed maximizing the probability of being galaxies, see Section \ref{sec:sample selection}, contaminants still appear in the sample and the deep learning code makes a great job in identifying spurious objects.
The number of NRS is nearly constant with  redshift, as shown in Figure  \ref{fig:noRelRel_FIG}, while the number of Reliable stamps decreases with increasing redshift.

\begin{figure*}
\centering
\includegraphics[width=1\textwidth]{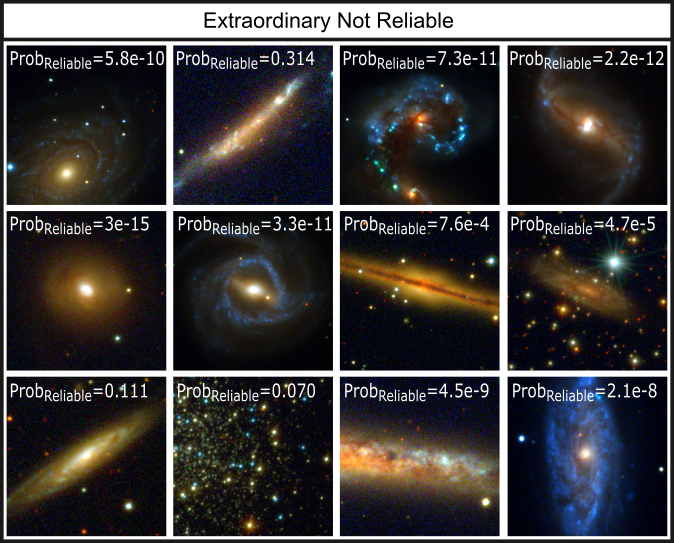}
    \caption{
 Example of extraordinary not reliable stamps. Large objects, whose projected radius is larger than the image stamp; star-forming galaxies, where more than one clump can be identified as an independent object during the catalogue extraction \citep{Almeida-Fernandes2022}; irregular galaxies as NGC 4038/NGC 4039; and dense regions of stars, maybe galactic clusters, can be found among the not reliable stamps. The probability of being a reliable stamp is given in the top left of each panel.}
 \label{fig:extraordinarytrash}

\end{figure*}

On the other hand, peculiar galaxies, especially if with clumpy star formation, or galaxies with a projected size larger than  the stamp might fall in the category of NRS, as shown in Figure \ref{fig:extraordinarytrash}. Somehow, the deep learning method not only allows us to identify unwanted objects, but it also helps in finding peculiar objects, of high interest/relevance. 

\subsection{Morphology as a probe of galaxy evolution and large scale structure formation}

Galaxies evolve through time via different mechanisms: major and minor mergers, secular evolution, harassment, stripping, and strangulation  \cite{Gunn,Salamanca,Quilis,Kronberger,Byrd,Bournaud}. Many of these processes are 
environment-dependent, i.e., they can occur only in clusters of galaxies (strangulation), or they are more likely in the field or groups (e.g., mergers). In general, it is now believed that minor mergers are more common than major mergers and that they are the main responsible for mass build-up in galaxies \citep{Bournaud2007}.
Different evolutionary scenarios leave specific imprints on the galaxy morphology; i.e. major mergers tend to disrupt the stellar orbits, resulting in pressure dominated systems, characterized by an elliptical shape. On the other hand, secular processes, or environmentally driven mechanisms, as ram-pressure stripping, affect more the gaseous component, chasing the star formation. Moreover, a morphology-density relation had been proven in  the last decades \citep{Dressler, Dressler80,Capellari2011,Buitrago2013}, where ETG inhabit the densest regions of the Universe, while spiral and irregular galaxies are more 
common in the fields.
Galaxy morphologies are a powerful proof of galaxy evolution as well as structure formation, as we will discuss in the next subsections. 
In here we use only objects classified as reliable stamps with $photo_z$ $odds > 0.4$ and $r < 17$ mag.
The magnitude have been corrected for galactic extinction using   the Clayton, Cardelli and Mathis \citep{Cardelli1898} dust law.

\subsubsection{Correlation between morphology and colors }
\label{subsec:characterization}

\begin{figure*}
    \centering
    \includegraphics[width=0.5\textwidth]{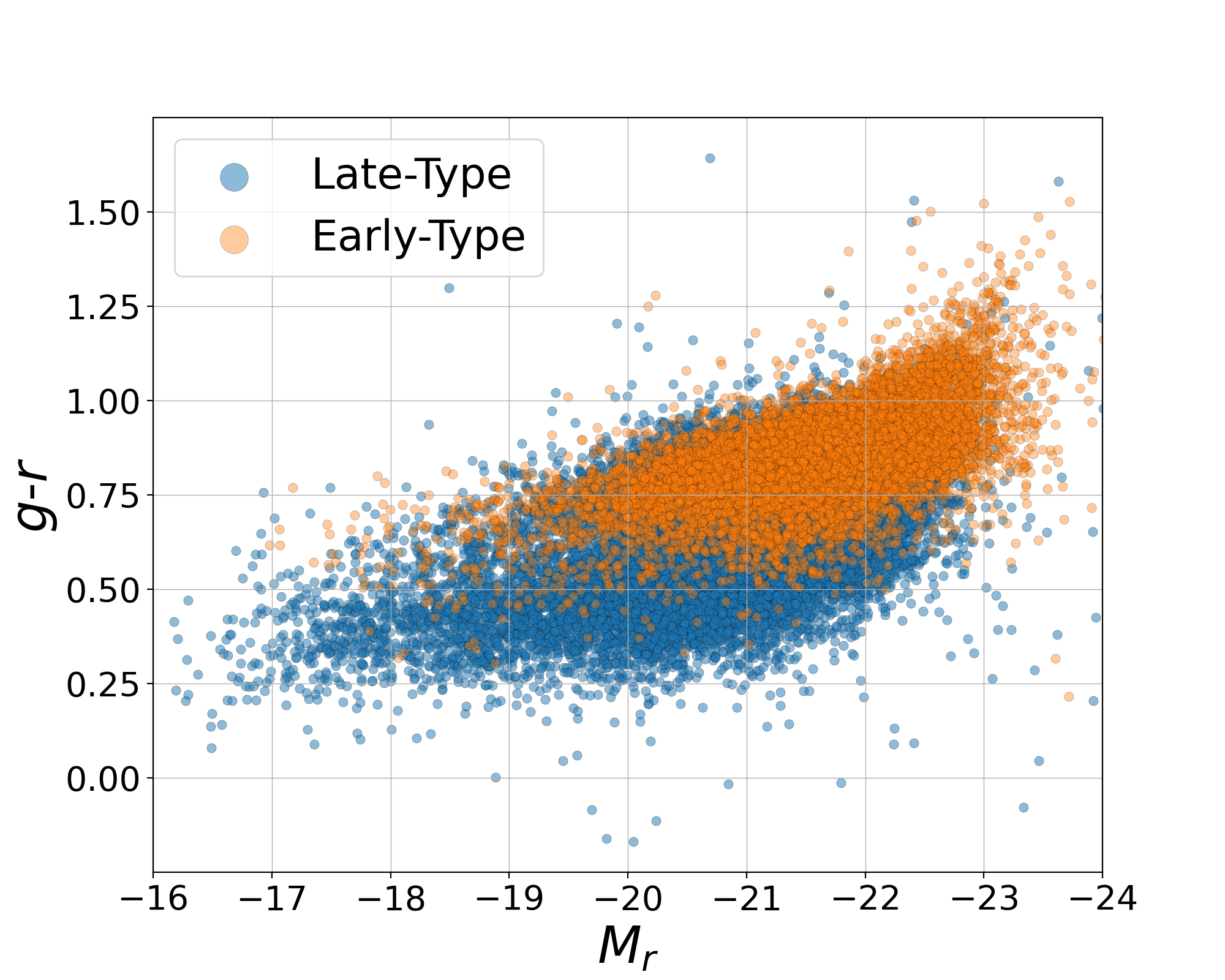}\includegraphics[width=0.45\textwidth]{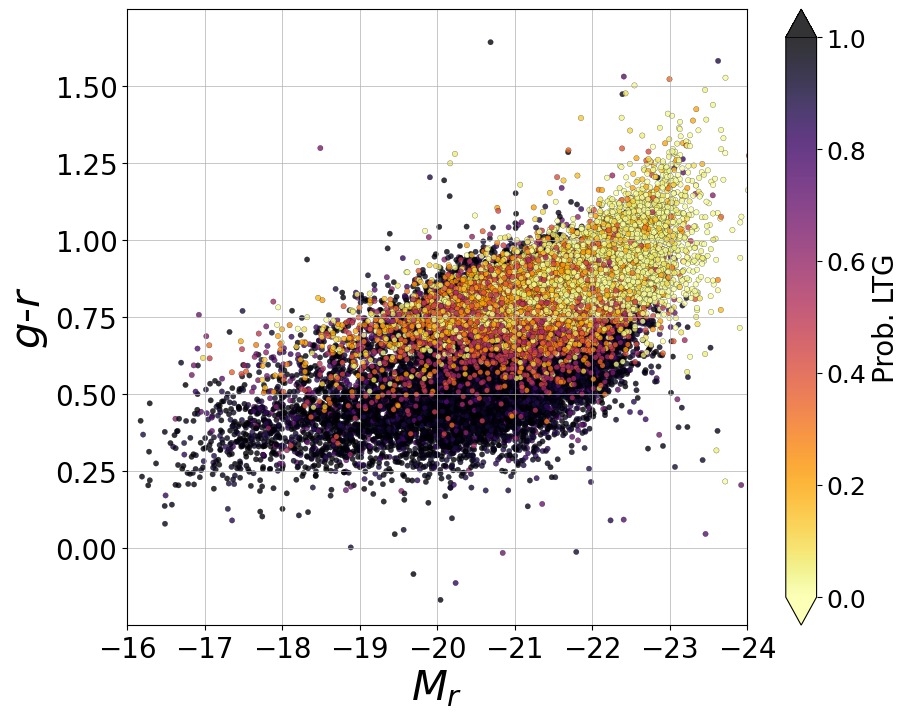}
    	\caption{Colour-magnitude diagram. (g-r) colour versus the absolute magnitude in r-band, calculated using standard cosmological parameters and the luminosity distance ($D_{L}$) estimated from the photometric redshift. The {\it left} panel shows the bin classification, while the {\it right} panel, is colour coded according to the probability of being a late-type galaxy.}  
    \label{fig:colour-mag}
    
\end{figure*}

Figure\,\ref{fig:colour-mag} shows the color-magnitude diagram (g-r colour vs r-band absolute magnitude), color coded according to the galaxies' morphologies. The left panel presents the dual classification, where elliptical galaxies are shown in orange and spiral galaxies, in light-blue, while in the right panel the colour scale shows the probability of being a spiral galaxy. It can be seen that elliptical (quiescent) galaxies inhabit the red sequence 
while spiral (star-forming) galaxies are mostly found in the 'blue cloud' as expected according to their dominant stellar populations \cite[see, for example,][]{Wong2012,limadias2021,Khanday2022}. 
Interestingly, in the right panels it is possible to see that the probability of being spiral increases nearly from 0 to 1, going from the red cloud to the blue sequence, in a  continuous manner. The intermediate region, where the probabilities range around $0.5$, is known as Green Valley (see, for example, \citealp{Zibetti2007}) and it has been largely studied as a region of transition, where late-type galaxies could be quenching their star formation, turning into late-type galaxies, or early-type galaxies could be 'rejuvenating', due to some interaction with other galaxies or accretion of gas \citep{Smethurst2015}. 
The morphology seems to be reflecting this transformation since the quenching is 'removing' the spiral arms, decreasing the probability of being a spiral galaxy. On the other side, a sparkle of star formation in an early-type galaxy could create clumps, or star-forming regions, that would increase the probability of being a spiral galaxy.

\subsubsection{Morphology-density relation}

There is a  connection between the environment a galaxy live in  and its morphology \citep{Dressler80}, but both the galaxy stellar population and environment evolve with time. While the galaxy stellar population is related to the galaxy mass  (more massive galaxies are more metal rich at a given time, \citet{Leaman2013} and gas content, the morphology is more related to the environment (spiral galaxies tend to live in low density environment, ellipticals in the center of galaxy clusters). Yet, a merger, whose probability is dictated by the environment a galaxy lives in, would affect both the galaxy mass and stellar population. Note that 20\% of high mass ($M_{*}$ $\ge$ $10^{9.5}$ $M_{\odot}$) galaxies have experienced a major merger since $z \simeq 6$ \citep{Ventou2017} and minor mergers, accretions, fly-bys are very common in the history of the Universe.

\begin{figure}
	\centering

\includegraphics[width=1.\linewidth]
{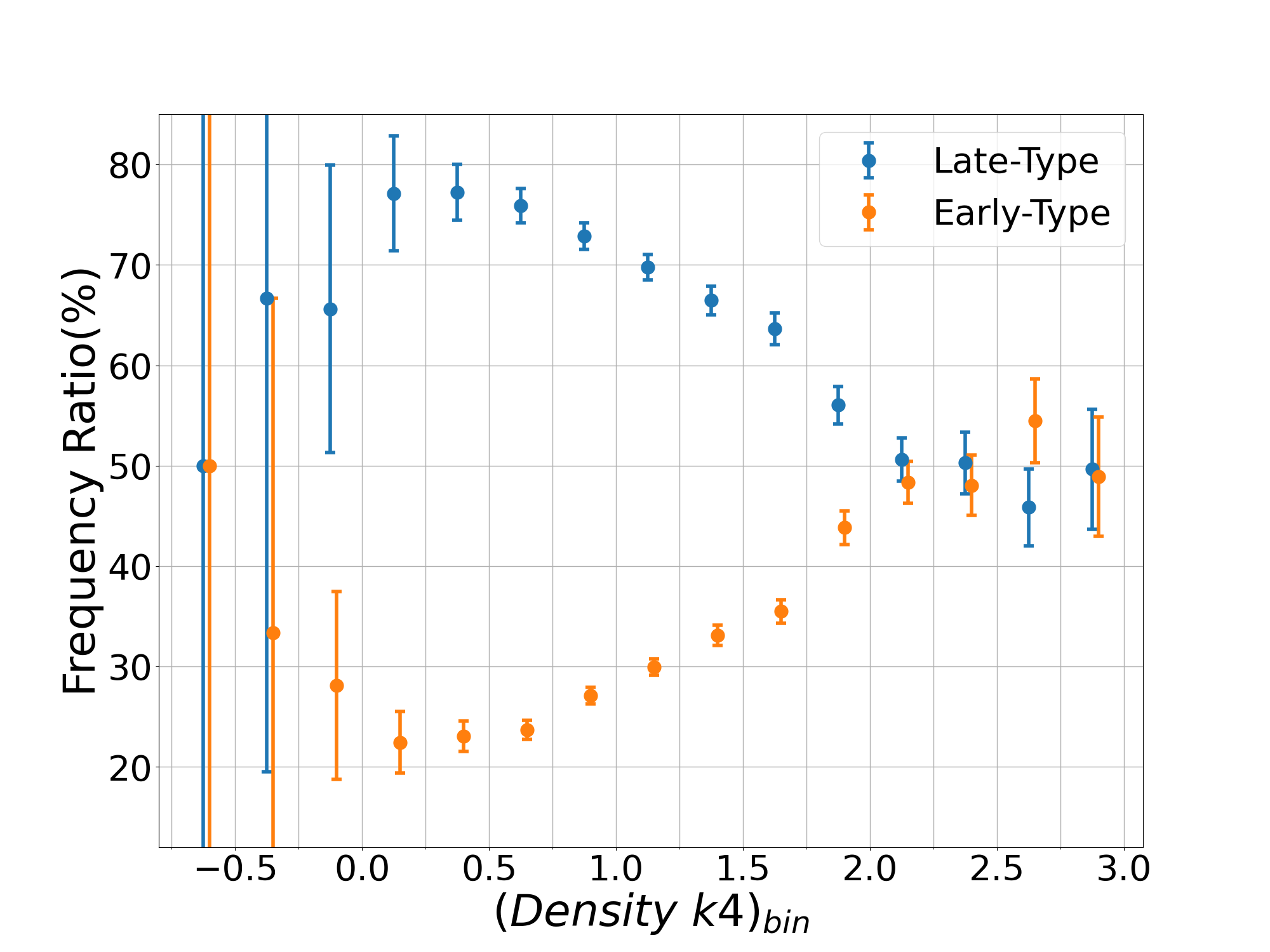}\\
\includegraphics[width=0.5\textwidth]
{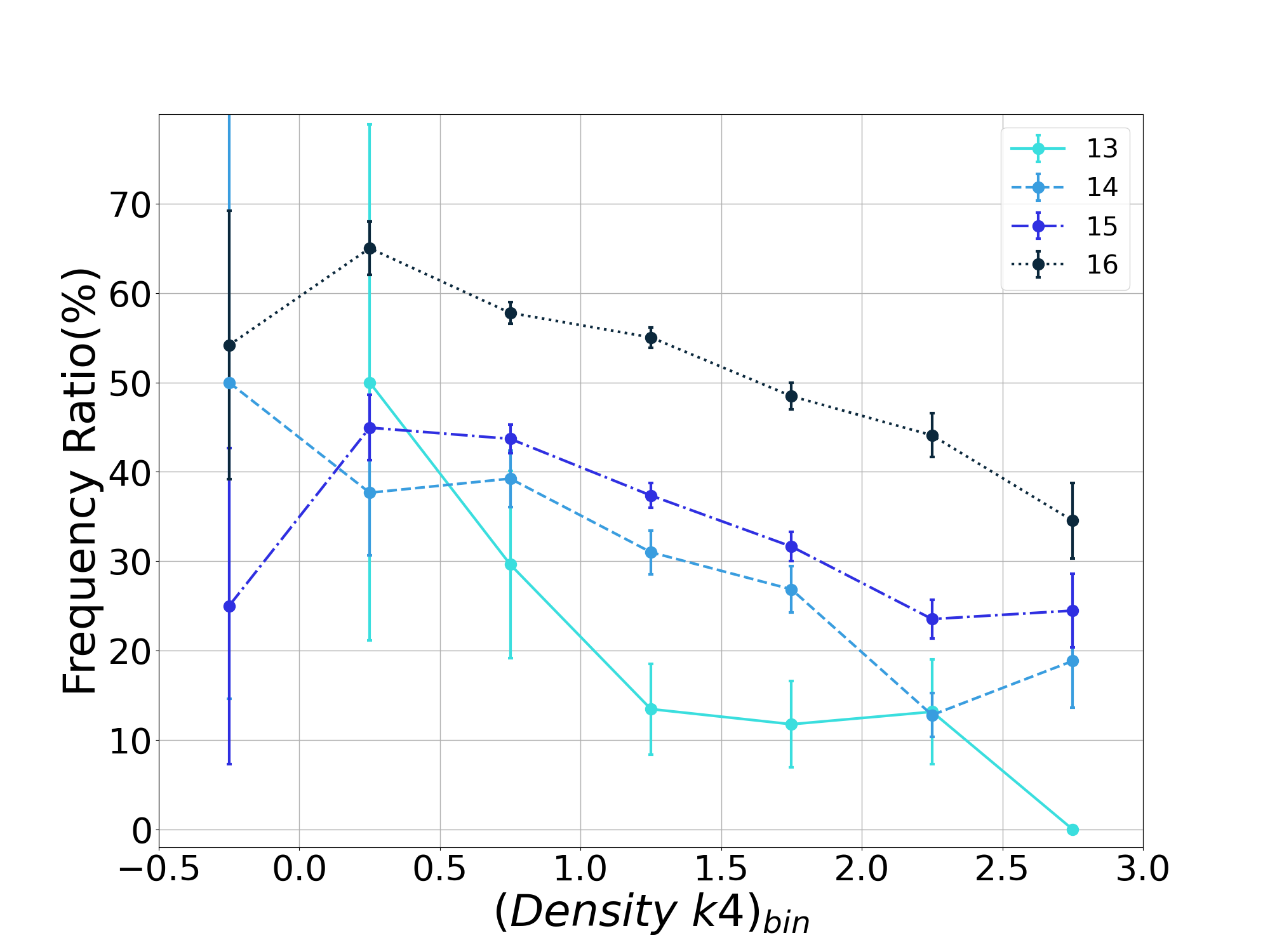}\\
\includegraphics[width=0.5\textwidth]{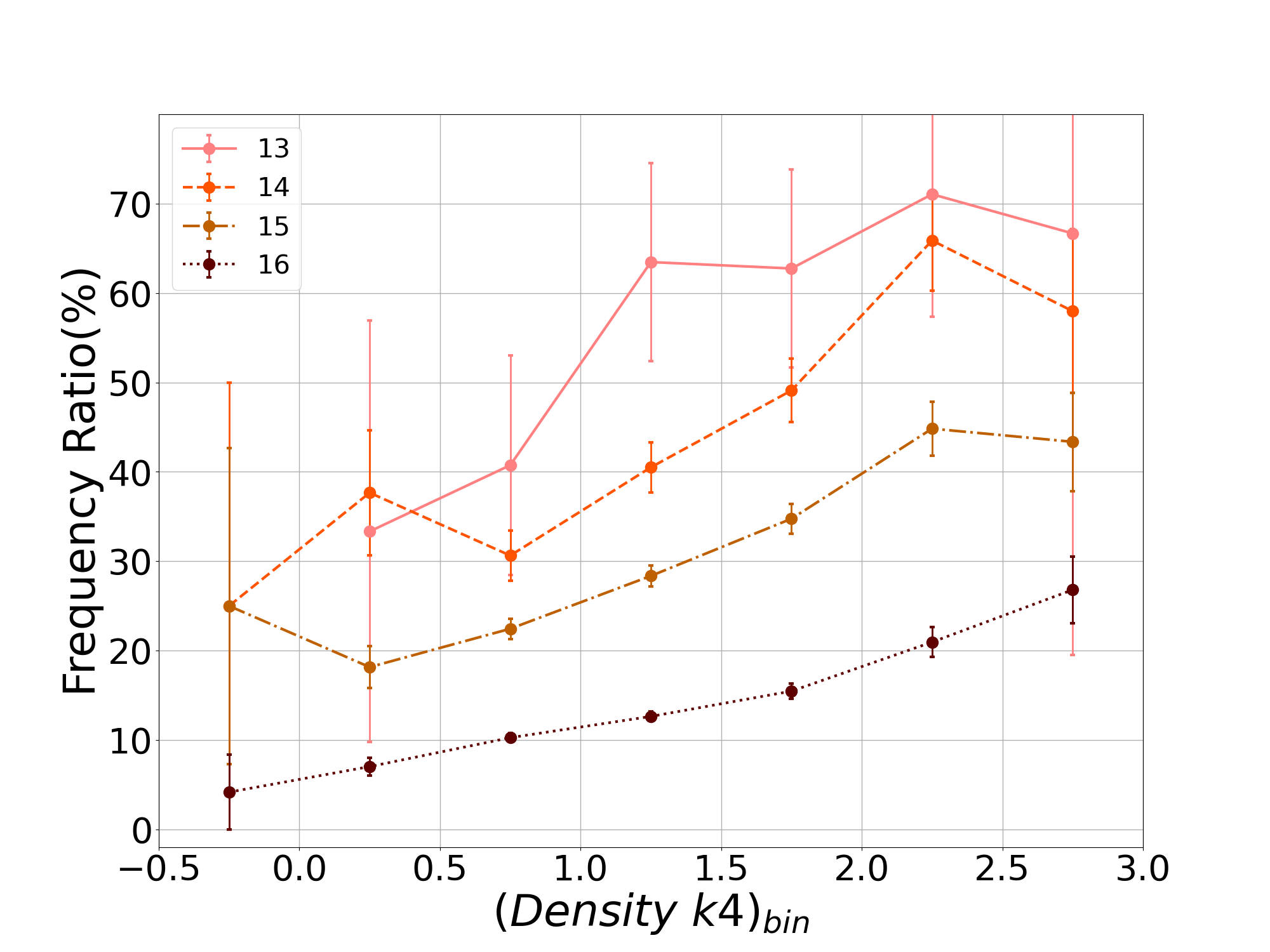}
        
        	\caption{Morphology-density relation. Normalized fraction of late and early-type galaxies, with a probability of belonging to a given class higher than $0.9$, for increasing density bin. The k4 estimator traces the  local densities. The {\it top} panel shows the total distribution, while in the {\it middle} panel, it is divided in  different magnitudes for Late-type galaxies, which  are drawn in a scale of blue. In the {\it bottom} panel early-type galaxies are colored in orange. For easier comparison, lines of equal  magnitude have the same style.}
    

	\label{fig:env}
\end{figure}

We use a K-Nearest Neighbor method, with $k=4,5,10$ \citep{Baldry2010}, to recover the projected density of the environment a galaxy live in, where $k=4,5$ refers to local environments, while $k=10$ is related to larger scales. 
Specifically, the density ($\Sigma_k$) at any given k is:
\begin{equation}
    \Sigma_k = \frac{k}{\pi D^{2}_{k}} \frac{1}{\psi(D))}, 
\end{equation}
where $k$ is the $k$ nearest neighbour, D is the comoving distance and $\psi(D)$ is a selection function to correct for the Malmquist bias \cite[e.g.][]{Santana-Silva2020}.  
Figure \ref{fig:env} presents the number density of late ($Prob_{late type}>0.9$) and early type ($Prob_{early type}>0.9$) galaxies for increasing  $k=4$ density measures. The left panel of Figure  \ref{fig:env}, presents all galaxies with magnitude $r\le17$, while the right panel split them into magnitude bins (represented by different line shapes, see Figure legend).   
Early-type galaxies are identified by  orange/red lines,  while late-type galaxies are shown as cyan/blue lines. The morphological classification provided in this work clearly reflects the morphology density relation, with early-type galaxies occupying the densest regions, and late-type galaxies being the dominant population in the field/low-density regions, see left panel in Figure \ref{fig:env}. When looking at the magnitude dependence of the morphology-density relation, we see that it holds for different magnitudes bins, where  the number density of early-type galaxies increases with increasing densities, while the opposite trend is found for late-type galaxies.
Finally, we observe that the crossover density is lower for more
luminous objects, indicating a correlation between lower densities and higher luminosities.

\begin{figure}
    
    \centering
    \includegraphics[width=1.\linewidth]{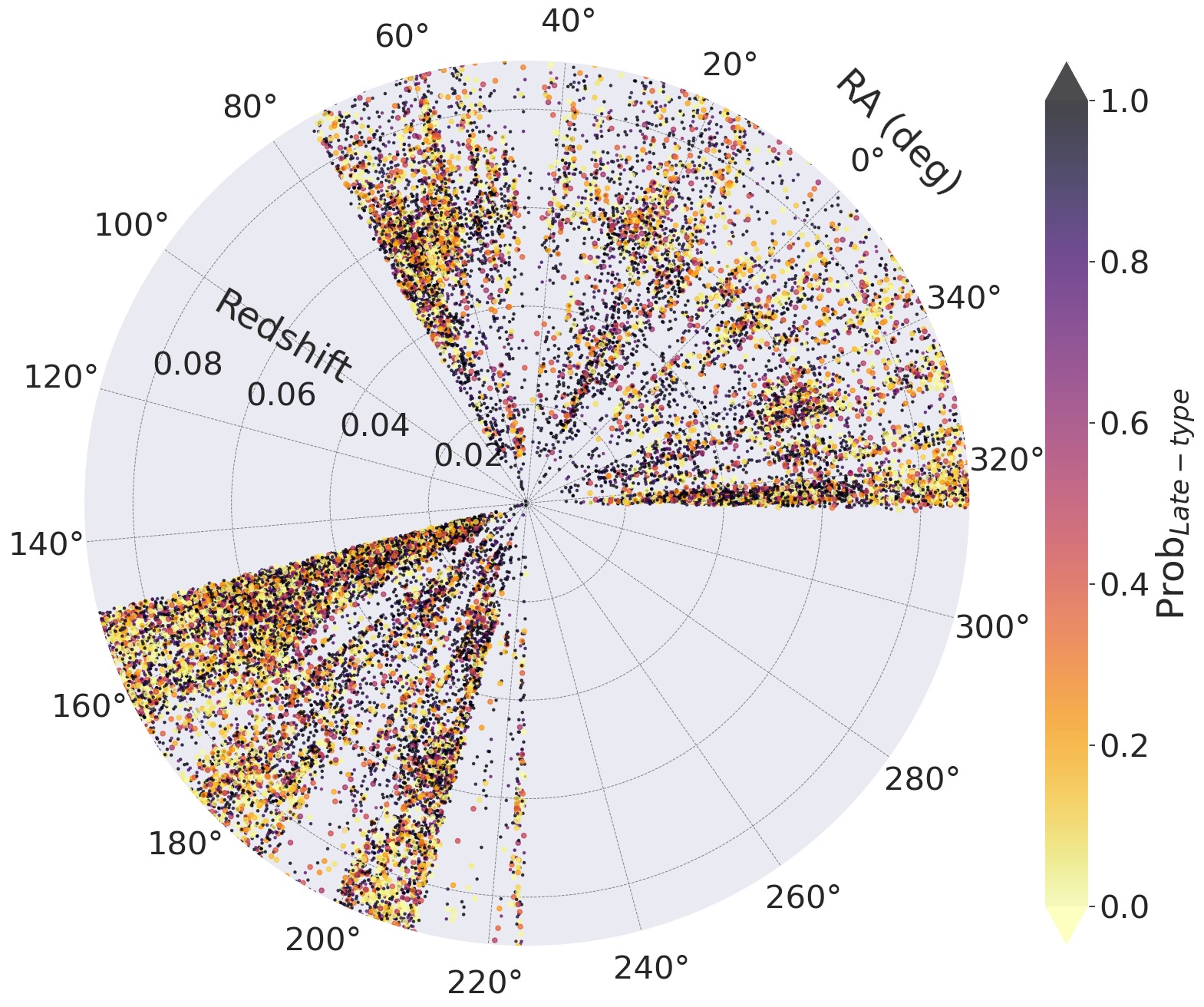}	
    
    \caption{Redshift distribution up to z = 0.08 for galaxies colour coded according to their probability of being late-type galaxies galaxies. Galaxies with dual classification as early-types are drawn as larger circles, for a better vision.}
    \label{fig:polar}    
    
\end{figure}

\subsubsection{Large scale structure as traced by galaxy morphology}

Galaxies trace the large-scale structure of the Universe, yet they account only for  $\simeq 20\%$ of the total matter  \cite[e.g.][]{Planck}, and their physics is  affected by non-gravitational mechanisms such as baryonic effects, radiation pressure, feedback, etc. 
A simple but powerful tool that can bridge the gap between galaxies and the DM distribution is the Halo Model \citep{HaloModel}. 
One of the consequences of that description is that galaxy abundances and their properties (such as stellar mass, color, morphology and star formation rate) can be traced back to the DM halos and sub-halos, as well as their properties (such as mass, age, concentration and spin) \cite{HaloGalaxy}.
From a large-scale structure perspective, the correlation function of the DM halos is related to the correlation function of the DM particles by the halo abundance, bias, and halo density profile.
To a good approximation, more massive halos are less abundant and are more highly biased with respect to the DM field, but other halo properties such  as concentration, age, and even spin (angular momentum) also play an important role \citep{SecBias}. 
Galaxies that populate halos and their sub-halos  inherit those properties, including their bias -- but they can also bring additional information that is not manifested in the halo properties, and which are indicative of the baryonic mechanisms such as ram-pressure stripping. 
Galaxy morphology is one of the additional indicators that can help distinguish between different types of halos and their environments, leading to a more accurate and precise description of the correlation functions of those tracers.

Figure \ref{fig:polar} shows the redshift distribution, up to $z \simeq 0.08$ \citep{bamford2009galaxy} for  galaxies colour coded according to their probability of being late type, in order to characterize how
morphologies evolve over time. Early type galaxies, plotted with larger symbles,  are generally more clustered. The presence of galaxy clusters is emphasized by the Finger of Gods effect, caused by the peculiar velocities of galaxies that
deviate from the Hubble flow.

\section{Discussion and concluding remarks}
\label{sec:discussion}

\subsection{Comparison to other surveys}
\label{sec:comp}

\citet{Vega-Ferrero2021} used DES galaxies with reliable morphological classification to assess whether CNNs are able to detect features that human eyes do not. To do that, they simulate the appearance that well morphologically classified DES galaxies would display at high redshifts, making them fainter and smaller. They find that, despite some of the features that distinguish ETGs from LTGs vanish after the simulation, the models are still able to correctly classify galaxies with an accuracy greater than 97\%. The main conclusion of that work is that it is possible to correctly classify galaxies from faint and small size images using CNNs models, satisfying the following conditions: final apparent magnitude below  $m_r(z) < 22.5$, and the size of the final image  larger than 32$\times$32 pixels. DES data
(DES DR1, \cite{Abbott2018a}) has a  median co-added catalog depth of 
$m_r = 24.08$ at signal-to-noise ratio S/N = 10, with a
pixel size of 0.2636

In comparison, S-PLUS has a scale of 0.55 "/pixel and a depth in r-band of $m_r=19.6$ at signal-to-noise ratio S/N = 10 \citep{Almeida-Fernandes2022}, resulting in  lower resolution when compared with DES data, as clear from Figure \ref{fig:example_classes}.  

\begin{figure}
    \centering
    \includegraphics[width=0.495\textwidth]{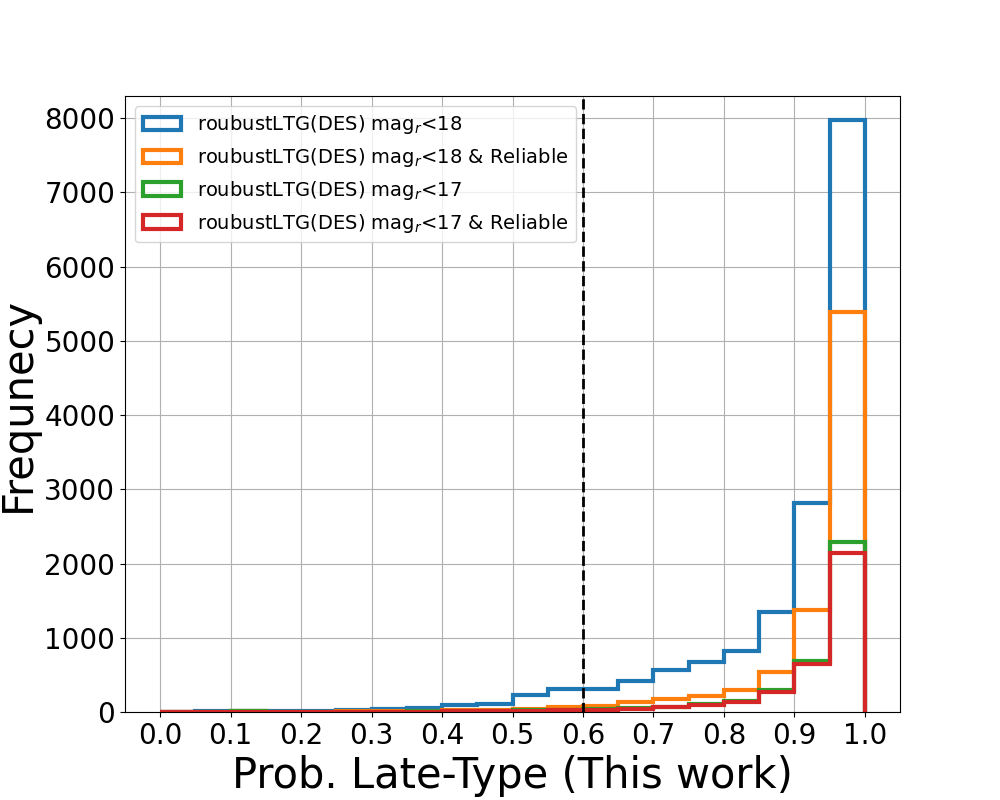}\\
    \includegraphics[width=0.495\textwidth]{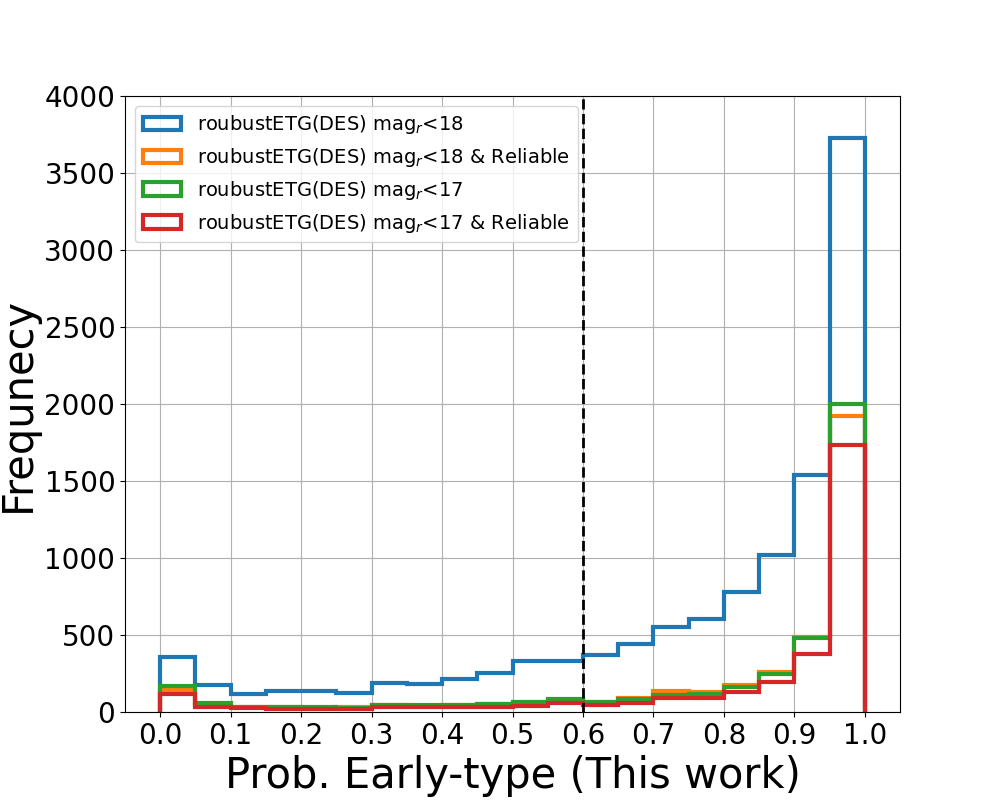}\\
     \includegraphics[width=0.495\textwidth]{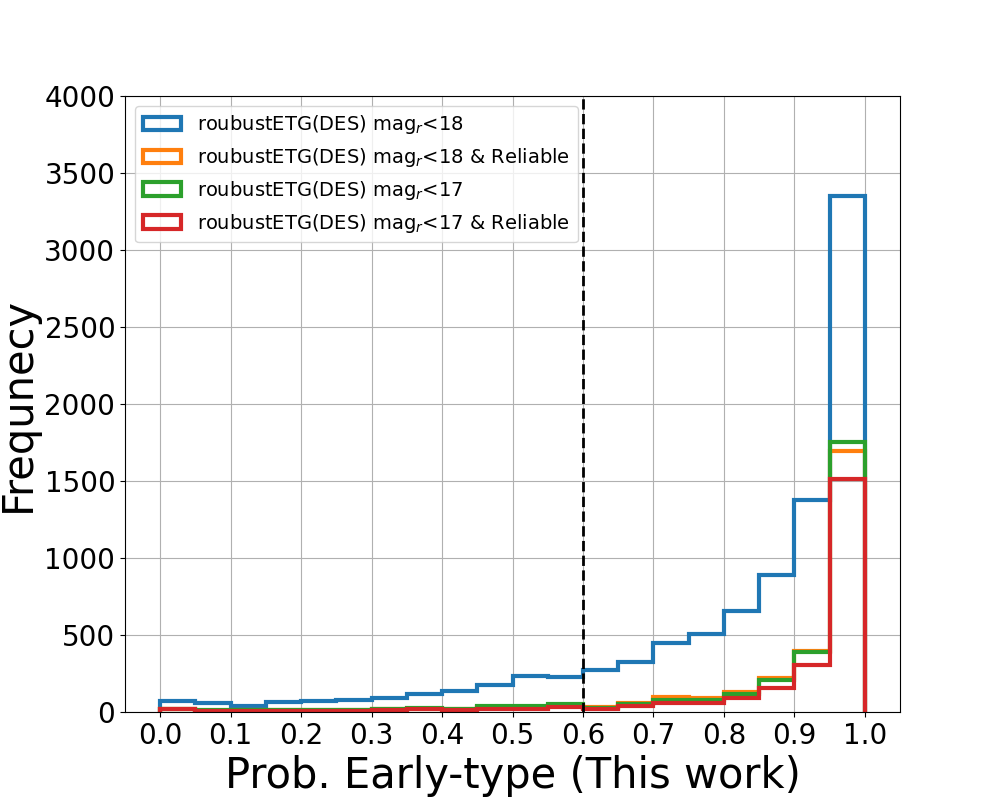}

    \caption{In this figure we present a histogram showing the proportion of galaxies that were classified in accordance with our classification. The {\it top panel} represents the galaxies classified as roubust LTG by DES and the {\it middle} one represents those classified as roubust ETG by DES. The {\it bottom panel} is like the {middle one}, but for robust ETG with $b/a > 0.7$, see text. The histograms were made using the probability of belonging to the corresponding classes obtained in this work, with the dashed line being the threshold used in our classification, in other words, every galaxy that stands in the right side of this line was classified equally by both works.}

    \label{fig:DESxSPLUS}
    
\end{figure}

S-PLUS DR3 and DES DR1 overlap, see Figure \ref{fig:footprint}, resulting in a combined catalogue from \cite{Vega-Ferrero2021} and this work of 36183 galaxies, brighter than $m_r<18.0$ and with a mean redshift of $z_{ml}=0.11847$. Comparing the classification presented in this work,  considering the depth of the DES images, allows us to investigate the goodness of the classification and the advantages of combining the results of the two DL codes, i.e. studying the reliable early and late-type classification.  
In Figure \ref{fig:DESxSPLUS}, top panel,  we present a histogram of the probability of being late-type galaxies obtained in this work, for galaxies classified as 'robust spirals' ($FLAG_{LTG}==5$)  in \cite{Vega-Ferrero2021}. The dashed line indicates the threshold used in this work, in other words, every galaxy that stands in the right side of this line is classified as a spiral in both works. The blue histogram shows the distribution of the probability of being a LTG for all 'robust spiral' galaxies  and presents the larger discrepancy with \cite{Vega-Ferrero2021}. The Orange histogram shows the probability of being LTG for all 'robust spiral' stamps classified as reliable according to the second DL model, see Section \ref{sec:Non reliable Stamps}. The green and red histograms represent all 'robust spiral'  galaxies brighter than $r<17$ mag, and among them all the ones classified as reliable stamps, respectively. 
The middle panel shows the same comparison for early-type galaxies. There is a non negligible fraction of galaxies with zero probability of being early-type galaxies in this work, but classified as elliptical in 
\cite{Vega-Ferrero2021}. In the bottom panel, we reproduce the same plot, now including only objects with $b/a>0.7$. This choice drastically decreases the number of discrepant classifications.
Similar results are obtained when performing the same comparison with \citet{Cheng_2023}.

\begin{figure}
    \centering
   \includegraphics[width=0.495\textwidth]{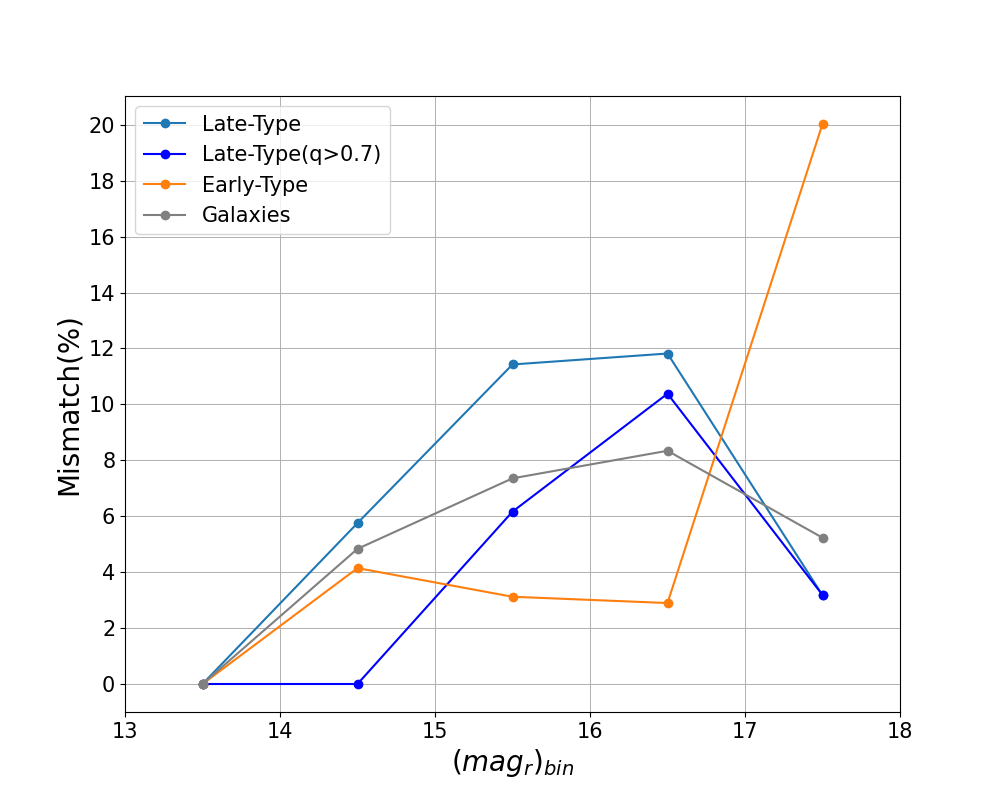}
    \caption{Fraction of objects with different classification between this work and \citet{Vega-Ferrero2021} for different magnitudes bin. Galaxies classified as late types in this work and as early types in VF21 are shown by the cyan line, while the blue line present the same selection but excluding edge-on objects, i.e. by imposing that the axis ratio $q=b/a>0.7$. In orange is shown the behavior of objects classifies as early-type galaxies in this work and as spiral in VG21. The grey line shows the global mismatch (the sum of the cyan and orange line), which indicate that the discrepancy in the classification in the two works increases with decreasing magnitude, as expected given the lower resolution and depth of S-PLUS in comparison with DES. We note than the total number of galaxies that are classified as 'robust'  in \citet{Vega-Ferrero2021} and as reliable stamps in this work decreases after $r = 17$, causing the improvement of the match at $r = 18$.}

    \label{fig:mismatch}
    
\end{figure}

In Figure \ref{fig:mismatch} we present the fraction of misclassified objects for  different magnitudes bins, and in Figures available in the appendix\footnote{The appendix is presented as an online supplementary material.} it is possible to find examples of objects classified differently in the two papers. It is noticeable that in many cases of galaxies classified as early types in this work and late types  in \citet{Vega-Ferrero2021}, they are multiple object images, Low Surface Brightness, bulge dominated spiral galaxies, or faint/compact spiral galaxies, see Section \ref{sec:ES class}. On the other hand, objects classified as late types in S-PLUS and early types in \citet{Vega-Ferrero2021} are often disk dominated (edge-on)  lenticular galaxies or merger/disturbed systems.

In conclusion, the classification presented in this work is in agreement with  \cite{Vega-Ferrero2021} with an average confidence level of $\simeq 92\%$ up to  $r<18$, for ETG and $\simeq 96\%$, for LTG, up to  $r<17$. The mismatch for ETG increases to $20\%$ for objects fainter than $r\simeq17$ as a result of the fading of the spiral arms in the S-PLUS images.  On the other hand, the mismatch  for LTG is mostly caused by the association of disk-dominated lenticular galaxies or edge-on red spirals \citep{Sodre2013} to this class in this work, while there is a perfect agreement between the two classifications when considering only objects with $q=B/A>0.7$ and $r<14.5$, see blue line in Figure \ref{fig:mismatch} and in the histogram presented in the appendix A. Implications from these results are further discusses in Section \ref{sec:S0formation}. Moreover, a visual inspection of the differently classified objects, see the panel figures in appendix A, reveals interesting objects resulting from a different structure of the DL codes and image depth and resolution, highlighting the importance of a diverse, open and collaborative scientific environment.

\subsection{Combining Morphology and precise photometric redshifts with narrow band surveys: where do galaxies live?}

\begin{figure}
    \centering
     \includegraphics[width=0.49\textwidth]{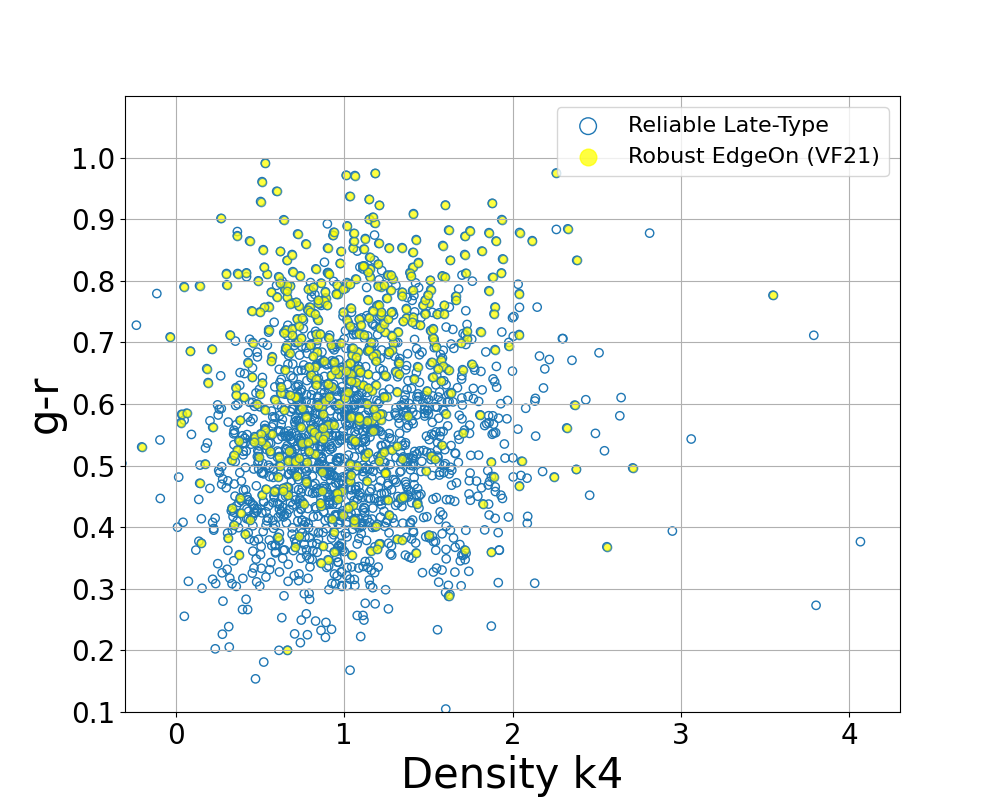}
     \includegraphics[width=0.49\textwidth]{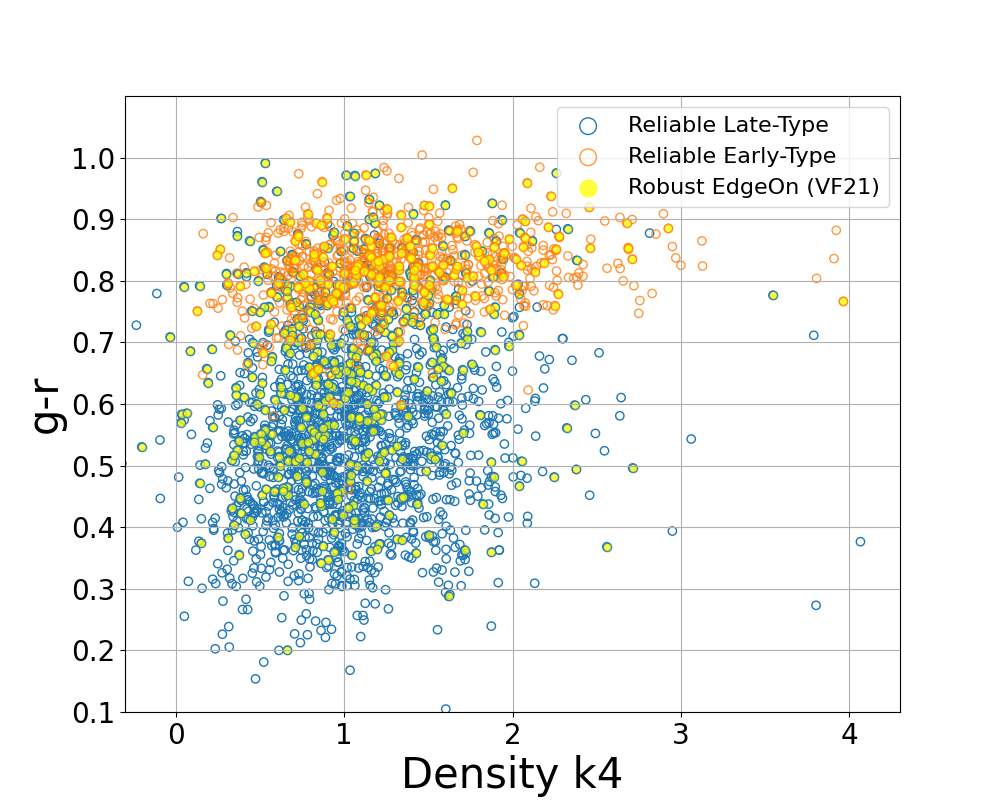}\\
    \caption{{\it Top:} (g-r) colour vs density k4 distribution for galaxies with $Prob_{LTG}>0.9$ (reliable Late-Type);  and {\it bottom:} including galaxies with $Prob_{ETG}>0.9$ (reliable Early-Type). The filled yellow circles are  classified as robust Edge-on in \citet{Vega-Ferrero2021}. }

    \label{fig:colour_density}
    
\end{figure}

The relation between galaxies' morphology, their mass and stellar population properties, and the environment they live in has been studied in great details and in a wealth of works \citep{Paulino-Afonso2019,Coccato2020}, as well as its redshift evolution \citep{GonzalezDelgado2015}.
Recent works show that the  bulge growth, measured as  bulge over total light ratio, is directly connected with the  quenching of the  star formation \citep{Paulino-Afonso2019,Dimauro2022,werner2022}. Group pre-processing is also found to play an important role in galaxies' star formation quenching and morphological evolution \citep{GonzalesDelgado2022, Brambila2023}.

S-PLUS photometric system allows to retrieve reliable photometric redshifts with a scatter of 0.023 \citep{Lima2022}, and  to recover reliable density estimates (Lopes da Silva et al. in prep.).
In Figure \ref{fig:colour_density} we show the (g-r) colour vs k4-density measures for reliable stamps  with $Prob_{Spiral} > 0.9$ (blue open dots), with $Prob_{Elliptical} > 0.9$  (red open dots), and galaxies classified as edge-on by \citet{Vega-Ferrero2021} as filled yellow circles. The galaxy environment is more correlated to its morphology, than its colour, see Figure \ref{fig:env}.  ETGs have $(g-r)  > 0.7 $. The bottom-left panel shows that the majority of late-type galaxies with   $(g-r)  > 0.7 $ are classified as edge-on in \citet{Vega-Ferrero2021}. As shown in Figure \ref{fig:example_classes} disk-dominated lenticular galaxies can be associated to the late-type class, explaining the red colour of the disky-late-type galaxies. Moreover, edge-on star-forming spiral galaxies might suffer reddening, due to the presence of dust clouds surrounding the disk \citep{bamford2009galaxy,Sodre2013}.

Figure \ref{fig:colour_density} points out that ETG have red colours, $(g-r) >0.7$ and are more common in denser environments. If color is a proxy for galaxy stellar population, these findings would suggest that both the quenching of the stellar population and the environment are connected with the early type morphology. On the other hand, late-type galaxy span a range of colour and their number seems to be more connected to the environment they live in, see Figure \ref{fig:env}.

\subsection{S0 galaxies formation scenarios and a physically motivated morphological classification}

\label{sec:S0formation}

Lenticular galaxies  are characterized by a hybrid morphology, with a bulge and a disk as spiral galaxies, but without spiral arms, as elliptical galaxies. 
\cite{VanDenBergh1990} suggests that the 'S0 classification type comprises
a number of physically quite distinct types of objects that exhibit only superficial
morphological similarities'.
Recent studies based on observations \citep{Fraser-McKelvie2018,Coccato2020} and on simulations \citep{Deeley2021} showed that this class of objects is, indeed, composed by  two or more sub-groups, formed via different physical mechanisms that lead to a similar morphology.
Specifically, stripped spiral galaxies could be the progenitors of disk-dominated lenticular galaxies, if the gas and arms of the spiral galaxies would be removed by interactions with the cluster environment, or by harassment in a group environment or generally pestering in all environments \citep{Cortesi2013b,Jaffe,Johnston2021}. Another group of lenticular galaxies could be the result of major or minor mergers and multiple accretions \citep{Tapia2017}. Others, low mass S0s, could be pristine galaxies, formed at redshift $z \simeq 2$ from mergers of the galaxy stellar/gaseous clumps \citep{Saha2018}, or the result of secular evolution \citep{Mishra2018}.

The discrepancy between the probability of being LTG predicted in this work and in \cite{Vega-Ferrero2021} decreases when only objects with $B/A > 0.7$ are considered, see Figure \ref{fig:DESxSPLUS}. Moreover, there is an extended population of late-type galaxies with red colors  and high probability of being edge-on systems \citep{Vega-Ferrero2021}, see Figure \ref{fig:colour_density}.

Specifically, there are $126$ objects classified as spiral galaxies in this work and as robust ellipticals in\cite{Vega-Ferrero2021}, with $B/A\le0.7$ and $r_{petro}<17$ mag. At a visual inspection they are all consistent with being disk-dominated S0 galaxies \citep{Coccato2020}, or edge-one reddened spiral galaxies \citep{bamford2009galaxy,Sodre2013} and their average colour is $(g-r)\simeq0.85$.

On the other hand, as discussed in Section \ref{sec:ES class}, a fraction of galaxies classified as early-type galaxies in this work is comprised by bulge dominated lenticular galaxies. 

The multiple origin of the S0-like isophotal profile seems to be depicted by the DL algorithm used in this work. This topic will be further studied in a folow-up work, where the DL classification will be correlated with galaxies' bulge-to-total light profiles.

\subsection{Summary}

In this study, we employ a Deep Learning architecture among the top-ranking techniques for image classification of ETG/LTG while also introducing a model to predict the stamps that contain reliable information to be classified. Our method presented several innovations compared to the Bom21 model, including the possibility of objects that are not classified either as ETG or LTG. 

Furthermore, we also make use of the precise photometric redshifts derived with 12 bands present in the S-PLUS. We recover the color diagrams for the morphological types and examine the local environment and density of ETG/LTG. Additionally, we assess large-scale structure traced by morphology.
As a result, we provide a novel Value Added Catalogue (VAC) of galaxy morphologies, in the full footprint of S-PLUS DR3 which includes areas never explored for other galaxy morphology catalogues.
The catalogue is composed by the results of two DL methods, which, for every stamp, recover the probability of having a given morphology, and of being a reliable stamp, as detailed below.

\subsubsection{A novel Valued Added morphology classification catalogue for the southern hemisphere}
\label{sec:value_added_1}

In order to mediate between the variety of galaxies  morphologies and the binary classification applied in this work, we allow for an independent classification into early and late-type galaxies, i.e. the sum of the probability of belonging to each class does not sum up to one, see Section \ref{sec:ETG/LTG Model}. As a consequence of this choice, some objects can be either classified to belong to both groups (using the binary classification) nor to any, see Section \ref{sec:ES class}. The study of these two peculiar types of objects allows us to identify bulge dominated lenticular or spiral galaxies  ($Amb_{1}$), as well as compact, flocculent, star-forming  galaxies, see Figure \ref{fig:example_classes}.
Finally, this catalogue of galaxy morphologies covers areas of the Southern Sky for which there is no release of morphological catalogues, for our knowledge, see Figure \ref{fig:footprint}.

\subsubsection{A novel parameter to assign a probability of being a reliable stamp}

An interesting correlation is found when comparing the number of galaxies with low probability of being LTG or ETG ($Amb_{0})$, with the probability of being a reliable stamp, see Section \ref{sec:Non reliable Stamps}. 
In fact,  the majority of objects with no bin classification as early nor late-type galaxies (see previous Section \ref{sec:value_added_1})  have a low probability of being  reliable stamps, see Figure \ref{fig:ambiguity_reliability}. Figure \ref{fig:DESxSPLUS} reveals that selecting only reliable stamps decreases  the discrepancy with \citet{Vega-Ferrero2021} classification into ETG and LTG, especially for faint objects ($m_r>17$). 
Moreover, as shown in Figure \ref{fig:extraordinarytrash}, among the non reliable stamps, there are extraordinary objects, as the Antenna Galaxy, which will be identified and studied in a follow-up work.

\section*{Data Availability}
We publicly release our Value Added Catalogue (VAC) in the S-PLUS data base (\url{splus.cloud}).

\section*{Acknowledgements}


The S-PLUS project, including the T80-South robotic telescope and the S-PLUS scientific survey, was founded as a partnership between the Funda\c{c}\~{a}o de Amparo \`{a} Pesquisa do Estado de S\~{a}o Paulo (FAPESP), the Observat\'{o}rio Nacional (ON), the Federal University of Sergipe (UFS), and the Federal University of Santa Catarina (UFSC), with important financial and practical contributions from other collaborating institutes in Brazil, Chile (Universidad de La Serena), and Spain (Centro de Estudios de F\'{\i}sica del Cosmos de Arag\'{o}n, CEFCA). We further acknowledge financial support from the São Paulo Research Foundation (FAPESP), the Brazilian National Research Council (CNPq), the Coordination for the Improvement of Higher Education Personnel (CAPES), the Carlos Chagas Filho Rio de Janeiro State Research Foundation (FAPERJ), and the Brazilian Innovation Agency (FINEP).

The authors who are members of the S-PLUS collaboration are grateful for the contributions from CTIO staff in helping in the construction, commissioning and maintenance of the T80-South telescope and camera. We are also indebted to Rene Laporte and INPE, as well as Keith Taylor, for their important contributions to the project. From CEFCA, we particularly would like to thank Antonio Mar\'{i}n-Franch for his invaluable contributions in the early phases of the project, David Crist{\'o}bal-Hornillos and his team for their help with the installation of the data reduction package \textsc{jype} version 0.9.9, C\'{e}sar \'{I}\~{n}iguez for providing 2D measurements of the filter transmissions, and all other staff members for their support with various aspects of the project.


CMdO and LSJ acknowledge funding for this work from FAPESP grants 2019/26492-3, 2019/11910-4, 2019/10923-5 and 2009/54202-8. GS, CMdO and LS acknowledge  support, respectively, from CNPq grants 309209/2019-6, 115795/2020-0 and 304819/201794. NM acknowledges the University of São Paulo PUB grant 83-1 of 2020.  
A. C. acknowledge the financial support provided by FAPERJ grants E-26/200.607 e 210.371/2022(270993).

CRB acknowledges the financial support from CNPq (316072/2021-4) and from FAPERJ (grants 201.456/2022 and 210.330/2022) and the FINEP contract 01.22.0505.00 (ref. 1891/22). KK acknowledges full financial support from ANID through FONDECYT Postdoctrorado Project 3200139, Chile. 

The authors made use of multi GPU Sci-Mind machines developed and tested for Artificial Intelligence and would like to thank P. Russano and M. Portes de Albuquerque for all the support in infrastructure matters. 

The authors made use and acknowledge TOPCAT \footnote{\url{http://www.starlink.ac.uk/topcat/ (TOPCAT)}} tool to analyse the data and astrotools \citep{astrotools} to visualize the objects.
For complementary visual inspection and some panels the authors made use of small cut outs images from the Legacy Survey. The Legacy Surveys consist of three individual and complementary projects: the Dark Energy Camera Legacy Survey (DECaLS; Proposal ID \#2014B-0404; PIs: David Schlegel and Arjun Dey), the Beijing-Arizona Sky Survey (BASS; NOAO Prop. ID \#2015A-0801; PIs: Zhou Xu and Xiaohui Fan), and the Mayall z-band Legacy Survey (MzLS; Prop. ID \#2016A-0453; PI: Arjun Dey). DECaLS, BASS and MzLS together include data obtained, respectively, at the Blanco telescope, Cerro Tololo Inter-American Observatory, NSF’s NOIRLab; the Bok telescope, Steward Observatory, University of Arizona; and the Mayall telescope, Kitt Peak National Observatory, NOIRLab. The Legacy Surveys project is honored to be permitted to conduct astronomical research on Iolkam Du’ag (Kitt Peak), a mountain with particular significance to the Tohono O’odham Nation.
The authors thanks F. Ferrari and J. Crosset for inspiring discussions and suggestions.








\bibliographystyle{mnras} 
\bibliography{bibliografia}

\begin{thebibliography}{}
\makeatletter
\relax
\def\mn@urlcharsother{\let\do\@makeother \do\$\do\&\do\#\do\^\do\_\do\%\do\~}
\def\mn@doi{\begingroup\mn@urlcharsother \@ifnextchar [ {\mn@doi@}
  {\mn@doi@[]}}
\def\mn@doi@[#1]#2{\def\@tempa{#1}\ifx\@tempa\@empty \href
  {http://dx.doi.org/#2} {doi:#2}\else \href {http://dx.doi.org/#2} {#1}\fi
  \endgroup}
\def\mn@eprint#1#2{\mn@eprint@#1:#2::\@nil}
\def\mn@eprint@arXiv#1{\href {http://arxiv.org/abs/#1} {{\tt arXiv:#1}}}
\def\mn@eprint@dblp#1{\href {http://dblp.uni-trier.de/rec/bibtex/#1.xml}
  {dblp:#1}}
\def\mn@eprint@#1:#2:#3:#4\@nil{\def\@tempa {#1}\def\@tempb {#2}\def\@tempc
  {#3}\ifx \@tempc \@empty \let \@tempc \@tempb \let \@tempb \@tempa \fi \ifx
  \@tempb \@empty \def\@tempb {arXiv}\fi \@ifundefined
  {mn@eprint@\@tempb}{\@tempb:\@tempc}{\expandafter \expandafter \csname
  mn@eprint@\@tempb\endcsname \expandafter{\@tempc}}}

\bibitem[\protect\citeauthoryear{{Abbott} et~al.,}{{Abbott}
  et~al.}{2018}]{Abbott2018a}
{Abbott} T.~M.~C.,  et~al., 2018, \mn@doi [\apjs] {10.3847/1538-4365/aae9f0},
  \href {https://ui.adsabs.harvard.edu/abs/2018ApJS..239...18A} {239, 18}

\bibitem[\protect\citeauthoryear{Abdel-Hamid, Mohamed, Jiang, Deng, Penn  \&
  Yu}{Abdel-Hamid et~al.}{2014}]{abdel2014convolutional}
Abdel-Hamid O.,  Mohamed A.-r.,  Jiang H.,  Deng L.,  Penn G.,   Yu D.,  2014,
  IEEE/ACM Transactions on audio, speech, and language processing, 22, 1533

\bibitem[\protect\citeauthoryear{{Almeida-Fernandes}
  et~al.,}{{Almeida-Fernandes} et~al.}{2022}]{Almeida-Fernandes2022}
{Almeida-Fernandes} F.,  et~al., 2022, \mn@doi [\mnras]
  {10.1093/mnras/stac284}, \href
  {https://ui.adsabs.harvard.edu/abs/2022MNRAS.511.4590A} {511, 4590}

\bibitem[\protect\citeauthoryear{{Ann}, {Seo}  \& {Ha}}{{Ann}
  et~al.}{2015}]{Ann2015}
{Ann} H.~B.,  {Seo} M.,   {Ha} D.~K.,  2015, \mn@doi [\apjs]
  {10.1088/0067-0049/217/2/27}, \href
  {https://ui.adsabs.harvard.edu/abs/2015ApJS..217...27A} {217, 27}

\bibitem[\protect\citeauthoryear{{Arag{\'o}n-Salamanca}}{{Arag{\'o}n-Salamanca}}{2008}]{Salamanca}
{Arag{\'o}n-Salamanca} A.,  2008, in {M.~Bureau, E.~Athanassoula, \& B.~Barbuy}
  ed.,  IAU Symposium Vol. 245, IAU Symposium. pp 285--288 (\mn@eprint {arXiv}
  {0710.2481}), \mn@doi{10.1017/S1743921308017924}

\bibitem[\protect\citeauthoryear{{Axelrod}}{{Axelrod}}{2006}]{Axelrod2006}
{Axelrod} T.~S.,  2006, in {Gabriel} C.,  {Arviset} C.,  {Ponz} D.,   {Enrique}
  S.,  eds,  Astronomical Society of the Pacific Conference Series Vol. 351,
  Astronomical Data Analysis Software and Systems XV. p.~103

\bibitem[\protect\citeauthoryear{{Baldry}, {Glazebrook}, {Brinkmann},
  {Ivezi{\'c}}, {Lupton}, {Nichol}  \& {Szalay}}{{Baldry}
  et~al.}{2004}]{baldry2004}
{Baldry} I.~K.,  {Glazebrook} K.,  {Brinkmann} J.,  {Ivezi{\'c}} {\v{Z}}.,
  {Lupton} R.~H.,  {Nichol} R.~C.,   {Szalay} A.~S.,  2004, \mn@doi [\apj]
  {10.1086/380092}, \href
  {https://ui.adsabs.harvard.edu/abs/2004ApJ...600..681B} {600, 681}

\bibitem[\protect\citeauthoryear{{Baldry} et~al.,}{{Baldry}
  et~al.}{2010}]{Baldry2010}
{Baldry} I.~K.,  et~al., 2010, \mn@doi [\mnras]
  {10.1111/j.1365-2966.2010.16282.x}, \href
  {https://ui.adsabs.harvard.edu/abs/2010MNRAS.404...86B} {404, 86}

\bibitem[\protect\citeauthoryear{Bamford et~al.,}{Bamford
  et~al.}{2009}]{bamford2009galaxy}
Bamford S.~P.,  et~al., 2009, Monthly Notices of the Royal Astronomical
  Society, 393, 1324

\bibitem[\protect\citeauthoryear{{Bernardi}, {Dom{\'\i}nguez S{\'a}nchez},
  {Brownstein}, {Drory}  \& {Sheth}}{{Bernardi} et~al.}{2019}]{Bernardi2019}
{Bernardi} M.,  {Dom{\'\i}nguez S{\'a}nchez} H.,  {Brownstein} J.~R.,  {Drory}
  N.,   {Sheth} R.~K.,  2019, \mn@doi [\mnras] {10.1093/mnras/stz2413}, \href
  {https://ui.adsabs.harvard.edu/abs/2019MNRAS.489.5633B} {489, 5633}

\bibitem[\protect\citeauthoryear{{Bertin} \& {Arnouts}}{{Bertin} \&
  {Arnouts}}{1996}]{Bertin1996}
{Bertin} E.,  {Arnouts} S.,  1996, \mn@doi [\aaps] {10.1051/aas:1996164}, \href
  {https://ui.adsabs.harvard.edu/abs/1996A&AS..117..393B} {117, 393}

\bibitem[\protect\citeauthoryear{{Bom}, {Makler}, {Albuquerque}  \& {Brand
  t}}{{Bom} et~al.}{2017}]{Bom2017}
{Bom} C.~R.,  {Makler} M.,  {Albuquerque} M.~P.,   {Brand t} C.~H.,  2017,
  \mn@doi [\aap] {10.1051/0004-6361/201629159}, \href
  {https://ui.adsabs.harvard.edu/abs/2017A&A...597A.135B} {597, A135}

\bibitem[\protect\citeauthoryear{{Bom} et~al.,}{{Bom} et~al.}{2021}]{Bom2021}
{Bom} C.~R.,  et~al., 2021, \mn@doi [\mnras] {10.1093/mnras/stab1981}, \href
  {https://ui.adsabs.harvard.edu/abs/2021MNRAS.507.1937B} {507, 1937}

\bibitem[\protect\citeauthoryear{Bom et~al.,}{Bom
  et~al.}{2022}]{bom2022developing}
Bom C.,  et~al., 2022, Monthly Notices of the Royal Astronomical Society, 515,
  5121

\bibitem[\protect\citeauthoryear{{Bournaud}, {Jog}  \& {Combes}}{{Bournaud}
  et~al.}{2005}]{Bournaud}
{Bournaud} F.,  {Jog} C.~J.,   {Combes} F.,  2005, \mn@doi [\aap]
  {10.1051/0004-6361:20042036}, \href
  {http://adsabs.harvard.edu/abs/2005A%26A...437...69B} {437, 69}

\bibitem[\protect\citeauthoryear{Bournaud, Jog  \& Combes}{Bournaud
  et~al.}{2007}]{Bournaud2007}
Bournaud F.,  Jog C.~J.,   Combes F.,  2007, \mn@doi [Astronomy {\&}
  Astrophysics] {10.1051/0004-6361:20078010}, 476, 1179

\bibitem[\protect\citeauthoryear{{Brambila}, {Lopes}, {Ribeiro}  \&
  {Cortesi}}{{Brambila} et~al.}{2023}]{Brambila2023}
{Brambila} D.,  {Lopes} P. A.~A.,  {Ribeiro} A. L.~B.,   {Cortesi} A.,  2023,
  \mn@doi [\mnras] {10.1093/mnras/stad1233}, \href
  {https://ui.adsabs.harvard.edu/abs/2023MNRAS.tmp.1248B} {}

\bibitem[\protect\citeauthoryear{{Buitrago}, {Trujillo}, {Conselice}  \&
  {H{\"a}u{\ss}ler}}{{Buitrago} et~al.}{2013}]{Buitrago2013}
{Buitrago} F.,  {Trujillo} I.,  {Conselice} C.~J.,   {H{\"a}u{\ss}ler} B.,
  2013, \mn@doi [\mnras] {10.1093/mnras/sts124}, \href
  {https://ui.adsabs.harvard.edu/abs/2013MNRAS.428.1460B} {428, 1460}

\bibitem[\protect\citeauthoryear{{Buta}}{{Buta}}{2011}]{Buta2011}
{Buta} R.~J.,  2011, \mn@doi [arXiv e-prints] {10.48550/arXiv.1102.0550}, \href
  {https://ui.adsabs.harvard.edu/abs/2011arXiv1102.0550B} {p. arXiv:1102.0550}

\bibitem[\protect\citeauthoryear{{Byrd} \& {Valtonen}}{{Byrd} \&
  {Valtonen}}{1990}]{Byrd}
{Byrd} G.,  {Valtonen} M.,  1990, \mn@doi [\apj] {10.1086/168362}, \href
  {http://adsabs.harvard.edu/abs/1990ApJ...350...89B} {350, 89}

\bibitem[\protect\citeauthoryear{{Calvi}, {Poggianti}, {Fasano}  \&
  {Vulcani}}{{Calvi} et~al.}{2012}]{Calvi2012}
{Calvi} R.,  {Poggianti} B.~M.,  {Fasano} G.,   {Vulcani} B.,  2012, \mn@doi
  [\mnras] {10.1111/j.1745-3933.2011.01168.x}, \href
  {https://ui.adsabs.harvard.edu/abs/2012MNRAS.419L..14C} {419, L14}

\bibitem[\protect\citeauthoryear{{Cappellari} et~al.,}{{Cappellari}
  et~al.}{2011}]{Capellari2011}
{Cappellari} M.,  et~al., 2011, \mn@doi [\mnras]
  {10.1111/j.1365-2966.2011.18600.x}, \href
  {https://ui.adsabs.harvard.edu/abs/2011MNRAS.416.1680C} {416, 1680}

\bibitem[\protect\citeauthoryear{{Cardelli}, {Clayton}  \& {Mathis}}{{Cardelli}
  et~al.}{1989}]{Cardelli1898}
{Cardelli} J.~A.,  {Clayton} G.~C.,   {Mathis} J.~S.,  1989, \mn@doi [\apj]
  {10.1086/167900}, \href
  {https://ui.adsabs.harvard.edu/abs/1989ApJ...345..245C} {345, 245}

\bibitem[\protect\citeauthoryear{Cardoso}{Cardoso}{2022}]{astrotools}
Cardoso N.~M.,  2022, Astrotools: Web-based astronomical tools,
  \mn@doi{10.5281/zenodo.7268504}, \url
  {https://doi.org/10.5281/zenodo.7268504}

\bibitem[\protect\citeauthoryear{{Cenarro} et~al.,}{{Cenarro}
  et~al.}{2019}]{Cenarro2019}
{Cenarro} A.~J.,  et~al., 2019, \mn@doi [\aap] {10.1051/0004-6361/201833036},
  \href {https://ui.adsabs.harvard.edu/abs/2019A&A...622A.176C} {622, A176}

\bibitem[\protect\citeauthoryear{{Cheng}, {Li}, {Conselice},
  {Arag{\'o}n-Salamanca}, {Dye}  \& {Metcalf}}{{Cheng}
  et~al.}{2019}]{cheng2019}
{Cheng} T.-Y.,  {Li} N.,  {Conselice} C.~J.,  {Arag{\'o}n-Salamanca} A.,  {Dye}
  S.,   {Metcalf} R.~B.,  2019, arXiv e-prints, \href
  {https://ui.adsabs.harvard.edu/abs/2019arXiv191104320C} {p. arXiv:1911.04320}

\bibitem[\protect\citeauthoryear{{Cheng} et~al.,}{{Cheng}
  et~al.}{2020}]{Cheng2020}
{Cheng} T.-Y.,  et~al., 2020, \mn@doi [\mnras] {10.1093/mnras/staa501}, \href
  {https://ui.adsabs.harvard.edu/abs/2020MNRAS.493.4209C} {493, 4209}

\bibitem[\protect\citeauthoryear{{Cheng} et~al.,}{{Cheng}
  et~al.}{2023}]{Cheng_2023}
{Cheng} T.~Y.,  et~al., 2023, \mn@doi [\mnras] {10.1093/mnras/stac3228}, \href
  {https://ui.adsabs.harvard.edu/abs/2023MNRAS.518.2794C} {518, 2794}

\bibitem[\protect\citeauthoryear{Choi, Fazekas, Sandler  \& Cho}{Choi
  et~al.}{2017}]{choi2017convolutional}
Choi K.,  Fazekas G.,  Sandler M.,   Cho K.,  2017, in 2017 IEEE International
  Conference on Acoustics, Speech and Signal Processing (ICASSP). pp 2392--2396

\bibitem[\protect\citeauthoryear{{Coccato} et~al.,}{{Coccato}
  et~al.}{2020}]{Coccato2020}
{Coccato} L.,  et~al., 2020, \mn@doi [\mnras] {10.1093/mnras/stz3592}, \href
  {https://ui.adsabs.harvard.edu/abs/2020MNRAS.492.2955C} {492, 2955}

\bibitem[\protect\citeauthoryear{Conselice}{Conselice}{2014}]{conselice2014}
Conselice C.~J.,  2014, \mn@doi [Ann. Rev. Astron. Astrophys.]
  {10.1146/annurev-astro-081913-040037}, 52, 291

\bibitem[\protect\citeauthoryear{{Cooray} \& {Sheth}}{{Cooray} \&
  {Sheth}}{2002}]{HaloModel}
{Cooray} A.,  {Sheth} R.,  2002, \mn@doi [\physrep]
  {10.1016/S0370-1573(02)00276-4}, \href
  {https://ui.adsabs.harvard.edu/abs/2002PhR...372....1C} {372, 1}

\bibitem[\protect\citeauthoryear{{Cortesi} et~al.,}{{Cortesi}
  et~al.}{2013}]{Cortesi2013b}
{Cortesi} A.,  et~al., 2013, \mn@doi [\mnras] {10.1093/mnras/stt529}, \href
  {https://ui.adsabs.harvard.edu/abs/2013MNRAS.432.1010C} {432, 1010}

\bibitem[\protect\citeauthoryear{{Crossett}, {Pimbblet}, {Stott}  \&
  {Jones}}{{Crossett} et~al.}{2014}]{Crossett2014}
{Crossett} J.~P.,  {Pimbblet} K.~A.,  {Stott} J.~P.,   {Jones} D.~H.,  2014,
  \mn@doi [\mnras] {10.1093/mnras/stt2065}, \href
  {https://ui.adsabs.harvard.edu/abs/2014MNRAS.437.2521C} {437, 2521}

\bibitem[\protect\citeauthoryear{{Deeley}, {Drinkwater}, {Sweet}, {Bekki},
  {Couch}, {Forbes}  \& {Dolfi}}{{Deeley} et~al.}{2021}]{Deeley2021}
{Deeley} S.,  {Drinkwater} M.~J.,  {Sweet} S.~M.,  {Bekki} K.,  {Couch} W.~J.,
  {Forbes} D.~A.,   {Dolfi} A.,  2021, \mn@doi [\mnras]
  {10.1093/mnras/stab2007}, \href
  {https://ui.adsabs.harvard.edu/abs/2021MNRAS.508..895D} {508, 895}

\bibitem[\protect\citeauthoryear{Deng, Dong, Socher, Li, Li  \& Fei-Fei}{Deng
  et~al.}{2009}]{deng2009imagenet}
Deng J.,  Dong W.,  Socher R.,  Li L.-J.,  Li K.,   Fei-Fei L.,  2009, in 2009
  IEEE conference on computer vision and pattern recognition. pp 248--255

\bibitem[\protect\citeauthoryear{{Desai} et~al.,}{{Desai}
  et~al.}{2007}]{desai07}
{Desai} V.,  et~al., 2007, \mn@doi [\apj] {10.1086/513310}, \href
  {http://adsabs.harvard.edu/abs/2007ApJ...660.1151D} {660, 1151}

\bibitem[\protect\citeauthoryear{{Dimauro} et~al.,}{{Dimauro}
  et~al.}{2022}]{Dimauro2022}
{Dimauro} P.,  et~al., 2022, \mn@doi [\mnras] {10.1093/mnras/stac884}, \href
  {https://ui.adsabs.harvard.edu/abs/2022MNRAS.513..256D} {513, 256}

\bibitem[\protect\citeauthoryear{{Dressler}}{{Dressler}}{1980}]{Dressler80}
{Dressler} A.,  1980, \mn@doi [\apj] {10.1086/157753}, \href
  {http://adsabs.harvard.edu/abs/1980ApJ...236..351D} {236, 351}

\bibitem[\protect\citeauthoryear{{Dressler} et~al.,}{{Dressler}
  et~al.}{1997}]{Dressler}
{Dressler} A.,  et~al., 1997, \mn@doi [\apj] {10.1086/304890}, \href
  {http://adsabs.harvard.edu/abs/1997ApJ...490..577D} {490, 577}

\bibitem[\protect\citeauthoryear{{Edelen}}{{Edelen}}{1969}]{Edelen1969}
{Edelen} D. G.~B.,  1969, \mn@doi [\apss] {10.1007/BF00649593}, \href
  {https://ui.adsabs.harvard.edu/abs/1969Ap&SS...3...56E} {3, 56}

\bibitem[\protect\citeauthoryear{{Farias}, {Ortiz}, {Damke}, {Jaque Arancibia}
  \& {Solar}}{{Farias} et~al.}{2020}]{Farias2020}
{Farias} H.,  {Ortiz} D.,  {Damke} G.,  {Jaque Arancibia} M.,   {Solar} M.,
  2020, \mn@doi [Astronomy and Computing] {10.1016/j.ascom.2020.100420}, \href
  {https://ui.adsabs.harvard.edu/abs/2020A&C....3300420F} {33, 100420}

\bibitem[\protect\citeauthoryear{{Fraser-McKelvie}, {Arag{\'o}n-Salamanca},
  {Merrifield}, {Tabor}, {Bernardi}, {Drory}, {Parikh}  \&
  {Argudo-Fern{\'a}ndez}}{{Fraser-McKelvie} et~al.}{2018}]{Fraser-McKelvie2018}
{Fraser-McKelvie} A.,  {Arag{\'o}n-Salamanca} A.,  {Merrifield} M.,  {Tabor}
  M.,  {Bernardi} M.,  {Drory} N.,  {Parikh} T.,   {Argudo-Fern{\'a}ndez} M.,
  2018, \mn@doi [\mnras] {10.1093/mnras/sty2563}, \href
  {https://ui.adsabs.harvard.edu/abs/2018MNRAS.481.5580F} {481, 5580}

\bibitem[\protect\citeauthoryear{{Freeman}, {Izbicki}, {Lee}, {Newman},
  {Conselice}, {Koekemoer}, {Lotz}  \& {Mozena}}{{Freeman}
  et~al.}{2013}]{Freeman2013}
{Freeman} P.~E.,  {Izbicki} R.,  {Lee} A.~B.,  {Newman} J.~A.,  {Conselice}
  C.~J.,  {Koekemoer} A.~M.,  {Lotz} J.~M.,   {Mozena} M.,  2013, \mn@doi
  [\mnras] {10.1093/mnras/stt1016}, \href
  {https://ui.adsabs.harvard.edu/abs/2013MNRAS.434..282F} {434, 282}

\bibitem[\protect\citeauthoryear{Gehrels \& and}{Gehrels \& and}{2015}]{WFIRST}
Gehrels N.,  and D.~S.,  2015, \mn@doi [Journal of Physics: Conference Series]
  {10.1088/1742-6596/610/1/012007}, 610, 012007

\bibitem[\protect\citeauthoryear{Glazebrook, Jacobs, Collett, More  \&
  McCarthy}{Glazebrook et~al.}{2017}]{glazebrook2017}
Glazebrook K.,  Jacobs C.,  Collett T.,  More A.,   McCarthy C.,  2017, \mn@doi
  [Monthly Notices of the Royal Astronomical Society] {10.1093/mnras/stx1492},
  471, 167

\bibitem[\protect\citeauthoryear{{Gonz{\'a}lez Delgado} et~al.,}{{Gonz{\'a}lez
  Delgado} et~al.}{2015}]{GonzalezDelgado2015}
{Gonz{\'a}lez Delgado} R.~M.,  et~al., 2015, \mn@doi [\aap]
  {10.1051/0004-6361/201525938}, \href
  {https://ui.adsabs.harvard.edu/abs/2015A&A...581A.103G} {581, A103}

\bibitem[\protect\citeauthoryear{{Gonz{\'a}lez Delgado} et~al.,}{{Gonz{\'a}lez
  Delgado} et~al.}{2022}]{GonzalesDelgado2022}
{Gonz{\'a}lez Delgado} R.~M.,  et~al., 2022, \mn@doi [\aap]
  {10.1051/0004-6361/202244030}, \href
  {https://ui.adsabs.harvard.edu/abs/2022A&A...666A..84G} {666, A84}

\bibitem[\protect\citeauthoryear{Goodfellow, Bengio  \& Courville}{Goodfellow
  et~al.}{2016}]{Goodfellow-et-al-2016}
Goodfellow I.,  Bengio Y.,   Courville A.,  2016, Deep Learning.
MIT Press

\bibitem[\protect\citeauthoryear{{Grosb{\o}l} \& {Dottori}}{{Grosb{\o}l} \&
  {Dottori}}{2012}]{Grosbol2012}
{Grosb{\o}l} P.,  {Dottori} H.,  2012, \mn@doi [\aap]
  {10.1051/0004-6361/201118099}, \href
  {https://ui.adsabs.harvard.edu/abs/2012A&A...542A..39G} {542, A39}

\bibitem[\protect\citeauthoryear{{Gunn} \& {Gott}}{{Gunn} \&
  {Gott}}{1972}]{Gunn}
{Gunn} J.~E.,  {Gott} III J.~R.,  1972, \mn@doi [\apj] {10.1086/151605}, \href
  {http://adsabs.harvard.edu/abs/1972ApJ...176....1G} {176, 1}

\bibitem[\protect\citeauthoryear{Hannun, Rajpurkar, Haghpanahi, Tison, Bourn,
  Turakhia  \& Ng}{Hannun et~al.}{2019}]{hannun2019cardiologist}
Hannun A.~Y.,  Rajpurkar P.,  Haghpanahi M.,  Tison G.~H.,  Bourn C.,  Turakhia
  M.~P.,   Ng A.~Y.,  2019, Nature medicine, 25, 65

\bibitem[\protect\citeauthoryear{Hausen \& Robertson}{Hausen \&
  Robertson}{2020}]{hausen2020morpheus}
Hausen R.,  Robertson B.~E.,  2020, The Astrophysical Journal Supplement
  Series, 248, 20

\bibitem[\protect\citeauthoryear{{Herschel}}{{Herschel}}{1864}]{Herschel1864}
{Herschel} J. F.~W.,  1864, Philosophical Transactions of the Royal Society of
  London Series I, \href
  {https://ui.adsabs.harvard.edu/abs/1864RSPT..154....1H} {154, 1}

\bibitem[\protect\citeauthoryear{{Holincheck} et~al.,}{{Holincheck}
  et~al.}{2016}]{Holincheck2016}
{Holincheck} A.~J.,  et~al., 2016, \mn@doi [\mnras] {10.1093/mnras/stw649},
  \href {https://ui.adsabs.harvard.edu/abs/2016MNRAS.459..720H} {459, 720}

\bibitem[\protect\citeauthoryear{{Jacobs} et~al.,}{{Jacobs}
  et~al.}{2019}]{jacobs2019}
{Jacobs} C.,  et~al., 2019, \mn@doi [\mnras] {10.1093/mnras/stz272}, \href
  {https://ui.adsabs.harvard.edu/\#abs/2019MNRAS.484.5330J} {484, 5330}

\bibitem[\protect\citeauthoryear{{Jaff{\'e}}, {Smith}, {Candlish}, {Poggianti},
  {Sheen}  \& {Verheijen}}{{Jaff{\'e}} et~al.}{2015}]{Jaffe}
{Jaff{\'e}} Y.~L.,  {Smith} R.,  {Candlish} G.~N.,  {Poggianti} B.~M.,  {Sheen}
  Y.-K.,   {Verheijen} M.~A.~W.,  2015, \mn@doi [\mnras]
  {10.1093/mnras/stv100}, \href
  {http://adsabs.harvard.edu/abs/2015MNRAS.448.1715J} {448, 1715}

\bibitem[\protect\citeauthoryear{{Johnston} et~al.,}{{Johnston}
  et~al.}{2021}]{Johnston2021}
{Johnston} E.~J.,  et~al., 2021, \mn@doi [\mnras] {10.1093/mnras/staa2838},
  \href {https://ui.adsabs.harvard.edu/abs/2021MNRAS.500.4193J} {500, 4193}

\bibitem[\protect\citeauthoryear{{Kelly} \& {McKay}}{{Kelly} \&
  {McKay}}{2004}]{Kelly2004}
{Kelly} B.~C.,  {McKay} T.~A.,  2004, \mn@doi [\aj] {10.1086/380934}, \href
  {https://ui.adsabs.harvard.edu/abs/2004AJ....127..625K} {127, 625}

\bibitem[\protect\citeauthoryear{{Khanday}, {Saha}, {Iqbal}, {Dhiwar}  \&
  {Pahwa}}{{Khanday} et~al.}{2022}]{Khanday2022}
{Khanday} S.~A.,  {Saha} K.,  {Iqbal} N.,  {Dhiwar} S.,   {Pahwa} I.,  2022,
  \mn@doi [\mnras] {10.1093/mnras/stac2009}, \href
  {https://ui.adsabs.harvard.edu/abs/2022MNRAS.515.5043K} {515, 5043}

\bibitem[\protect\citeauthoryear{{Knabel} et~al.,}{{Knabel}
  et~al.}{2020}]{Knabel2020}
{Knabel} S.,  et~al., 2020, \mn@doi [\aj] {10.3847/1538-3881/abb612}, \href
  {https://ui.adsabs.harvard.edu/abs/2020AJ....160..223K} {160, 223}

\bibitem[\protect\citeauthoryear{{Kronberger}, {Kapferer}, {Ferrari},
  {Unterguggenberger}  \& {Schindler}}{{Kronberger} et~al.}{2008}]{Kronberger}
{Kronberger} T.,  {Kapferer} W.,  {Ferrari} C.,  {Unterguggenberger} S.,
  {Schindler} S.,  2008, \mn@doi [\aap] {10.1051/0004-6361:20078904}, \href
  {http://adsabs.harvard.edu/abs/2008A%26A...481..337K} {481, 337}

\bibitem[\protect\citeauthoryear{{Lang}}{{Lang}}{2014}]{unwise}
{Lang} D.,  2014, \mn@doi [\aj] {10.1088/0004-6256/147/5/108}, \href
  {https://ui.adsabs.harvard.edu/abs/2014AJ....147..108L} {147, 108}

\bibitem[\protect\citeauthoryear{{Lanusse}, {Ma}, {Li}, {Collett}, {Li},
  {Ravanbakhsh}, {Mandelbaum}  \& {P{\'o}czos}}{{Lanusse}
  et~al.}{2018}]{lanusse2018cmu}
{Lanusse} F.,  {Ma} Q.,  {Li} N.,  {Collett} T.~E.,  {Li} C.-L.,  {Ravanbakhsh}
  S.,  {Mandelbaum} R.,   {P{\'o}czos} B.,  2018, \mn@doi [\mnras]
  {10.1093/mnras/stx1665}, \href
  {https://ui.adsabs.harvard.edu/\#abs/2018MNRAS.473.3895L} {473, 3895}

\bibitem[\protect\citeauthoryear{{Leaman} et~al.,}{{Leaman}
  et~al.}{2013}]{Leaman2013}
{Leaman} R.,  et~al., 2013, \mn@doi [\apj] {10.1088/0004-637X/767/2/131}, \href
  {https://ui.adsabs.harvard.edu/abs/2013ApJ...767..131L} {767, 131}

\bibitem[\protect\citeauthoryear{Li, Ding  \& Sun}{Li
  et~al.}{2018}]{li2018remaining}
Li X.,  Ding Q.,   Sun J.-Q.,  2018, Reliability Engineering and System Safety,
  172, 1

\bibitem[\protect\citeauthoryear{{Lima-Dias} et~al.,}{{Lima-Dias}
  et~al.}{2021}]{limadias2021}
{Lima-Dias} C.,  et~al., 2021, \mn@doi [\mnras] {10.1093/mnras/staa3326}, \href
  {https://ui.adsabs.harvard.edu/abs/2021MNRAS.500.1323L} {500, 1323}

\bibitem[\protect\citeauthoryear{{Lima} et~al.,}{{Lima}
  et~al.}{2022}]{Lima2022}
{Lima} E.~V.~R.,  et~al., 2022, \mn@doi [Astronomy and Computing]
  {10.1016/j.ascom.2021.100510}, \href
  {https://ui.adsabs.harvard.edu/abs/2022A&C....3800510L} {38, 100510}

\bibitem[\protect\citeauthoryear{Lintott et~al.,}{Lintott
  et~al.}{2008}]{lintott2008}
Lintott C.~J.,  et~al., 2008, \mn@doi [Monthly Notices of the Royal
  Astronomical Society] {10.1111/j.1365-2966.2008.13689.x}, 389, 1179

\bibitem[\protect\citeauthoryear{Lintott et~al.,}{Lintott
  et~al.}{2010}]{lintott2010}
Lintott C.,  et~al., 2010, \mn@doi [Monthly Notices of the Royal Astronomical
  Society] {10.1111/j.1365-2966.2010.17432.x}, 410, 166

\bibitem[\protect\citeauthoryear{Liu, Jiang, He, Chen, Liu, Gao  \& Han}{Liu
  et~al.}{2019}]{liu2019variance}
Liu L.,  Jiang H.,  He P.,  Chen W.,  Liu X.,  Gao J.,   Han J.,  2019, arXiv,
  pp arXiv--1908

\bibitem[\protect\citeauthoryear{Lu, Wang  \& Zhou}{Lu
  et~al.}{2017}]{lu2017simultaneous}
Lu J.,  Wang G.,   Zhou J.,  2017, IEEE Transactions on Image Processing, 26,
  4042

\bibitem[\protect\citeauthoryear{{Ma} et~al.,}{{Ma} et~al.}{2019}]{Ma2019}
{Ma} Z.,  et~al., 2019, \mn@doi [\apjs] {10.3847/1538-4365/aaf9a2}, \href
  {https://ui.adsabs.harvard.edu/abs/2019ApJS..240...34M} {240, 34}

\bibitem[\protect\citeauthoryear{Madireddy, Li, Ramachandra, Balaprakash  \&
  Habib}{Madireddy et~al.}{2019}]{madireddy2019}
Madireddy S.,  Li N.,  Ramachandra N.,  Balaprakash P.,   Habib S.,  2019,
  Modular Deep Learning Analysis of Galaxy-Scale Strong Lensing Images
  (\mn@eprint {arXiv} {1911.03867})

\bibitem[\protect\citeauthoryear{{Margalef-Bentabol}, {Conselice}, {Mortlock},
  {Hartley}, {Duncan}, {Ferguson}, {Dekel}  \& {Primack}}{{Margalef-Bentabol}
  et~al.}{2016}]{margalef16}
{Margalef-Bentabol} B.,  {Conselice} C.~J.,  {Mortlock} A.,  {Hartley} W.,
  {Duncan} K.,  {Ferguson} H.~C.,  {Dekel} A.,   {Primack} J.~R.,  2016,
  \mn@doi [\mnras] {10.1093/mnras/stw1451}, \href
  {https://ui.adsabs.harvard.edu/abs/2016MNRAS.461.2728M} {461, 2728}

\bibitem[\protect\citeauthoryear{{Margalef-Bentabol}, {Huertas-Company},
  {Charnock}, {Margalef-Bentabol}, {Bernardi}, {Dubois}, {Storey-Fisher}  \&
  {Zanisi}}{{Margalef-Bentabol} et~al.}{2020}]{margalef2020}
{Margalef-Bentabol} B.,  {Huertas-Company} M.,  {Charnock} T.,
  {Margalef-Bentabol} C.,  {Bernardi} M.,  {Dubois} Y.,  {Storey-Fisher} K.,
  {Zanisi} L.,  2020, \mn@doi [\mnras] {10.1093/mnras/staa1647}, \href
  {https://ui.adsabs.harvard.edu/abs/2020MNRAS.496.2346M} {496, 2346}

\bibitem[\protect\citeauthoryear{Mendes de Oliveira
  et~al.,}{Mendes de Oliveira et~al.}{2019}]{mendes_de_oliveira2019}
Mendes de Oliveira C.,  et~al., 2019, \mn@doi [Monthly Notices of the Royal
  Astronomical Society] {10.1093/mnras/stz1985}, 489, 241

\bibitem[\protect\citeauthoryear{Metcalf et~al.,}{Metcalf
  et~al.}{2019}]{challenge01}
Metcalf R.~B.,  et~al., 2019, Astronomy \& Astrophysics, 625, A119

\bibitem[\protect\citeauthoryear{{Mishra}, {Wadadekar}  \& {Barway}}{{Mishra}
  et~al.}{2018}]{Mishra2018}
{Mishra} P.~K.,  {Wadadekar} Y.,   {Barway} S.,  2018, \mn@doi [\mnras]
  {10.1093/mnras/sty1107}, \href
  {https://ui.adsabs.harvard.edu/abs/2018MNRAS.478..351M} {478, 351}

\bibitem[\protect\citeauthoryear{{Montero-Dorta} et~al.,}{{Montero-Dorta}
  et~al.}{2020}]{SecBias}
{Montero-Dorta} A.~D.,  et~al., 2020, \mn@doi [\mnras]
  {10.1093/mnras/staa1624}, \href
  {https://ui.adsabs.harvard.edu/abs/2020MNRAS.496.1182M} {496, 1182}

\bibitem[\protect\citeauthoryear{Moreno-Torres, S{\'a}ez  \&
  Herrera}{Moreno-Torres et~al.}{2012}]{moreno2012study}
Moreno-Torres J.~G.,  S{\'a}ez J.~A.,   Herrera F.,  2012, IEEE Transactions on
  Neural Networks and Learning Systems, 23, 1304

\bibitem[\protect\citeauthoryear{{Mortlock} et~al.,}{{Mortlock}
  et~al.}{2013}]{mortlock13}
{Mortlock} A.,  et~al., 2013, \mn@doi [\mnras] {10.1093/mnras/stt793}, \href
  {https://ui.adsabs.harvard.edu/abs/2013MNRAS.433.1185M} {433, 1185}

\bibitem[\protect\citeauthoryear{{Nair} \& {Abraham}}{{Nair} \&
  {Abraham}}{2010}]{Nair2010}
{Nair} P.~B.,  {Abraham} R.~G.,  2010, \mn@doi [\apjs]
  {10.1088/0067-0049/186/2/427}, \href
  {https://ui.adsabs.harvard.edu/abs/2010ApJS..186..427N} {186, 427}

\bibitem[\protect\citeauthoryear{{Nakazono} et~al.,}{{Nakazono}
  et~al.}{2021a}]{splus_star_gal}
{Nakazono} L.,  et~al., 2021a, \mn@doi [\mnras] {10.1093/mnras/stab1835}, \href
  {https://ui.adsabs.harvard.edu/abs/2021MNRAS.507.5847N} {507, 5847}

\bibitem[\protect\citeauthoryear{{Nakazono} et~al.,}{{Nakazono}
  et~al.}{2021b}]{Nakazono2021}
{Nakazono} L.,  et~al., 2021b, \mn@doi [\mnras] {10.1093/mnras/stab1835}, \href
  {https://ui.adsabs.harvard.edu/abs/2021MNRAS.507.5847N} {507, 5847}

\bibitem[\protect\citeauthoryear{{Niemack}, {Jimenez}, {Verde}, {Menanteau},
  {Panter}  \& {Spergel}}{{Niemack} et~al.}{2009}]{GALEX}
{Niemack} M.~D.,  {Jimenez} R.,  {Verde} L.,  {Menanteau} F.,  {Panter} B.,
  {Spergel} D.,  2009, \mn@doi [\apj] {10.1088/0004-637X/690/1/89}, \href
  {https://ui.adsabs.harvard.edu/abs/2009ApJ...690...89N} {690, 89}

\bibitem[\protect\citeauthoryear{{Ostrovski} et~al.,}{{Ostrovski}
  et~al.}{2017}]{Ostrovski2017}
{Ostrovski} F.,  et~al., 2017, \mn@doi [\mnras] {10.1093/mnras/stw2958}, \href
  {https://ui.adsabs.harvard.edu/abs/2017MNRAS.465.4325O} {465, 4325}

\bibitem[\protect\citeauthoryear{{Paulino-Afonso} et~al.,}{{Paulino-Afonso}
  et~al.}{2019}]{Paulino-Afonso2019}
{Paulino-Afonso} A.,  et~al., 2019, \mn@doi [\aap]
  {10.1051/0004-6361/201935137}, \href
  {https://ui.adsabs.harvard.edu/abs/2019A&A...630A..57P} {630, A57}

\bibitem[\protect\citeauthoryear{{Petrillo} et~al.,}{{Petrillo}
  et~al.}{2017}]{petrillo2017}
{Petrillo} C.~E.,  et~al., 2017, arXiv: 1702.07675, \href
  {http://adsabs.harvard.edu/abs/2017arXiv170207675P} {}

\bibitem[\protect\citeauthoryear{{Petrillo} et~al.,}{{Petrillo}
  et~al.}{2019a}]{petrillo2019b}
{Petrillo} C.~E.,  et~al., 2019a, \mn@doi [\mnras] {10.1093/mnras/sty2683},
  \href {https://ui.adsabs.harvard.edu/\#abs/2019MNRAS.482..807P} {482, 807}

\bibitem[\protect\citeauthoryear{{Petrillo} et~al.,}{{Petrillo}
  et~al.}{2019b}]{petrillo2019a}
{Petrillo} C.~E.,  et~al., 2019b, \mn@doi [\mnras] {10.1093/mnras/stz189},
  \href {https://ui.adsabs.harvard.edu/\#abs/2019MNRAS.484.3879P} {484, 3879}

\bibitem[\protect\citeauthoryear{{Planck Collaboration} et~al.,}{{Planck
  Collaboration} et~al.}{2020}]{Planck}
{Planck Collaboration} et~al., 2020, \mn@doi [\aap]
  {10.1051/0004-6361/201833910}, \href
  {https://ui.adsabs.harvard.edu/abs/2020A&A...641A...6P} {641, A6}

\bibitem[\protect\citeauthoryear{{Povi{\'c}} et~al.,}{{Povi{\'c}}
  et~al.}{2015}]{Povic2015}
{Povi{\'c}} M.,  et~al., 2015, \mn@doi [\mnras] {10.1093/mnras/stv1663}, \href
  {https://ui.adsabs.harvard.edu/abs/2015MNRAS.453.1644P} {453, 1644}

\bibitem[\protect\citeauthoryear{{Quilis}, {Moore}  \& {Bower}}{{Quilis}
  et~al.}{2000}]{Quilis}
{Quilis} V.,  {Moore} B.,   {Bower} R.,  2000, \mn@doi [Science]
  {10.1126/science.288.5471.1617}, \href
  {http://adsabs.harvard.edu/abs/2000Sci...288.1617Q} {288, 1617}

\bibitem[\protect\citeauthoryear{{Saha} \& {Cortesi}}{{Saha} \&
  {Cortesi}}{2018}]{Saha2018}
{Saha} K.,  {Cortesi} A.,  2018, \mn@doi [\apjl] {10.3847/2041-8213/aad23a},
  \href {http://adsabs.harvard.edu/abs/2018ApJ...862L..12S} {862, L12}

\bibitem[\protect\citeauthoryear{{S{\'a}nchez}, {Cardiel}, {Verheijen},
  {Pedraz}  \& {Covone}}{{S{\'a}nchez} et~al.}{2007}]{Sanchez2007}
{S{\'a}nchez} S.~F.,  {Cardiel} N.,  {Verheijen} M.~A.~W.,  {Pedraz} S.,
  {Covone} G.,  2007, \mn@doi [\mnras] {10.1111/j.1365-2966.2007.11335.x},
  \href {https://ui.adsabs.harvard.edu/abs/2007MNRAS.376..125S} {376, 125}

\bibitem[\protect\citeauthoryear{{Santana-Silva} et~al.,}{{Santana-Silva}
  et~al.}{2020}]{Santana-Silva2020}
{Santana-Silva} L.,  et~al., 2020, \mn@doi [\mnras] {10.1093/mnras/staa2757},
  \href {https://ui.adsabs.harvard.edu/abs/2020MNRAS.498.5183S} {498, 5183}

\bibitem[\protect\citeauthoryear{{Sarkar} \& {Pandey}}{{Sarkar} \&
  {Pandey}}{2020}]{Sarkar2020}
{Sarkar} S.,  {Pandey} B.,  2020, \mn@doi [\mnras] {10.1093/mnras/staa2236},
  \href {https://ui.adsabs.harvard.edu/abs/2020MNRAS.497.4077S} {497, 4077}

\bibitem[\protect\citeauthoryear{{Shamir}, {Holincheck}  \& {Wallin}}{{Shamir}
  et~al.}{2013}]{Shamir2013}
{Shamir} L.,  {Holincheck} A.,   {Wallin} J.,  2013, \mn@doi [Astronomy and
  Computing] {10.1016/j.ascom.2013.09.002}, \href
  {https://ui.adsabs.harvard.edu/abs/2013A&C.....2...67S} {2, 67}

\bibitem[\protect\citeauthoryear{{Shao}, {Disseau}, {Yang}, {Hammer}, {Puech},
  {Rodrigues}, {Liang}  \& {Deng}}{{Shao} et~al.}{2015}]{Shao2015}
{Shao} X.,  {Disseau} K.,  {Yang} Y.~B.,  {Hammer} F.,  {Puech} M.,
  {Rodrigues} M.,  {Liang} Y.~C.,   {Deng} L.~C.,  2015, \mn@doi [\aap]
  {10.1051/0004-6361/201525796}, \href
  {https://ui.adsabs.harvard.edu/abs/2015A&A...579A..57S} {579, A57}

\bibitem[\protect\citeauthoryear{{Simmons} et~al.,}{{Simmons}
  et~al.}{2017}]{Simmons2017}
{Simmons} B.~D.,  et~al., 2017, \mn@doi [\mnras] {10.1093/mnras/stw2587}, \href
  {https://ui.adsabs.harvard.edu/abs/2017MNRAS.464.4420S} {464, 4420}

\bibitem[\protect\citeauthoryear{{Skrutskie} et~al.,}{{Skrutskie}
  et~al.}{2006}]{2mass}
{Skrutskie} M.~F.,  et~al., 2006, \mn@doi [\aj] {10.1086/498708}, \href
  {https://ui.adsabs.harvard.edu/abs/2006AJ....131.1163S} {131, 1163}

\bibitem[\protect\citeauthoryear{{Smethurst} et~al.,}{{Smethurst}
  et~al.}{2015}]{Smethurst2015}
{Smethurst} R.~J.,  et~al., 2015, \mn@doi [\mnras] {10.1093/mnras/stv161},
  \href {https://ui.adsabs.harvard.edu/abs/2015MNRAS.450..435S} {450, 435}

\bibitem[\protect\citeauthoryear{{Sodr{\'e}}, {Ribeiro da Silva}  \&
  {Santos}}{{Sodr{\'e}} et~al.}{2013}]{Sodre2013}
{Sodr{\'e}} L.,  {Ribeiro da Silva} A.,   {Santos} W.~A.,  2013, \mn@doi
  [\mnras] {10.1093/mnras/stt1188}, \href
  {https://ui.adsabs.harvard.edu/abs/2013MNRAS.434.2503S} {434, 2503}

\bibitem[\protect\citeauthoryear{{Spiekermann}}{{Spiekermann}}{1992}]{Spiekermann1992}
{Spiekermann} G.,  1992, \mn@doi [\aj] {10.1086/116215}, \href
  {https://ui.adsabs.harvard.edu/abs/1992AJ....103.2102S} {103, 2102}

\bibitem[\protect\citeauthoryear{{Storrie-Lombardi}, {Lahav}, {Sodre}  \&
  {Storrie-Lombardi}}{{Storrie-Lombardi} et~al.}{1992}]{Storrie-Lombardi1992}
{Storrie-Lombardi} M.~C.,  {Lahav} O.,  {Sodre} L.,   {Storrie-Lombardi} L.~J.,
   1992, \mn@doi [\mnras] {https://doi.org/10.1093/mnras/259.1.8P}, 259, 8

\bibitem[\protect\citeauthoryear{Tan \& Le}{Tan \& Le}{2019}]{efficientnet}
Tan M.,  Le Q.,  2019, in International Conference on Machine Learning. pp
  6105--6114

\bibitem[\protect\citeauthoryear{Tan, Chen, Pang, Vasudevan, Sandler, Howard
  \& Le}{Tan et~al.}{2019}]{tan2019mnasnet}
Tan M.,  Chen B.,  Pang R.,  Vasudevan V.,  Sandler M.,  Howard A.,   Le Q.~V.,
   2019, in Proceedings of the IEEE Conference on Computer Vision and Pattern
  Recognition. pp 2820--2828

\bibitem[\protect\citeauthoryear{{Tapia}, {Eliche-Moral}, {Aceves},
  {Rodr{\'\i}guez-P{\'e}rez}, {Borlaff}  \& {Querejeta}}{{Tapia}
  et~al.}{2017}]{Tapia2017}
{Tapia} T.,  {Eliche-Moral} M.~C.,  {Aceves} H.,  {Rodr{\'\i}guez-P{\'e}rez}
  C.,  {Borlaff} A.,   {Querejeta} M.,  2017, \mn@doi [\aap]
  {10.1051/0004-6361/201628821}, \href
  {https://ui.adsabs.harvard.edu/abs/2017A&A...604A.105T} {604, A105}

\bibitem[\protect\citeauthoryear{{Tohill}, {Bamford}  \& {Conselice}}{{Tohill}
  et~al.}{2023}]{tohill2023}
{Tohill} C.-B.,  {Bamford} S.,   {Conselice} C.,  2023, \mn@doi [arXiv
  e-prints] {10.48550/arXiv.2302.11482}, \href
  {https://ui.adsabs.harvard.edu/abs/2023arXiv230211482T} {p. arXiv:2302.11482}

\bibitem[\protect\citeauthoryear{{Tyson}}{{Tyson}}{2002}]{Tyson2002}
{Tyson} J.~A.,  2002, in {Tyson} J.~A.,  {Wolff} S.,  eds,  Society of
  Photo-Optical Instrumentation Engineers (SPIE) Conference Series Vol. 4836,
  Survey and Other Telescope Technologies and Discoveries. pp 10--20
  (\mn@eprint {arXiv} {astro-ph/0302102}), \mn@doi{10.1117/12.456772}

\bibitem[\protect\citeauthoryear{{Vaucouleurs}}{{Vaucouleurs}}{1959}]{Vaucouleurs1959}
{Vaucouleurs} G.,  1959, \mn@doi [Handbuch der Physik]
  {10.1007/978-3-642-45932-0_7}, \href
  {https://ui.adsabs.harvard.edu/abs/1959HDP....53..275V} {11, 275}

\bibitem[\protect\citeauthoryear{Vecchiotti, Vesperini, Principi, Squartini  \&
  Piazza}{Vecchiotti et~al.}{2018}]{vecchiotti2018convolutional}
Vecchiotti P.,  Vesperini F.,  Principi E.,  Squartini S.,   Piazza F.,  2018,
  in , Multidisciplinary Approaches to Neural Computing.
Springer, pp 161--170

\bibitem[\protect\citeauthoryear{Vega-Ferrero et~al.,}{Vega-Ferrero
  et~al.}{2021}]{Vega-Ferrero2021}
Vega-Ferrero J.,  et~al., 2021, \mn@doi [Monthly Notices of the Royal
  Astronomical Society] {10.1093/mnras/stab594}, 506, 1927

\bibitem[\protect\citeauthoryear{{Ventou} et~al.,}{{Ventou}
  et~al.}{2017}]{Ventou2017}
{Ventou} E.,  et~al., 2017, \mn@doi [\aap] {10.1051/0004-6361/201731586}, \href
  {https://ui.adsabs.harvard.edu/abs/2017A&A...608A...9V} {608, A9}

\bibitem[\protect\citeauthoryear{{Vulcani}, {Poggianti}, {Fritz}, {Fasano},
  {Moretti}, {Calvi}  \& {Paccagnella}}{{Vulcani} et~al.}{2015}]{vulcani2015}
{Vulcani} B.,  {Poggianti} B.~M.,  {Fritz} J.,  {Fasano} G.,  {Moretti} A.,
  {Calvi} R.,   {Paccagnella} A.,  2015, \mn@doi [\apj]
  {10.1088/0004-637X/798/1/52}, \href
  {https://ui.adsabs.harvard.edu/abs/2015ApJ...798...52V} {798, 52}

\bibitem[\protect\citeauthoryear{{Walmsley} et~al.,}{{Walmsley}
  et~al.}{2020}]{Walmsley2020}
{Walmsley} M.,  et~al., 2020, \mn@doi [\mnras] {10.1093/mnras/stz2816}, \href
  {https://ui.adsabs.harvard.edu/abs/2020MNRAS.491.1554W} {491, 1554}

\bibitem[\protect\citeauthoryear{{Wang}, {Cappellari}, {Peng}  \&
  {Graham}}{{Wang} et~al.}{2020}]{Wang2020}
{Wang} B.,  {Cappellari} M.,  {Peng} Y.,   {Graham} M.,  2020, \mn@doi [\mnras]
  {10.1093/mnras/staa1325}, \href
  {https://ui.adsabs.harvard.edu/abs/2020MNRAS.495.1958W} {495, 1958}

\bibitem[\protect\citeauthoryear{{Wechsler} \& {Tinker}}{{Wechsler} \&
  {Tinker}}{2018}]{HaloGalaxy}
{Wechsler} R.~H.,  {Tinker} J.~L.,  2018, \mn@doi [\araa]
  {10.1146/annurev-astro-081817-051756}, \href
  {https://ui.adsabs.harvard.edu/abs/2018ARA&A..56..435W} {56, 435}

\bibitem[\protect\citeauthoryear{{Werner}, {Hatch}, {Muzzin}, {van der Burg},
  {Balogh}, {Rudnick}  \& {Wilson}}{{Werner} et~al.}{2022}]{werner2022}
{Werner} S.~V.,  {Hatch} N.~A.,  {Muzzin} A.,  {van der Burg} R.~F.~J.,
  {Balogh} M.~L.,  {Rudnick} G.,   {Wilson} G.,  2022, \mn@doi [\mnras]
  {10.1093/mnras/stab3484}, \href
  {https://ui.adsabs.harvard.edu/abs/2022MNRAS.510..674W} {510, 674}

\bibitem[\protect\citeauthoryear{{Willett} et~al.,}{{Willett}
  et~al.}{2013}]{Willett2013}
{Willett} K.~W.,  et~al., 2013, \mn@doi [\mnras] {10.1093/mnras/stt1458}, \href
  {https://ui.adsabs.harvard.edu/abs/2013MNRAS.435.2835W} {435, 2835}

\bibitem[\protect\citeauthoryear{{Wjeisinghe}, {Hopkins}, {Kelly}, {Welikala}
  \& {Connolly}}{{Wjeisinghe} et~al.}{2010}]{Wjeisinghe2010}
{Wjeisinghe} D.~B.,  {Hopkins} A.~M.,  {Kelly} B.~C.,  {Welikala} N.,
  {Connolly} A.~J.,  2010, \mn@doi [\mnras] {10.1111/j.1365-2966.2010.16424.x},
  \href {https://ui.adsabs.harvard.edu/abs/2010MNRAS.404.2077W} {404, 2077}

\bibitem[\protect\citeauthoryear{{Wong} et~al.,}{{Wong}
  et~al.}{2012}]{Wong2012}
{Wong} O.~I.,  et~al., 2012, \mn@doi [\mnras]
  {10.1111/j.1365-2966.2011.20159.x}, \href
  {https://ui.adsabs.harvard.edu/abs/2012MNRAS.420.1684W} {420, 1684}

\bibitem[\protect\citeauthoryear{{Wu}}{{Wu}}{2020}]{Wu2020}
{Wu} J.~F.,  2020, \mn@doi [\apj] {10.3847/1538-4357/abacbb}, \href
  {https://ui.adsabs.harvard.edu/abs/2020ApJ...900..142W} {900, 142}

\bibitem[\protect\citeauthoryear{{York} et~al.,}{{York}
  et~al.}{2000}]{York2000}
{York} D.~G.,  et~al., 2000, \mn@doi [\aj] {10.1086/301513}, \href
  {https://ui.adsabs.harvard.edu/abs/2000AJ....120.1579Y} {120, 1579}

\bibitem[\protect\citeauthoryear{Zaborowski et~al.,}{Zaborowski
  et~al.}{2022}]{zaborowski2022identification}
Zaborowski E.,  et~al., 2022, arXiv preprint arXiv:2210.10802

\bibitem[\protect\citeauthoryear{{Zibetti}, {M{\'e}nard}, {Nestor}, {Quider},
  {Rao}  \& {Turnshek}}{{Zibetti} et~al.}{2007}]{Zibetti2007}
{Zibetti} S.,  {M{\'e}nard} B.,  {Nestor} D.~B.,  {Quider} A.~M.,  {Rao} S.~M.,
    {Turnshek} D.~A.,  2007, \mn@doi [\apj] {10.1086/511300}, \href
  {https://ui.adsabs.harvard.edu/abs/2007ApJ...658..161Z} {658, 161}

\bibitem[\protect\citeauthoryear{{Zwicky}}{{Zwicky}}{1940}]{Zwicky1940}
{Zwicky} F.,  1940, \mn@doi [Physical Review] {10.1103/PhysRev.58.478}, \href
  {https://ui.adsabs.harvard.edu/abs/1940PhRv...58..478Z} {58, 478}

\bibitem[\protect\citeauthoryear{{van den Bergh}}{{van den
  Bergh}}{1990}]{VanDenBergh1990}
{van den Bergh} S.,  1990, \mn@doi [\apj] {10.1086/168213}, \href
  {https://ui.adsabs.harvard.edu/abs/1990ApJ...348...57V} {348, 57}

\bibitem[\protect\citeauthoryear{van~den Bergh}{van~den
  Bergh}{1998}]{vandenbergh1998}
van~den Bergh S.,  1998, Galaxy Morphology and Classification, 1st ed. 1998
  edn.
Cambridge University Press

\makeatother
\end{thebibliography}
\appendix

\section{Comparison with VF2021}

In this study, we crossmatch the ETG classification from this work with the classification as robust late or early-type galaxies ($FLAG_{LTG}=5$ and $FLAG_{LTG}=4$, respectively) from \cite{Vega-Ferrero2021}. There are 27450 objects in common ($12\le mag_r \le 18$), of which 12295 are classified as reliable stamps in this work.

\begin{figure}
    \centering
    \includegraphics[width=0.495\textwidth]{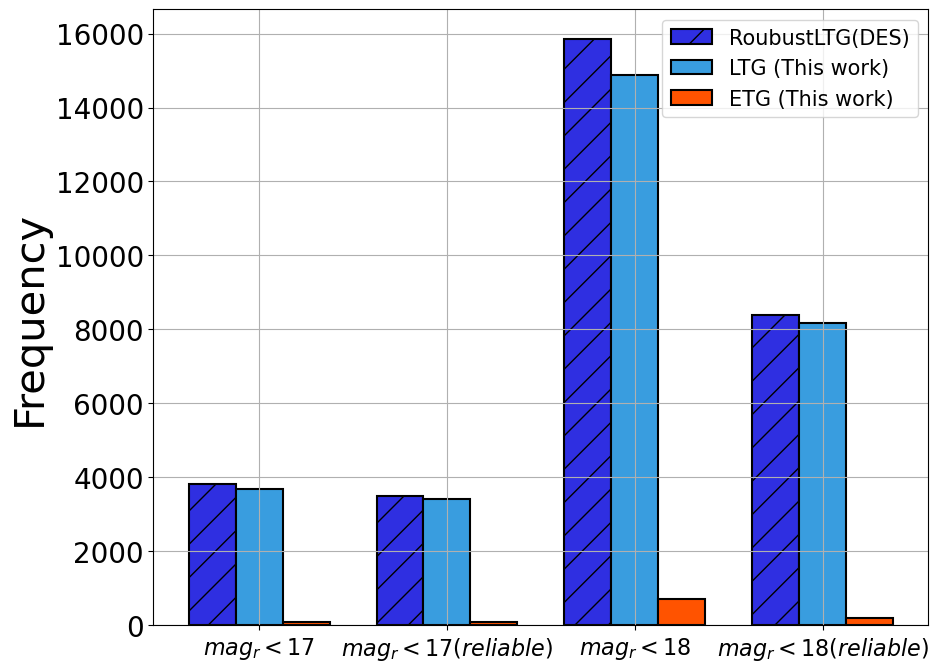}\\
    \includegraphics[width=0.495\textwidth]{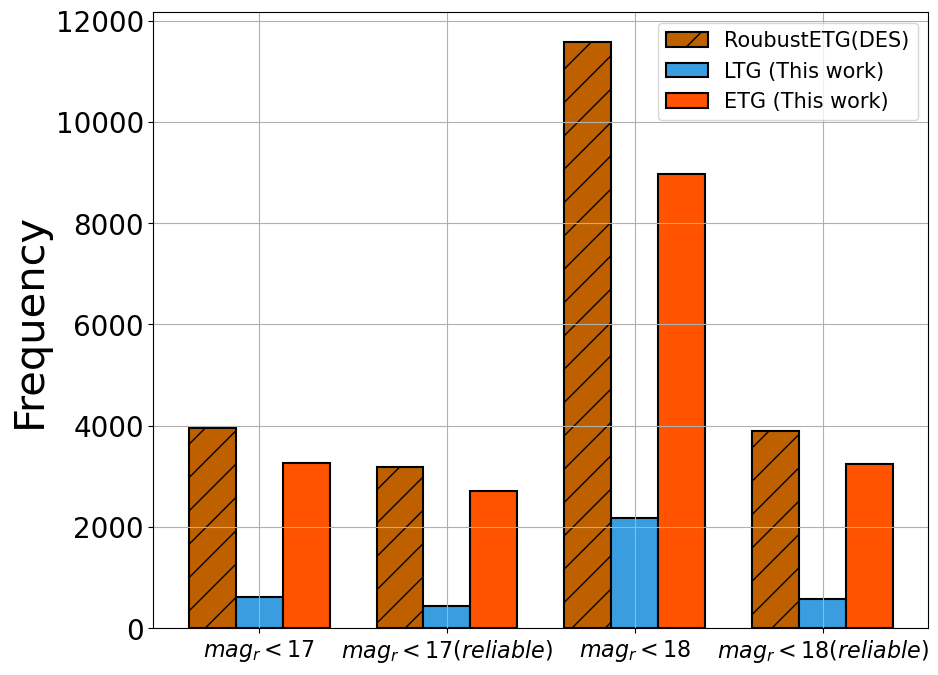}    
    \caption{{\it Top:} histogram comparing the number of galaxies classified as robust late type in \citet{Vega-Ferrero2021} presented in dark blue and their respective classification in this work, light blue for late types and orange for early types. {\it Bottom}: as in the upper panel, but for robust early type galaxies. The comparison includes only reliable stamps.}    
    \label{fig:mismatch_2}
\end{figure}

Figure \ref{fig:mismatch_2} (top panel) present a comparison of objects classified as robust LTG in \cite{Vega-Ferrero2021}, dark blue histogram, for different magnitudes bins and for reliable and not reliable stamps. The light blue and orange histograms classify those objects as late and early-type galaxies respectively, as obtained in this work. Figure \ref{fig:mismatch_2} present the same comparison for objects classified as robust ETG in \cite{Vega-Ferrero2021}.
The mismatch (number of objects classified differently by the two methods) is lower for  reliable stamps in both magnitude bins and for both classifications (early and late types). Since in this work the probability of being early and late-type galaxies is obtained independently and does not always sum up to one, it is possible to notice that the mismatch is slightly higher for ETG, i.e. objects classified as early types in \cite{Vega-Ferrero2021} and as late types in this work. Implications from these findings are discussed in Section \ref{sec:discussion}.

\begin{figure*}
    \centering
    \includegraphics[width=0.8\textwidth]{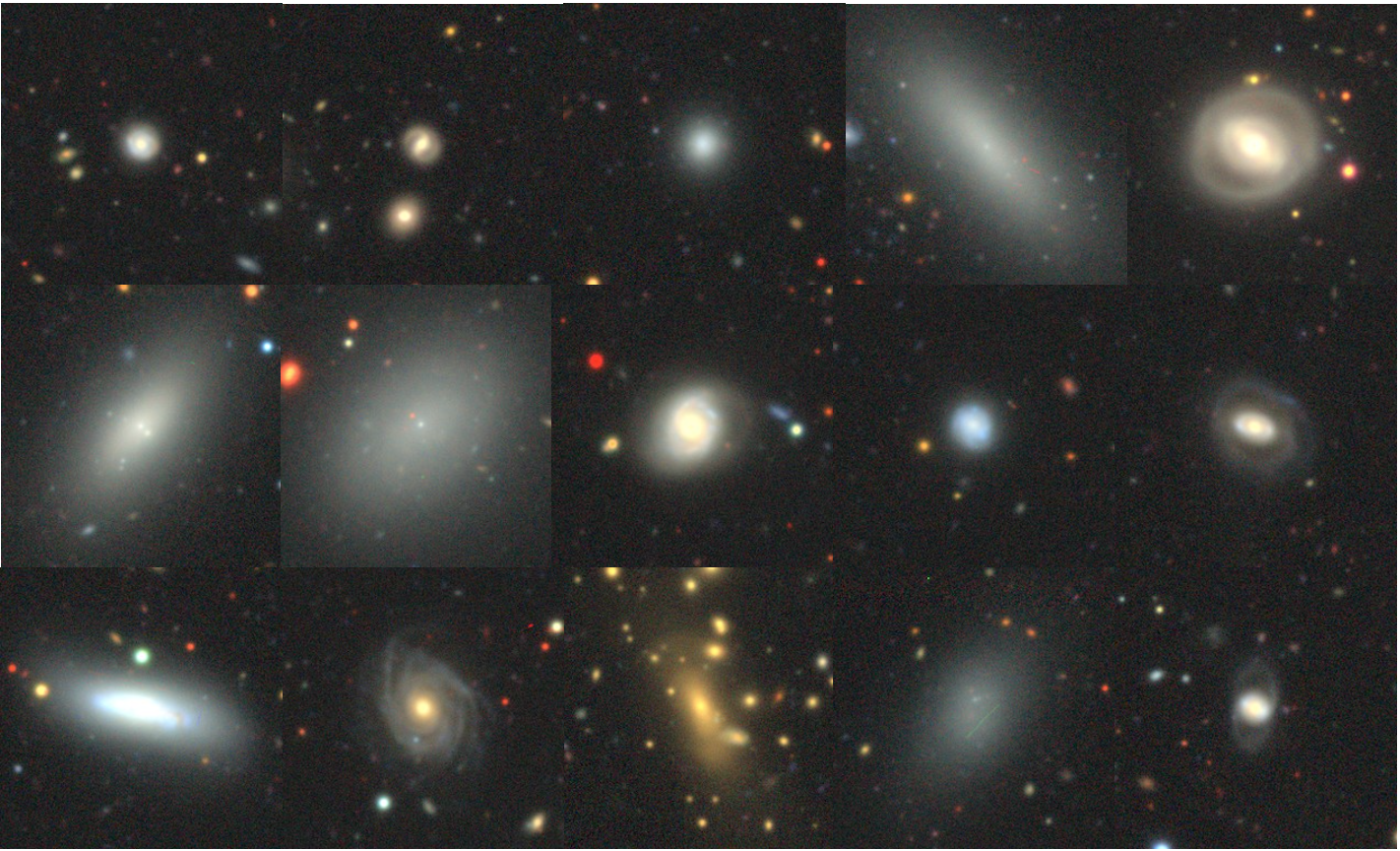}\\
   
    \caption{Examples of images from Legacy Survey of galaxies classified as early types in this work and as late types in \citet{Vega-Ferrero2021}.}

    \label{fig:comp_des}
    
\end{figure*}

\begin{figure*}
    \centering
    \includegraphics[width=0.8\textwidth]{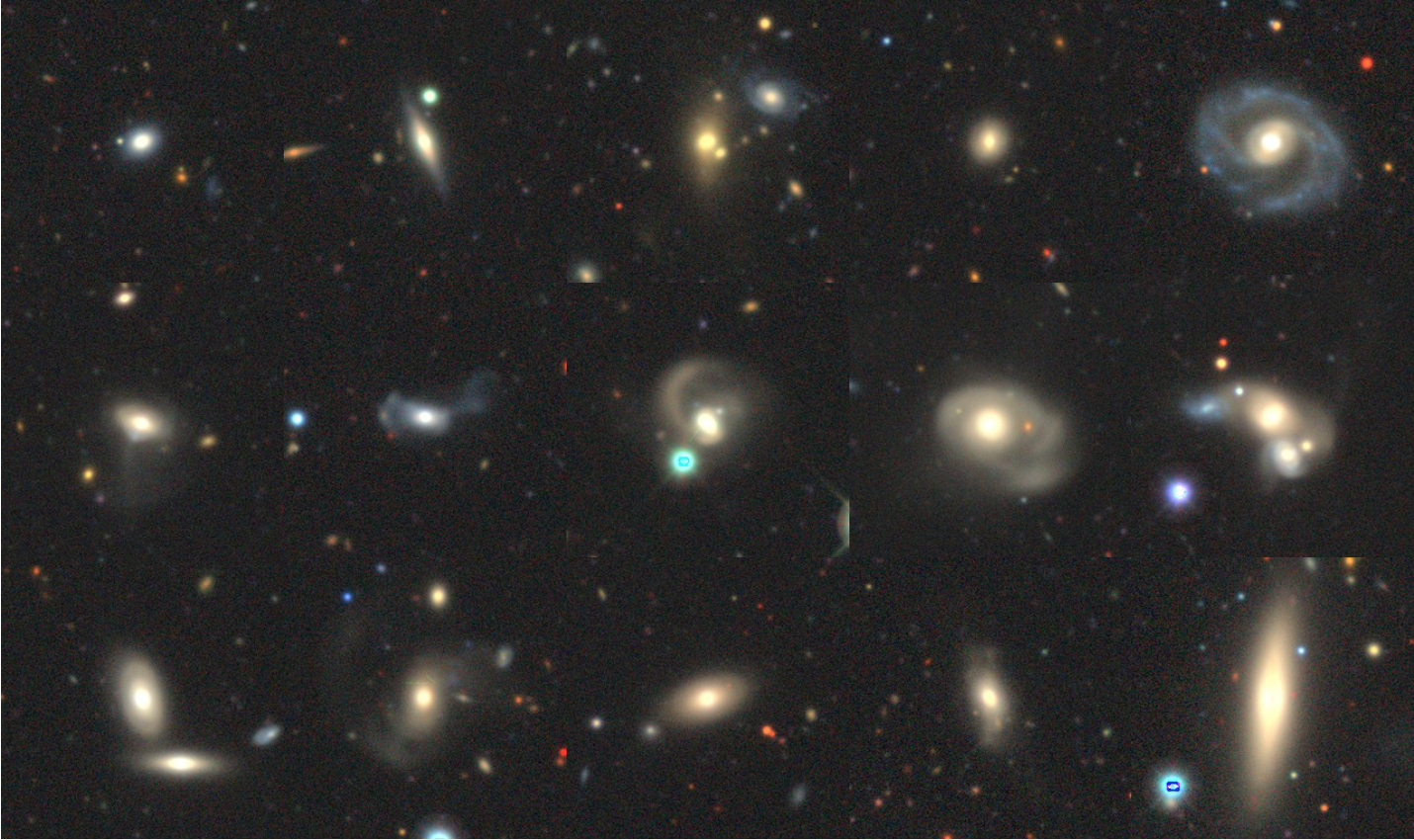}\\
     
    \caption{Examples of images of galaxies from Legacy Survey classified as late types in this work and as early types in \citet{Vega-Ferrero2021}.}

    \label{fig:comp_des_S}
    
\end{figure*}

Figures \ref{fig:comp_des} and \ref{fig:comp_des_S} show examples of objects classified as reliable ETG in this work and as robust late types in \cite{Vega-Ferrero2021}, and the opposite case, respectively.
As expected, some spiral galaxies might be classified as ETG as seen through the T80 telescope, due to the lack of resolution (especially for small objects) and depth (for faint outer spiral arms) in comparison with DES data. Another interesting case of  mismatch is low surface brightness galaxies and merging systems.

For the opposite case, for objects classified as late types in this  work and early types in \cite{Vega-Ferrero2021}, the majority of cases are edge-on objects, red spiral galaxies \citep{bamford2009galaxy, Sodre2013}, and disk
 lenticular galaxies. Several cases of  disturbed, merging systems are also present.

\section{Catalogue description}

The full catalogue was made applying the selection criteria established at the beginning of subsection 2.1.1 in S-PLUS Data Release 3. On the other hand, the catalogue separated for the training contained galaxies with Petrosian magnitude up to 17, thus we separated the full catalog into two groups. The first one is the DR3-Blind containing 46763 objects with $r_{petro}<17$ and the second is the DR3-Extended that incorporates all the 161635 objects separated by the selection criteria with $r_{petro}<18$. 

The catalogue has 47 columns in total from which ID, RA, DEC can be used to identify the correspondent objects in DR3. In case any other column is desired one can use these columns to make a match using the Splus Cloud Website. It also contains information about 5 broad bands $\{u,g,r,i,z\}$ together with the photometric redshift which is also available in S-PLUS data. 

Concerning the classification of those objects we count with 9 columns containing important information from the model calculations. We have 4 columns with the probability estimated by the LTG/ETG Model and the Reliable Stamp Model, together with 4 additional columns with the correspondent binary probability for those columns. The binary probability is given in terms of the threshold selected, so if the value in column \textit{ProbLTG} is greater than $0.6001742$ the column \textit{ProbLTG\_bin} will be equal to 1 and so on. With the binary probability we made the column \textit{Classification} that separates the objects into the classes LTG, ETG, AMB1 and AMB0 discussed through this work. See table \ref{tab:catalog_description} for more details concerning the selections made in the probabilities for the classification. It's worth mentioning that the \textit{Classification} column is made only using the LTG/ETG binary probabilities, so it has reliable and not reliable stamps in it. If only the reliable stamps are desired they can be easily selected using the \textit{ProbRel\_bin} column.

\begin{table*}
\small
\caption{\label{tab:catalog_description} Selections in the probabilities for the classification}

\begin{tabular}{llc}
\hline\hline
Column & Description & Selections  \\

\hline 
ProbNRel & model probability of being a not reliable stamp & -
\\
ProbRel & model probability of being a reliable stamp & -
\\
ProbLTG & model probability of being a Late-type Galaxy & -
\\
ProbETG & model probability of being an Early-type Galaxy & -

\\
\hline

ProbNRel\_bin & binary probability of being a Not Reliable stamp given in terms of the threshold &  
\begin{tabular}{cc}
1 \textbf{if} ProbNRel > $0.53675460815429$
\\
0 \textbf{if} ProbNRel < $0.53675460815429$
\end{tabular}
\\
\\
ProbRel\_bin & binary probability of being a Reliable stamp given in terms of the threshold &  
\begin{tabular}{cc}
1 \textbf{if} ProbRel > $0.53675460815429$
\\
0 \textbf{if} ProbRel < $0.53675460815429$
\end{tabular}

\\
\\ 

ProbLTG\_bin & binary probability of being a Late-type galaxy given in terms of the threshold &  
\begin{tabular}{cc}
1 \textbf{if} ProbLTG > $0.60017424821853$
\\
0 \textbf{if} ProbLTG < $0.60017424821853$
\end{tabular}
\\
\\
ProbETG\_bin & binary probability of being an Early-type galaxy given in terms of the threshold &  
\begin{tabular}{cc}
1 \textbf{if} ProbETG > $0.60017424821853$
\\
0 \textbf{if} ProbETG < $0.60017424821853$
\end{tabular}

\\
\hline

Classification & Classification of the objects based on the binary classification for the LTG/ETG model & 
\begin{tabular}{cccc}
LTG \textbf{if} (ProbLTG\_bin = 1) \textbf{and} (ProbETG\_bin = 0) 
\\
ETG \textbf{if} (ProbLTG\_bin = 0) \textbf{and} (ProbETG\_bin = 1)
\\
AMB1 \textbf{if} (ProbLTG\_bin = 1) \textbf{and} (ProbETG\_bin = 1)
\\
AMB0 \textbf{if} (ProbLTG\_bin = 0) \textbf{and} (ProbETG\_bin = 0)
\end{tabular}
\\
\hline \hline
\end{tabular}
\end{table*}

\end{document}